\documentclass[12pt,letterpaper]{article}
\usepackage[usenames, dvipsnames]{xcolor}
\usepackage{jcapmod}

\usepackage{tocloft}
\usepackage[]{todonotes}
\usepackage{verbatim}
\usepackage{physics}
\usepackage{amsmath}
\usepackage{bbold}
\usepackage{mathrsfs}
\usepackage{cleveref}
\usepackage[shortlabels]{enumitem}
\usepackage{pgf}
\usepackage{tikz-cd}
\usepackage{leftindex}
\usetikzlibrary{shapes,arrows}
\usetikzlibrary{calc}
\usepackage{pgfplots}
\pgfplotsset{compat=1.12}
\usepackage{tikz-3dplot}

\usepackage{amsthm}

\usepackage{tikz}
\usepackage{setspace,caption}
\usepackage{lipsum}
\usetikzlibrary{matrix,arrows,calc}
\makeatletter
\newsavebox\myboxA
\newsavebox\myboxB
\newlength\mylenA

\definecolor{cornellRed}{HTML}{B31B1B}
\definecolor{cornellBlue}{HTML}{0068AC}
\definecolor{cornellGreen}{HTML}{6EB43F}

\usepackage{bbm} 					
\usepackage{slashed} 				
\usepackage{graphicx}				
\usepackage{subcaption}			
\usepackage{psfrag}				
\usepackage{tensor}				
\usepackage{fouridx}				
\usepackage{bm}					
\usepackage{mdframed}				
\usepackage{multirow}				
\usepackage{soul}					
\usepackage{bbold}				
\usepackage{multicol}				
\usepackage{tikz-cd}
\usepackage{rotating}

\usetikzlibrary{arrows}

\tikzset{
commutative diagrams/.cd,
arrow style=tikz,
diagrams={>=latex}}

\usepackage{amsmath}
\usepackage{amssymb}
\usepackage{amsfonts}
\usepackage{mathtools}

\usepackage{feynmf}
\usepackage{marvosym}

\usepackage{import}

\newcommand*\xoverline[2][0.75]{%
    \sbox{\myboxA}{$\m@th#2$}%
    \setbox\myboxB\null
    \ht\myboxB=\ht\myboxA%
    \dp\myboxB=\dp\myboxA%
    \wd\myboxB=#1\wd\myboxA
    \sbox\myboxB{$\m@th\overline{\copy\myboxB}$}
    \setlength\mylenA{\the\wd\myboxA}
    \addtolength\mylenA{-\the\wd\myboxB}%
    \ifdim\wd\myboxB<\wd\myboxA%
       \rlap{\hskip 0.5\mylenA\usebox\myboxB}{\usebox\myboxA}%
    \else
        \hskip -0.5\mylenA\rlap{\usebox\myboxA}{\hskip 0.5\mylenA\usebox\myboxB}%
    \fi}
\makeatother

\definecolor{cobalt}{RGB}{44, 98, 120}
\definecolor{celadon}{rgb}{0.67, 0.88, 0.69}
\definecolor{dm}{cmyk}{.20, 0, .30, 0}
\definecolor{burgundy}{rgb}{0.5, 0.0, 0.13}
\definecolor{plotBlue}{RGB}{94, 130, 181}
\definecolor{bisque}{rgb}{1.0, 0.89, 0.77}

\newcommand{\af}[1]{{\color{purple}\bf [Alex: #1]}}

\hypersetup{
  colorlinks,
  citecolor=Violet,
  linkcolor=cobalt,
  urlcolor=Blue}

\DeclareSymbolFontAlphabet{\mathbb}{AMSb}
\NewDocumentCommand{\xrightarrows}{ O{}O{} }{%
\mathrel{%
\vcenter{\hbox{%
\begin{tikzpicture}
  \node[minimum width=1cm,minimum height=1ex,anchor=south,align=center] (a){\text{\vphantom{hg}#1}\\[0.5ex] \vphantom{hg}#2};
  \draw[<-] ([yshift=0.35ex]a.west) -- ([yshift=0.35ex]a.east);
  \draw[->] ([yshift=-0.35ex]a.west) -- ([yshift=-0.35ex]a.east);
\end{tikzpicture}
}}%
}%
}

\newif\iffastcompile

\fastcompilefalse

\iffastcompile
\newcommand{\mk}[1]{}
\else
\newcommand{\mk}[1]{\todo[color=burgundy!30, size=\scriptsize, bordercolor=burgundy!30]{MK: #1}}
\fi

\iffastcompile
\newcommand{\af}[1]{}
\else

\fi

\iffastcompile
\newcommand{\mc}[1]{}
\else
\newcommand{\mc}[1]{\todo[color=bisque!30, size=\scriptsize, bordercolor=bisque!30]{MC: #1}}
\fi

\ProvideTextCommandDefault{\Dbar}{%
\leavevmode\lower.5ex\rlap{\hskip-.07em\accent"16}D%
}

\usepackage{environ}
\usepackage{changepage}

\begin{document}
	\newcommand{\main}{.}
\begin{titlepage}

\setcounter{page}{1} \baselineskip=15.5pt \thispagestyle{empty}
\setcounter{tocdepth}{2}
\bigskip\

\vspace{1cm}
\begin{center}
{\large \bfseries Non-linear sigma model in string field theory}
\end{center}

\vspace{0.55cm}

\begin{center}
\scalebox{0.95}[0.95]
{{\fontsize{14}{30}\selectfont Alexander Frenkel$^{a,b}$, Manki Kim$^{a}$\vspace{0.25cm}}}

\end{center}

\begin{center}

\vspace{0.15 cm}
{\fontsize{11}{30}
\textsl{$^{a}$Leinweber Institute for Theoretical Physics at Stanford, 382 Via Pueblo, Stanford, CA 94305, USA}}\\
{\fontsize{11}{30}
\textsl{$^{b}$Simons Center for Geometry and Physics, Stony Brook University, Stony Brook, NY 11794, USA}\\
\vspace{0.25cm}}

\vskip .5cm
\end{center}

\vspace{0.8cm}
\noindent

\vspace{1.1cm}
We revisit the non-linear sigma model approach to string theory with the closed superstring field theory. We construct the string field theory around the non-linear sigma model background with the patch-by-patch description. We show that our string field theory action is invariant under the gauge transformation and solves the BV master equation, thereby providing a tool to study quantum gravitational effects in curved backgrounds in small $\alpha'$ and $g_s$ approximation. We illustrate how to use our results to study curved backgrounds in $\alpha'$ expansion by studying Calabi-Yau compactification in $\alpha'$ expansion. We draw connections between Tseytlin's approach to the non-linear sigma model and the string field theoretic approach. We comment on future directions.

\vspace{3.1cm}

\noindent\today

\end{titlepage}
\tableofcontents\newpage

\section{Introduction}
String theory provides powerful machinery for studying well-defined quantum gravitational amplitudes in simple backgrounds, such as 10-dimensional spacetime with supersymmetry. However, it remains poorly understood how to apply this machinery, even perturbatively, even at tree level, to general curved backgrounds. This state of affairs stands in stark contrast to the semiclassical gravitational path integral, where a complicated curved background can be broken down into well-understood, nearly flat local patches from which global effects may be resummed. Our goal in this paper is to demonstrate how this same patch-by-patch approach in semiclassical quantum gravity can be applied to string theory, at least to all orders in $\alpha'$ and $g_s$.

In the context of the RNS formulation of string perturbation theory, the lack of tools to study generic curved backgrounds can be traced back to the scarcity of known exact 2d conformal field theories (CFTs). In particular, it is not well understood how to construct CFTs without the help of high degrees of symmetry (e.g. supersymmetry or a current algebra). However, it should be noted that often the most physically interesting phenomena are not formed in a highly symmetric setting, nor is it guaranteed that the worldsheet theory will always be local and unitary, creating a barrier to accessing such phenomena using currently known CFT technology. More broadly, attempting to understand physical phenomena only through exact tools is severely self-limiting. We are therefore motivated to build beyond this exact approach, and look for tools that can function as controlled approximations to exact string backgrounds.

One proposed approach to string theory whose formulation does not a priori require exact CFTs is the non-linear sigma model (NLSM) \cite{Callan:1989nz}. The NLSM approach is pleasantly intuitive -- in order to study nontrivial backgrounds in the string theory target space, one directly deforms target space by introducing a non-trivial metric $G_{\mu \nu}(X)$ and dilaton profile $\Phi(X)$, which become field-dependent couplings on the worldsheet. The challenge, initially addressed in \cite{Tseytlin:1988rr,Tseytlin:1988tv,Tseytlin:2000mt} and recently reviewed in \cite{Ahmadain:2022eso,Ahmadain:2022tew}, is finding a prescription for dealing with off-shell target space configurations corresponding to non-CFT NLSMs on the worldsheet. These prescriptions suffer from limitations at the moment (we review some of them in \S\ref{ssec:Tseytlin-Review}). 

In light of this, we would like to revisit the non-linear sigma model (NLSM) in string field theory (SFT). Closed string field theory provides a very general framework for considering off-shell deformations on the worldsheet, valid to arbitrary order in $g_s$ and $\alpha'$, with no requirement that deformations of the worldsheet theory need to be unitary or normalizable. To test this proposed technology, our goal in this work is to demonstrate its use in constructing the worldsheet data for strings probing curved backgrounds perturbatively in $\alpha'$ and $g_s$. 

Despite the fact that the non-linear sigma model is a decades-old subject that has received much attention, there are a number of good reasons to revisit the subject with the newly constructed superstring field theory \cite{deLacroix:2017lif,Erler:2019loq,Sen:2024nfd}. String field theory provides a natural tool to beyond the limitations of the currently understood non-linear sigma model approach, as it can study D-instanton amplitudes \cite{Sen:2020cef,Sen:2020eck,Sen:2021qdk,Sen:2021tpp,Eniceicu:2022nay,Chakravarty:2022cgj,Eniceicu:2022xvk,Alexandrov:2021shf,Alexandrov:2021dyl,Alexandrov:2022mmy,Agmon:2022vdj,Chakrabhavi:2024szk}, loop amplitudes involving states that undergo renormalization \cite{Pius:2013sca,Pius:2014iaa}, Ramond-Ramond backgrounds \cite{Cho:2018nfn,Cho:2023mhw,Kim:2024dnw,Cho:2025coy}, and boundary terms of the action \cite{Stettinger:2024uus,Firat:2024kxq,Maccaferri:2025orz,Maccaferri:2025onc}\footnote{For the non-linear sigma model approach on the boundary terms, see \cite{Ahmadain:2024hgd,Ahmadain:2024uyo}.}, all of which are currently difficult to study in the textbook formulation of string perturbation theory. Furthermore, string field theory provides a systematic procedure to compute higher-order amplitudes. For example, such capabilities to compute higher-order amplitudes are crucial to better understand the flux vacua \cite{McAllister:2023vgy,McAllister:2024lnt}.\footnote{For recent progress on computing higher-order terms in the effective action in closed superstring, see \cite{Liu:2022bfg,Liu:2025uqu}.}

Formulating the non-linear sigma model in string field theory requires some thought. As currently formulated, string field theory requires a well-defined string background as input data, whose worldsheet CFT description is known. However, the lack of such CFT descriptions for general curved backgrounds was precisely the motivation for us to study the non-linear sigma model. 

To overcome this difficulty, we will combine the idea of local background independence of string field theory \cite{Sen:1993kb}, and the fact that a target space without singularities is locally flat. As the metric of any locally flat manifold enjoys the normal coordinate expansion
\begin{equation}
    g_{\mu\nu}(x)=\eta_{\mu\nu}-\frac{1}{3} R_{\nu\rho\nu\sigma}(0) x^\rho x^\sigma+\dots\,,
\end{equation}
one can therefore treat the local coordinate patch with the non-trivial metric as a non-trivial perturbative background solution of string field theory constructed from the free field CFT. Once the local form of the perturbative background solution is obtained, we can perturb around the non-trivial background solution to quantize states and compute quantum gravitational effects in string field theory.\footnote{For recent applications of string field theory to study conformal perturbation theory, see \cite{Scheinpflug:2023osi,Mazel:2024alu}.} 

With the local background solution that we construct, we are now positioned to study whether the local solution can be extended globally, given the topology of the target space that we input. In fact, as we will find, this problem of extending the solution globally is rather non-trivial. That is because the perturbative background solution has a non-trivial profile on the boundary of the local coordinate chart, and the local coordinate chart has a finite size. As the recent activities to understand the boundary action of string field theory suggest, when such a boundary exists, the gauge invariance and hence the solvability of the BV master equation break down. One way to fix the problem is to introduce the boundary action to cancel the breakdown of the gauge invariance , but we will not do so here, as our goal is to study the target space that does not necessarily have a boundary.

To restore the BRST invariance of the action and the solvability of the BV master equation, we shall use a very particular choice of partition of the unity of the target space. We shall first find a collection of open sets $U_i$ whose union comprises the target space $\cup_i U_i=\mathcal{M}.$ In each of $U_i,$ we can define local string field $\Psi^{(i)}$ with the local form of the action $S^{(i)}(\Psi^{(i)}).$ When $U_i\cap U_j\neq\phi,$ there is an ambiguity in defining what we mean by the string field $\Psi.$ We shall declare that $\Psi^{(i)}$ and $\Psi^{(j)}$ are related by gauge transformation. As one can see, the overlapping regions $U_i\cap U_j$ for $i\neq j$ are double-counted. To subtract the double-counted contributions, we can define a local string field $\Psi^{(i\cap j)}$ in $U_i\cap U_j$ which is an appropriate interpolation of $\Psi^{(i)}$ and  $\Psi^{(j)}.$ With $\Psi^{(i\cap j)}$ we can define the local form of the action $S^{(i\cap j)}$ which we can use to subtract the double-counted contributions. However, by doing so, we have subtracted too much, and need to add back contributions from $U_i\cap U_j\cap U_k$ for $i\neq j,~i\neq k,~j\neq k.$ By repeating this procedure, we find the string field theory action that is globally well defined
\begin{equation}
    S(\Psi)= \sum_i S^{(i)}(\Psi^{(i)})-\sum_{i\neq j} S^{(i\cap j)} (\Psi^{(i\cap j)})+\dots\,,
\end{equation}
which is both BRST invariant and solves the BV master equation as we show in \S\ref{sec:patch-by-patch}.

One cautionary remark is in order. Zero mass level string fields are related to the supergravity field by a non-trivial field redefinition \cite{Hull:2009mi,Mazel:2025fxj,Mamade:2025jbs,Mamade:2025htb}. Therefore, in finding the desired background solution in string field theory, one must carefully identify the characteristics of the solutions. One way is to decompose the string field into a representation of the diffeomorphism, as was shown in \cite{Mazel:2025fxj}. A different, but equivalent, way is to use various physical probes such as D-branes. We shall use the latter in this paper. 

With the well-defined global construction of the non-linear sigma model in string field theory, one can study curved string backgrounds order by order in $\alpha'$ and $g_s$ expansion. A crucial distinction from the low-energy supergravity approach is that we can compute quantum gravitational effects directly within string theory to build up higher-order solutions both in $\alpha'$ and $g_s$. A few notable interesting applications include: computations of $g_s$ and $\alpha'$ corrections in Calabi-Yau orientifold compactifications with or without Ramond-Ramond fluxes, formulation of string perturbation theory in cosmological backgrounds, perturbative study of black hole backgrounds in string theory. One main limitation of our work is that the topology and metric of the target space are required as input data. 

This work is organized as follows. In \S\ref{sec:review}, we set the conventions for the worldsheet CFT and string field theory. We then review string field theory and the BV quantization. In \S\ref{sec:patch-by-patch}, we construct the string field theoretic description of the non-linear sigma model by employing the patch-by-patch description. We show that our construction is well defined in the sense that the string field theory action we construct is invariant under the gauge transformation and solves the BV master equation. We also explain how to use the patch-by-patch description to quantize states and compute physical observables. In \S\ref{sec:example 1}, we use a circle compactification to illustrate the central ideas of the patch-by-patch description. In \S\ref{sec:example 2}, we study Calabi-Yau compactifications in the patch-by-patch description. We construct the background solution to the second order in $\alpha'$ expansion, and explain how to extend the solution beyond the second order. With the background solution, we find the vertex operator for the moduli fields in $\alpha'$ expansion. In \S\ref{sec:Wey and local coords}, we compare the SFT approach to NLSM we develop in this paper to the traditional RG-based approach, developed first by Tseytlin in \cite{Tseytlin:1988rr, Tseytlin:2000mt} and recently reviewed and extended in \cite{Ahmadain:2022tew, Ahmadain:2022eso, Ahmadain:2024hdp}. We re-interpret the RG approach in the context of SFT, allowing us to extend the regime of validity at the cost of additional computational complexity. In \S\ref{sec:conclusions}, we conclude with future directions.

Note that the work of \cite{Mazel:2025fxj} has a significant overlap with our work. 

\section{Review}\label{sec:review}
In this section, we shall summarize our conventions for the worldsheet CFT, review the basic elements of string field theory. For more comprehensive reviews on string field theory, see \cite{Sen:2024nfd,deLacroix:2017lif,Erbin:2021smf,Erler:2019loq,Erler:2019vhl}.

\subsection{Worldsheet convention}
As we shall study the non-linear sigma model in $\alpha'$ expansion, the leading order data will be provided by the free field CFT. As we shall cover the target space with an at tlas $\{U, \varphi\},$ for each open set $U_i$ we need a free field CFT, which we will denote by $\text{CFT}_i.$ For the matter field CFT, we have $(\mathcal{N},\overline{\mathcal{N}})=(1,1)$ supersymmetric free-field CFT with the central charge $(c,\bar{c})=(15,15).$ The ghost CFT contains $b,~c,~\beta,~\gamma,$ system. In the matter sector, we have 10 bosons $X^A,$ 10 left-moving fermions $\psi^A,$ and 10 right-moving fermions $\bar{\psi}^A.$ We write OPEs of the matter fields as
\begin{equation}
    X^A(x)X^B(0)\sim -\frac{\alpha'}{2} \eta^{AB}\log|z|^2\,,\quad \psi^A(z)\psi^B(0)\sim \frac{\eta^{AB}}{z}\,,
\end{equation}
\begin{equation}
    \bar{\psi}^A(\bar{z})\bar{\psi}^B(0)\sim\frac{\eta^{AB}}{\bar{z}}\,.
\end{equation}
We shall bosonize the $\beta,$ $\gamma$ ghosts as
\begin{equation}
    \beta=\partial\xi e^{-\phi}\,,\quad \gamma=\eta e^\phi\,.
\end{equation}
The OPEs of the ghost fields are given as
\begin{align}
    &c(z)b(0)\sim \frac{1}{z}\,,\quad \bar{c}(\bar{z})\bar{b}(0)\sim\frac{1}{\bar{z}}\,,\quad \xi(z)\eta(0)\sim\frac{1}{z}\,,\\
    &\partial\phi(z)\partial\phi(0)\sim -\frac{1}{z^2}\,,\quad e^{q_1\phi}(z)e^{q_2\phi}(0)\sim z^{-q_1q_2}e^{(q_1+q_2)\phi}(0)\,.
\end{align}

We shall now introduce spin fields. We denote the ten-dimensional chiral spin field by $\Sigma_\alpha$ and the anti-chiral spin field by $\Sigma^\alpha.$ We shall choose the GSO projection such that the following fields are projected in
\begin{equation}
    e^{-\phi/2}\Sigma_\alpha\,,\quad e^{-3\phi/2}\Sigma^\alpha\,.
\end{equation}

We normalize the ghost correlator as
\begin{equation}
    \langle 0| c_{-1} \bar{c}_{-1} c_0\bar{c}_0 c_1\bar{c}_1 e^{-2\phi} e^{-2\bar{\phi}}|0\rangle_i=- \int_{U_i} d^{10}X\,.
\end{equation}
where $U_i$ is an open coordinate patch. In bosonic string theory, we shall normalize the ghost correlator as
\begin{equation}
    \langle 0| c_{-1}\bar{c}_{-1} c_0\bar{c}_0 c_1\bar{c}_1 |0\rangle_i = \int_{U_i} d^{26}X\,.
\end{equation}

We write the BRST current as
\begin{equation}
    j_B=c\left( T_m -\frac{1}{2} (\partial\phi)^2 -\partial^2\phi-\eta\partial\xi \right) +\eta e^\phi T_F +bc\partial c -\eta \partial\eta be^{2\phi}\,,
\end{equation}
\begin{equation}
    \bar{j}_B=\bar{c} \bar{T}_m +\bar{b}\bar{c}\bar{\partial}\bar{c}\,,
\end{equation}
where 
\begin{equation}
    T_m=-\frac{1}{\alpha'}\partial X^A \partial X_A-\frac{1}{2}\psi_A\partial \psi^A\,,
\end{equation}
and
\begin{equation}
    T_F=i\sqrt{\frac{2}{\alpha'}}\psi^A\partial X_A\,.
\end{equation}
We write the BRST charge as
\begin{equation}
    Q_B:= \oint dz j_B+\oint d\bar{z} \bar{j}_B\,.
\end{equation}
We define the picture changing operator (PCO) as 
\begin{equation}
    \mathcal{X}:= \{ Q_B,\xi\}=c\partial \xi +e^\phi T_F-\partial \eta be^{2\phi}-\partial(\eta be^{2\phi})\,.
\end{equation}

\subsection{Lightning review of string field theory}
In this section, we shall review and collect the conventions for closed string field theories. As this draft mostly concerns the tree-level amplitudes, for practical calculations we will perform in the draft, the distinction between 1PI action and the BV master action is not present. However, we shall nevertheless review the conventions for the BV master action, as we will show that the non-linear sigma model in string field theory is BV quantizable. For review on string field theory, see, for example, \cite{deLacroix:2017lif,Erbin:2021smf,Erler:2019loq,Sen:2024nfd}.

In order to formulate string field theory, with the currently limited formulation of string field theory, one must start with a well-defined worldsheet CFT background. In general, the worldsheet CFT for the target space with non-trivial spacetime curvature is not well understood. Hence, a suitable approximation scheme is required to understand the underlying worldsheet theory. 

As we will study a class of string backgrounds in which a good notion of $\alpha'$ expansion exists, each open local coordinate chart $U_i$ of the target space can be well approximated as a deformation of free-field CFT by $\alpha'$ corrections. Therefore, we shall proceed as follows. We shall construct, for each local coordinate patch $U_i,$ a string field theory $\text{SFT}_i$ starting with the free field CFT for the local coordinate patch $U_i.$ With the string field theory description, we can now then find a non-trivial background solution in $\alpha'$ expansion. The non-trivial background solution $\Psi_{0}^{(i)}$ of $\text{SFT}_i$ then describes the worldsheet CFT for the local coordinate patch $U_i.$ As the worldsheet CFT is supposed to describe the whole spacetime, one must patch together $\text{SFT}_i$ and $\Psi_0^{(i)}.$ Provided that there exists well defined transition maps between $\text{SFT}_i$ and $\Psi^{(i)}_0,$ we can denote $\text{SFT}_i$ and $\Psi^{(i)}_0$ by $\text{SFT}|_{U_i}$ and $\Psi_0|_{U_i}.$ Once we found a globally well-defined background solution $\Psi_0,$ we can then formulate string field theory around the non-linear sigma model background, such that the string field $\delta\Psi$ formulated in the non-linear sigma model is now given as $\Psi_0+\delta\Psi$ of the original string field theory formulated with the collection of free field CFTs.  

As such, our description of string field theory around the non-linear sigma model background will be based on the string field theory formulated with the free field CFT for the local patch $U_i.$ So, we shall, in this section, spell out the conventions we shall use for $\text{SFT}_i.$ 

First, we shall start by defining the state spaces. Although, in principle, the state space for $\text{SFT}_i$ is different from that of $\text{SFT}_j,$ as all of the $\text{SFT}_i$ are constructed with the free field CFT, the state spaces are isomorphic to each other. Hence, to ease the notational complexity, we shall use the same symbols for the state spaces. We shall denote the Hilbert space of GSO even states in the small Hilbert space of the worksheet CFT with picture number $p$ by $\mathcal{H}_p\,,$ with its states $|s\rangle \in\mathcal{H}_p$ satisfying the following constraints
\begin{equation}
 b_0^-|s\rangle=L_0^-|s\rangle=0\,.
\end{equation}

To formulate heterotic string field theory, we shall use two sets of state spaces
\begin{equation}
    \hat{\mathcal{H}}:= \mathcal{H}_{-1}\oplus \mathcal{H}_{-1/2}\,,\quad \tilde{\mathcal{H}}:=\mathcal{H}_{-1}\oplus\mathcal{H}_{-3/2}\,.
\end{equation}
Similarly, to formulate closed type II string field theory, we shall use the following two sets of state spaces
\begin{equation}
    \hat{\mathcal{H}}:=\mathcal{H}_{-1,-1}\oplus \mathcal{H}_{-1,-1/2}\oplus \mathcal{H}_{-1/2,-1}\oplus \mathcal{H}_{-1/2,-1/2}\,,
\end{equation}
\begin{equation}
    \tilde{\mathcal{H}}:=\mathcal{H}_{-1,-1}\oplus\mathcal{H}_{-1,-3/2}\oplus\mathcal{H}_{-3/2,-1}\oplus\mathcal{H}_{-3/2,-3/2}\,.
\end{equation}
For the formulation of closed bosonic string field theory, as there is no picture number and Ramond states, we only need one ordinary set of state space. We shall denote states in $\hat{\mathcal{H}}$ by $\Psi^{(i)},$ and states in $\tilde{\mathcal{H}}$ by $\tilde{\Psi}^{(i)}.$

With the off-shell states in the state space, one can compute off-shell amplitudes in string perturbation theory. To construct the off-shell action, we shall divide the off-shell amplitudes that are computed with fundamental interaction vertices and the rest of the contributions that are computed with the Feynman diagrams with propagators. 

To gain intuition, let us first review on-shell amplitudes in string perturbation theory. On-shell amplitudes in string perturbation theory are computed by integrating CFT correlators over the moduli space of punctured Riemann surfaces. And, if constructing the off-shell action is possible, in the sense of identifying fundamental vertices, it should also be possible to divide the contributions to the on-shell amplitudes into the fundamental vertex contributions and the rest.  Therefore, the relevant question is then to identify which region of the moduli space corresponds to the fundamental vertices and the non-trivial Feynman diagrams. 

The main idea is that there are special points of the moduli space of punctured Riemann surfaces that describe degenerations of Riemann surfaces. There are two types of degeneration that can happen. First, a punctured Riemann surface degenerates into two punctured Riemann surfaces glued by joining two punctures, one from each Riemann surface. The other type of degeneration forces a punctured Riemann surface with a Riemann surface with one less genus and two more punctures where the additional punctures are glued. As one can see, both of the degenerations can be understood as non-trivial Feynman diagrams. We shall, therefore, excise small regions around such degeneration points in the moduli space and declare that the rest of the moduli space contributes to the fundamental vertex. Note that this excision procedure can be understood as a hard cutoff UV regulator in the worldsheet computation, where the size of the hard cutoffs, in general, depends on moduli. 

We are now ready to get back to the construction of the off-shell action. Off-shell amplitudes in string theory are computed with off-shell states which are not in general weight-zero conformal primaries. Therefore, to compute the off-shell amplitudes, we also need to specify the metric of the punctured Riemann surfaces, especially around punctures. This choice of metric can be conveniently phrased in terms of local coordinates around punctures. In addition to the metric, in the formulation of super string field theory based on the RNS formalism with PCOs, the off-shell amplitudes will also depend on the locations of PCOs. Therefore, we shall also specify the locations of PCOs. Lastly, to formulate a consistent string field theory that can be quantized in the BV formalism, not all choices of local coordinates and the PCO locations are allowed and one must impose boundary conditions. In particular, requiring that the local coordinates and the PCO locations on the boundary of the vertex regions should agree with those of the boundary of the Feynman region guarantees that the BV master equation is solved. Hence, the string field theory can be quantized in the path integral formulation. 

We shall now define precisely off-shell amplitudes. Let us denote the moduli space of a Riemann surface of genus g with $n$ NS punctures and $m$ R punctures by $\mathcal{M}_{g,n,m}.$ At a given point $q\in \mathcal{M}_{g,n,m},$ we can divide the punctured Riemann surface into $n+m$ disks $\{D_a\}$, one around each puncture, and $2g-2+n+m$ spheres $\{S_i\}$ with three holes that are glued along $3g-3+2n+2m$ circles $\{C_s\}$ \cite{deLacroix:2017lif}. We shall denote the local coordinates on $D_a$ by $w_a,$ and the local coordinates on $S_i$ by $z_i.$ The topology of the punctured Riemann surface at $m$ is then determined by transition functions between the local coordinates
\begin{equation}
    z_i=f_{ij}(z_j)\,,
\end{equation}
and
\begin{equation}
    z_i =g_{ia}(w_a)\,.
\end{equation}
Following \cite{deLacroix:2017lif}, we shall collectively denote the transition maps by
\begin{equation}
    \sigma_s =F_s(\tau_s)\,,
\end{equation}
for every $C_s,$ where $\sigma_s$ and $\tau_s$ denote the local coordinates on the left and right of $C_s.$ So far, we have not specified differential structure on the Riemann surface. To do so, we can declare, for example, that the transition map $g_{ia}(w_a)$ is inverse of the local coordinate map where viewing $z_i$ as a global coordinate such that the coordinate map maps a local coordinate patch $D_a$ to a unit disk with $|w_a|\leq1$. With the coordinate map, we can declare that the metric of the unit disk, for example, is 
\begin{equation}
    ds^2= |dw|^2\,.
\end{equation}
This also fixes the metric of the punctured Riemann surface around the punctures. 

As one can easily convince oneself, there are infinitely many distinct choices for the local coordinates and the metric. We can conveniently denote such choices collectively by a fiber bundle $\hat{P}_{g,n,m}\rightarrow \mathcal{M}_{g,n,m}\,,$ where each fiber is a space of local coordinates modulo the phase ambiguity. We shall then construct a fiber bundle $\tilde{P}_{g,n,m}\rightarrow \hat{P}_{g,n,m},$ whose fiber is a space of PCO locations. The off-shell amplitudes will then be constructed as integrals of $6g-6+2n+2m$ form over a section of $\hat{P}_{g,n,m}.$ To construct the p-form $\Omega_p^{(g,n,m)},$ we shall first define the Beltrami differential
\begin{equation}
    \mathcal{B}\left[\frac{\partial}{\partial t_i}\right]:= \sum_s\biggr[ \oint_{C_s} \frac{\partial F_s}{\partial t_i} d\sigma_s b(\sigma_s) +\oint_{C_s} \frac{\partial\bar{F}_s}{\partial t_i} d\bar{\sigma}_s \bar{b}(\bar{\sigma}_s) \biggr]-\sum_\alpha \frac{1}{\mathcal{X}(y_\alpha)} \frac{\partial y_\alpha}{\partial t_i} \partial\xi(y_\alpha)\,,
\end{equation}
where $t_i$ denotes the parameters that label the p dimensional subspace of $\tilde{P}_{g,n},$ and $\partial/\partial y_\alpha\in T_m \tilde{P}_{g,n}$ is a tangent vector projects to a trivial tangent vector in $T_p \hat{P}_{g,n}.$ Such $\partial/\partial y_\alpha$ describes the change of PCO locations. It is important to note that $1/\mathcal{X}$ is a formal expression, and the role of this operator is to remove an insertion of a PCO operator at $y_\alpha$ from the evaluation of a conformal correlator. We define the p-form $\Omega_{p}^{(g,n)}$ as
\begin{equation}
    \Omega_p^{(g,n)} :=\left\langle \mathfrak{B}_1dt_1\wedge \dots \wedge \mathfrak{B}_pdt_p \prod_\alpha \mathcal{X}(y_\alpha) \Psi^{-1}_1\dots\Psi^{-1}_n\otimes \Psi^{-\frac{1}{2}}_1\dots\Psi^{-\frac{1}{2}}_m \right\rangle_{\Sigma_{g,n}}\,,
\end{equation}
where the number of PCO insertions is determined to be
\begin{equation}
    2g-2+n+m/2\,.
\end{equation}
A few comments are in order. First, we assumed that all string fields are Grassmann even. This can be achieved by multiplying Grassmann odd c-numbers to a Grassman odd vertex operator. For type II string theories, we must also insert an appropriate number of anti-holomorphic PCOs. Also, it is important to note that we used states in $\hat{\mathcal{H}}$ to construct the p-form. Said differently, we shall treat the fields in $\tilde{\mathcal{H}}$ as essentially free fields.

Let us denote by $\overline{R}_{g,n,m}$ the section segments of $\tilde{P}_{g,n,m}$ corresponding to a fundamental interaction vertex of n NS fields and m R fields at genus g. As we previously explained, not every choice of $\overline{R}_{g,n,m}$ is allowed. To formulate the constraints, we shall define a few operations. Let us suppose that we have two section segments $\overline{R}_{g_1,n_1,m_1}$ and $\overline{R}_{g_2,n_2,m_2}.$ We can create a new section segment by gluing either NS punctures or R punctures. We shall denote a new section segment obtained by gluing two NS punctures, one from each section segment, by
\begin{equation}
    \overline{R}_{g_1,n_1,m_1} \circ\overline{R}_{g_2,n_2,m_2}\,.
\end{equation}
Similarly, we denote a new section segment created by gluing two R punctures, one from each segment, by
\begin{equation}
    \overline{R}_{g_1,n_1,m_1} \star\overline{R}_{g_2,n_2,m_2}\,.
\end{equation}
There is one more way to create a section segment by joining two punctures of a single segment. We denote this operation by
\begin{equation}
    \nabla_{NS} \overline{R}_{g,n,m}\,,
\end{equation}
and
\begin{equation}
    \nabla_R \overline{R}_{g,n,m}\,.
\end{equation}
As all of the new section segments are obtained by the plumbing fixture, to obtain the boundaries of the section segments, we can set $s=0.$ We shall denote such boundaries by
\begin{equation}
    \{\overline{R}_{g_1,n_1,m_1},\overline{R}_{g_2,n_2,m_2}\}:=\overline{R}_{g_1,n_1,m_1} \circ\overline{R}_{g_2,n_2,m_2}|_{s=0}\,, 
\end{equation}
\begin{equation}
    \{\overline{R}_{g_1,n_1,m_1};\overline{R}_{g_2,n_2,m_2}\}:=\overline{R}_{g_1,n_1,m_1} \star\overline{R}_{g_2,n_2,m_2}|_{s=0}\,, 
\end{equation}
\begin{equation}
    \Delta_{NS}\overline{R}_{g,n,m}:= \nabla_{NS} \overline{R}_{g,n,m}|_{s=0}\,,\quad \Delta_{R}\overline{R}_{g,n,m}:= \nabla_{R} \overline{R}_{g,n,m}|_{s=0}\,.
\end{equation}
We then require that the section segments satisfy the following equation
\begin{align}
\partial \overline{R}_{g,n,m}=&-\frac{1}{2} \sum_{g_1+g_2=g}\sum_{n_1+n_2=n+2}\sum_{m_1+m_2=m} \mathbf{S}\left[ \{\overline{R}_{g_1,n_1,m_1},\overline{R}_{g_2,n_2,m_2}\}\right]  \nonumber\\
&-\frac{1}{2} \sum_{g_1+g_2=g}\sum_{n_1+n_2=n}\sum_{m_1+m_2=m+2} \mathbf{S}\left[ \{\overline{R}_{g_1,n_1,m_1};\overline{R}_{g_2,n_2,m_2}\}\right]  \nonumber\\
&-\Delta_{NS}\overline{R}_{g-1,n+2,m}-\Delta_R\overline{R}_{g-1,n,m+2}\,,
\end{align}
where $\mathbf{S}$ denotes a sum over permutations of NS punctures and R punctures. 

We shall now define a string vertices as
\begin{equation}
    \{\Psi_1^{-1}\dots \Psi_n^{-1}\otimes\Psi_1^{-1/2}\dots\Psi_m^{-1/2}\}_g:= g_s^{2g} \int_{\overline{R}_{g,n,m}} \Omega^{(g,n,m)}_{6g-6+2n+2m} ( \Psi_1^{-1},\dots ,\Psi_n^{-1},\Psi_1^{-1/2},\dots,\Psi_m^{-1/2})\,,
\end{equation}
and
\begin{equation}
    \{\Psi_1^{-1}\dots \Psi_n^{-1}\otimes\Psi_1^{-1/2}\dots\Psi_m^{-1/2}\}:=\sum_{g=0}^\infty \{\Psi_1^{-1}\dots \Psi_n^{-1}\otimes\Psi_1^{-1/2}\dots\Psi_m^{-1/2} \}_{g}\,.
\end{equation}
Similarly, we define string brackets $[]_g$ and $[]$ by the following equalities
\begin{equation}
    \langle \Psi_0 |c_0^-|[\Psi_1\dots\Psi_n]_g\rangle=\{ \Psi_0\dots\Psi_n\}_g\,, 
\end{equation}
and
\begin{equation}
  \langle \Psi_0 |c_0^-|[\Psi_1\dots\Psi_n]\rangle=\{ \Psi_0\dots\Psi_n\}\,.
\end{equation}

Finally, we can now define the string field theory action
\begin{equation}\label{eqn:SFT action}
    S=\frac{1}{g_s^2}\left[-\frac{1}{2}\langle\tilde{\Psi}|c_0^-Q_B\mathcal{G} |\tilde{\Psi}\rangle+\langle\tilde{\Psi}|c_0^-Q_B|\Psi\rangle +\sum_n\frac{1}{n!} \{\Psi^n\} \right]\,.
\end{equation}
The string field theory action enjoys two very important properties. First, as we shall show in \S\ref{sec:patch-by-patch}, \eqref{eqn:SFT action} satisfies the BV master equation. Therefore, one can quantize string fields in the path-integral formulation in a consistent manner. Second, the action is invariant under the gauge transformation
\begin{equation}
    |\delta\Psi\rangle= Q_B|\Lambda\rangle +\sum_{n=0}^\infty \frac{1}{n!}\mathcal{G}[\Psi^n\Lambda]\,,\quad |\delta\tilde{\Psi}\rangle=Q_B|\tilde{\Lambda}\rangle+\sum_{n=0}^\infty \frac{1}{n!}[\Psi^n\Lambda]\,.
\end{equation}
We shall use this string field theory action to formulate the patch-by-patch description for the non-linear sigma model in string field theory.

\subsection{Action and BV quantization}\label{ssec:BV}
BRST-BV quantization (or simply BV for short) is an extension of the BRST formalism to gauged systems where the algebra of gauge generators does not close off-shell. Part of the power of the BV formalism is that one does not need to start with a gauge-invariant classical action $S_0$ in order to make sense of the theory -- it is enough to have a gauge-fixed action and corresponding BRST charge $\hat{s}$ (we use $\hat{s}$ for the target space BRST charge for now, to distinguish from the worldsheet BRST charge $Q_B$) that implicitly encodes the structure of the gauge symmetry. This is the situation we find ourselves in in the target space of string theory -- the full nature of the gauge group is only understood perturbatively, and all description we have of the theory (aside from duality, like to a CFT or some matrix integral) are gauge-fixed in some way. 

The formalism is as follows (following the great review in \cite{Fuster:2005eg}). The starting point is an action $S(\psi_{s})$, that depends on some collection of fields $\psi_{s}$. First, we double the field content of the theory to include a collection of antifields, $\bar{\psi}_{s}$, where the antifield $\bar{\psi}_{s}$ is the partner of (and has opposite grassmanality to) $\psi_{s}$, and introduce an `antibracket'  $(\cdot,\cdot)$ that acts on any functions $F$ or $G$ of the fields and antifields:
\begin{equation}
(F, G) := \sum_s \frac{\delta^R F}{\delta \psi_s}\frac{\delta^L G}{\delta \bar{\psi}_s} - \frac{\delta^R F}{\delta \bar{\psi}_s}\frac{\delta^L G}{\delta \psi_s},
\end{equation}
and the Laplacian
\begin{equation}
\Delta F := \sum_s \frac{\delta^L}{\delta \psi_s} \frac{\delta^R}{\delta \bar{\psi}_s} F.
\end{equation}

The field-anti-field content of the system is chosen precisely such that taking the antibracket with the action $S$ is equivalent to action by the BRST charge:
\begin{equation}
(S,F) = \hat{s} F.
\end{equation}
In the simplest case, for a given field $\psi_s$ this implies
\begin{equation}
(S,\psi_s) = -\frac{\delta^R S}{\delta \bar{\psi}_{s}} = \hat{s}\bar{\psi}_s.
\end{equation}
emphasizes the relationship between the antifields and the BRST ghost structure of the theory.

At the classical level, the only important condition for a consistent theory is that the action itself should be BRST invariant:
\begin{equation}
(S_0,S_0) = \hat{s} S_0 = 0.
\end{equation}
Once this condition is satisfied, the solutions to the equations of motion are BRST invariant and the classical values of gauge-invariant observables may be evaluated directly.

Once we wish to quantize the theory by inserting it into a path integral, there is no universal separation between the action and the measure $\mu(\psi_s\bar{\psi}_s)$ on field configuration space. The BV formalism treats them as a single object, with the measure serving as a quantum correction to the classical action $S_0$:
\begin{equation}
\begin{split}
&Z = \int \mu(\{\psi_s\}, \{\bar{\psi}_s\})\prod_s D\psi_s D\bar{\psi}_s e^{\frac{i}{\hbar} S_0(\psi_s, \bar{\psi}_s))} = \int \prod_s D\psi_s D\bar{\psi}_s e^{\frac{i}{\hbar}S},\\
& S = S_0 - i\hbar \log \mu = S_0 + \sum_{n \geq 1} \hbar^n S_n,
\end{split}
\end{equation}
where each term $S_n(\psi_s,\bar{\psi}_s)$ is itself independent of $\hbar$. That all powers of $\hbar$ appear in the Taylor expansion may be surprising, but this generically occurs because the constraints of gauge invariance on the full quantum theory may mean that the measure $\mu$ must depend on $\hbar$.

The quantized theory is a consistent gauge theory so long as all expectation values of BRST exact operators are gauge invariant. It is the central theorem of the BV construction that this is guaranteed so long as
\begin{equation}
\Delta e^{\frac{i}{\hbar}S} = \left((S,S) - \hbar \Delta S \right)e^{\frac{i}{\hbar}S} = 0 \implies (S,S) - \hbar \Delta S = 0.
\end{equation}
This is the BV master equation, and a theory built from some classical action $S_0$ and BRST charge $\hat{s}$ is consistently quantizable so long as this equation may be solved for $S$. In \S\ref{sssec:CSFT-BV-Review} we briefly review how Zwiebach's solution the BV master equation in the target space of closed string theory, and in the remainder of \S\ref{ssec:BV} we extend the construction to the patch-by-patch description.

\subsection{Review of BV formalism in CSFT}\label{sssec:CSFT-BV-Review}

In \cite{Zwiebach:1992ie}, Zwiebach showed how to apply the formalism of BV quantization to closed bosonic string field theory in order to make sense of the perturbative target space path integral. Amazingly, string field theory makes little distinction between the fields $\psi_{s}$ and the antifields $\bar{\psi}_s$ -- both flavors of field have corresponding vertex operators $\mathcal{V}_s$ or $\mathcal{V}_s^c$ they source on the worldsheet, and both the field and antifield contributions to the full quantum-corrected target space action may be computed as string vertices -- worldsheet correlators and integrals over moduli space. In particular, the list of components of the string field $\Psi$ includes all fields and antifields:
\begin{equation}
\Psi = \sum_s \psi_s \mathcal{V}_s + \bar{\psi}_s \mathcal{V}_s^c.
\end{equation}

The full classical action is
\begin{equation}
S_0 = \sum_n \frac{g_s^{2n-2}}{n!}\{\Psi^n\},
\end{equation}
and the linear gauge transformation is
\begin{equation}
\delta_{\Lambda} \Psi = Q_{B}\Lambda + \sum_n \frac{1}{n!} [\Psi^n, \Lambda].
\end{equation}
The antibracket of two string vertices $\{\Psi^n\}$ and $\{\Psi^m\}$ is expressed as another string vertex:
\begin{equation}
(\{\Psi^n\}, \{\Psi^m\}) = \{\Psi^n, [\Psi^m]\},
\end{equation}
so the antibracket of the action with itself is
\begin{equation}
(S,S) = \sum_n \sum_m \frac{1}{n! m!} \{\Psi^n, [\Psi^m]\}.
\end{equation}
It may be checked (as it is in \cite{Zwiebach:1992ie}) that both the gauge invariance and the closure of the antibracket is equivalent to the main identity:
\begin{equation}
\sum_m \frac{1}{m!(n-m)!}[\Psi^n,[\Psi^{n - m}]] = 0.
\end{equation}

Zwiebach's construction of the BV master equation was based on bosonic closed string theory \cite{Zwiebach:1992ie}. For a while, the extension of Zwiebach's construction of string field theory from bosonic string theory to superstring theory has been a challenging problem, notably due to the inclusion of the Ramond sector. In addition, one can conjure a na\"ive argument against the existence of the string field theoretic formulation of superstring theory. If it were to be possible to construct such a field theory, one can argue that it must be possible to construct a local covariant type IIB supergravity, which is not possible due to the self-dual 5-form Ramond-Ramond flux. A crucial insight of Ashoke Sen was that, as in the case of chiral field theories, one can add an extra set of fields to impose the stringy analogue of the self-duality condition. This led to the construction of consistent string field theory for closed superstring theories whose action, e.g., \eqref{eqn:SFT action}, solves the BV master equation \cite{Sen:2014pia,Sen:2014dqa,Sen:2015hha,Sen:2015uaa} based on the PCO formalism \cite{Sen:2015hia}. 
\section{Patch-by-patch description}\label{sec:patch-by-patch}
In this section, we shall construct the string field theory description of non-linear sigma models by the patch-by-patch description.
\subsection{Strategy}
In this section, we shall reiterate the strategy we outlined in the introduction in more detail.

To formulate string field theory, given our limited understanding, one first needs to identify a worldsheet CFT description around which we can perturb. Therefore, any attempt to use string field theory to systematically study the non-linear sigma model background may sound like an oxymoron, in that we already stated that the lack of systematic tools to find such a worldsheet description is our motivation.

However, we would like to stress that if the non-linear sigma model background can be understood as a deformation of a well-known worldsheet CFT background there is no need to start with the CFT description for the non-linear sigma model background. In particular, this was the main philosophy used in the series of recent work studying Ramond-Ramond backgrounds with non-trivial spacetime curvature in string field theory \cite{Cho:2018nfn,Cho:2023mhw,Kim:2024dnw,Cho:2025coy}.

Therefore, the right question for us to ask is whether there is any well-defined worldsheet CFT that can be deformed to describe non-linear sigma model backgrounds. Not surprisingly, the worldsheet CFT in question is the very first CFT we learn in string theory: the free field CFT with the correct central charge, combined with the ghost system. 

The intuition behind this claim is as follows. Smooth manifolds are known to be locally Euclidean.\footnote{In this work, we will only concern ourselves with smooth target spaces. However, the results reported in this paper can be generalized to spacetime with orbifold singularities.} Topologically, this means that a smooth manifold $\mathcal{M}$ of dimension $d$ is equipped with Atlas $\{ U,\varphi\}_\mathcal{M},$ where $U_i\subset \mathcal{M}$ is an open set and $\varphi$ is a homeomorphism $\varphi_i:= \varphi|_{U_i}: U_i \rightarrow \Bbb{R}^d.$ However, it should be stressed that the local Euclideanness does not imply that the local coordinate charts $U_i$ with $i\in \mathcal{I}$ have a trivial differential, more precisely Riemannian, structure. Instead, the coordinate map $\varphi_i$ locally trivializes the metric. Suppose that there is an open ball $U_p$ around a point $p\in\mathcal{M}.$ Then, if the Atlas is equipped with the coordinate map that generates the Riemann normal coordinates such that $\varphi_p(p)=0$, we have
\begin{equation}
    \varphi \cdot \left( g_{ab} dv^a \otimes dv^b \right) = \left( g_{ab}(p)-\frac{1}{3} R_{acbd}x^cx^d+\dots \right) dx^a\otimes dx^b\,,
\end{equation}
where $dv^a$ is a cotangent vector in $T_p^*\mathcal{M},$  $dx^a$ is a cotangent vector in $T_0^*\Bbb{R}^d.$ As a result, provided that the curvature is small, we can treat the local coordinate patch of a smooth manifold as a deformation of the flat space by the curvature term. 

In the context of string field theory, this is equivalent to treating each local coordinate chart of the spacetime as a non-trivial background of string field theory formulated around the flat spacetime. Therefore, provided that we can solve equations of motion of string field theory in $\alpha'$ expansion, we can at least systematically study local coordinate charts of non-linear sigma model backgrounds. 

We shall illustrate how to build the local background solution of $\text{SFT}_i.$ As string field theory action, in general, is poorly understood to perform general computations in a reasonable manner, having a perturbative expansion scheme is crucial. We shall work in a regime where the curvature of the target space is parametrically smaller than $\alpha'.$ To quantify how small the curvature is compared to $\alpha',$ we shall need some diffeomorphism invariant quantity. The simplest choice is to use Ricci scalar. However, Ricci scalar is not the best choice, given that many interesting vacuum solutions admit trivial Ricci scalars despite featuring non-trivial curvature. The next simplest choice is the Kretschmann invariant
\begin{equation}
    K:=R_{abcd} R^{abcd}\,.
\end{equation}
The Kretschmann invariant, in general, is non-trivial, even for vacuum solutions. Therefore, it is well suitable for the purpose of setting the parametric expansion. We will henceforth work in a limit where $\varepsilon:=\sqrt{K}\alpha'$ is parametrically small.\footnote{Note that for the target space with non-trivial fluxes, for example, one can use a different expansion scheme, e.g., in $K^{1/4}\alpha'^{1/2}.$ }

We shall now solve the background equations of motion in $\sqrt{K}\alpha'$ expansion at the leading order in $g_s.$ The equations of motion are given as
\begin{align}
    &Q_B|\Psi_0^{(i)}\rangle=Q_B\mathcal{G} |\tilde{\Psi}^{(i)}_0\rangle\,,\\
    &Q_B|\tilde{\Psi}^{(i)}_0\rangle+\sum_{n=2}^\infty \frac{1}{n!}[(\Psi_0^{(i)})^n]_0=0\,,
\end{align}
or equivalently
\begin{equation}
    Q_B|\Psi_0^{(i)}\rangle+\sum_{n=2}^\infty \frac{1}{n!} \mathcal{G}[(\Psi_0^{(i)})^n]_0=0\,.
\end{equation}
We shall expand $\Psi_0^{(i)}$ in $\varepsilon$
\begin{equation}
    \Psi_0^{(i)}:= \sum_n  \Psi_{0,n}^{(i)}\,,
\end{equation}
with 
\begin{equation}
    \mathcal{O}(\Psi_{0,n}^{(i)})=\mathcal{O}\left(\varepsilon^n\right)\,.
\end{equation}
The perturbative background equation is then, to the first few orders, given as
\begin{equation}
    Q_B|\Psi_{0,1}^{(i)}\rangle=0\,,
\end{equation}
\begin{equation}
    Q_B|\Psi_{0,2}^{(i)}\rangle+\frac{1}{2} \mathcal{G}[ (\Psi_{0,1}^{(i)})^2]_0=0\,,
\end{equation}
\begin{equation}
    Q_B|\Psi_{0,3}^{(i)}\rangle+\frac{1}{3!} \mathcal{G}[(\Psi_{0,1}^{(i)})^3]_0+\mathcal{G}[\Psi_{0,1}^{(i)}\otimes\Psi_{0,2}^{(i)}]_0=0\,,
\end{equation}
\begin{equation}
    \vdots
\end{equation}

We propose to use the following form for the first-order background solution
\begin{equation}
    \Psi_{0,1}^{(i)}= -\frac{1}{6\pi\alpha'}c\bar{c} R_{ACBD}X^CX^D\partial X^A \bar{\partial} X^B\,, 
\end{equation}
for bosonic string, and
\begin{equation}
    (\Psi_{0,1}^{(i)})^{-1,-1}=\frac{1}{12\pi} c\bar{c} R_{ACBD}X^CX^De^{-\phi}\psi^A  e^{-\bar{\phi}}\bar{\psi}^B\,,
\end{equation}
for type II string theories. 

At this point, we have not specified what it means to attach a polynomial in the worldsheet boson in the vertex operator. In principle, there can be two distinct prescriptions. We can either treat $X$ as the full quantum field or as the zero mode of the worldsheet boson. However, as we will show in \S\ref{app:diff2} and \S\ref{app:diff3}, treating $X$ as the zero mode leads to an erroneous conclusion. For example, to find the closed string background in which the dilaton remains fixed, the closed string background solution should be sensitive to the off-shell parameters associated to the open string diagrams. Therefore, we conclude that the correct treatment is to treat $X$ as the full quantum field.\footnote{We thank Michael Haack for the discussion on this problem.}

As one can check, the first-order background equation can be rewritten as
\begin{equation}
    R_{AB}=0\,,
\end{equation}
which agrees with the one-loop beta function computation of the non-linear sigma model, as it should. To proceed, we shall assume that there exists a perturbative background solution to an arbitrarily high order in the $\varepsilon$ expansion. 

In order to have a satisfactory description of the non-linear sigma model backgrounds in string field theory, not only do we need descriptions of local coordinate patches, but also a description of global topology. In particular, careful treatment of global topology is absolutely crucial as topology determines the quantization condition of states.

A crucial ingredient we shall use is the transition map constructed from the coordinate maps. Suppose that two local coordinate charts $U_i$ and $U_j$ have a non-trivial intersection $U_i\cap U_j\neq \phi.$ Then, $\varphi_i\cdot \varphi_j^{-1}$ furnishes a diffeomorphism between $U_i|_{U_j}$ and $U_j|_{U_i}.$ In the target space physics, diffeomorphism represents the gauge redundancy as physical observables do not depend on how a differential structure is represented. What this implies is that we can cover the target space with a set of local coordinate charts with the coordinate maps, and when coordinate charts overlap, we can declare that string field in one coordinate chart is gauge equivalent to the corresponding string field in the other coordinate chart
\begin{equation}\label{eqn:gluing top1}
    \Psi|_{U_i}|_{U_j}=\Psi|_{U_j}|_{U_i}+Q_B \Lambda + \sum_n \frac{1}{n!} [\Lambda\Psi^n]|_{U_j}|_{U_i}\,,
\end{equation}
where $\Lambda$ is properly related to the diffeomorphism $\varphi_i\cdot\varphi_j^{-1}.$

Note that, strictly speaking, the gauge transformation in string field theory is an infinitesimal transformation. We will henceforth treat $\Lambda$ as a weak string field of order $\epsilon.$ Therefore, one can understand the above gauge transformation as a covariant derivative of the spacetime fields. We shall assume that the target space we will study has no torsion. In terms of the gauge transformation, the torsionless condition can be phrased as follows. Let's suppose that there are four nearby local patches $(U_i,\varphi_i)$ for $i=1,\dots4.$ Then, we require that the gauge transformations due to the diffeomorphisms $\varphi_4\cdot \varphi_3^{-1}\cdot \varphi_3 \cdot \varphi_1^{-1}$ and $\varphi_4\cdot \varphi_2^{-1}\cdot \varphi_2 \cdot \varphi_1^{-1}$ are equivalent. Without loss of generality, we shall assume 
\begin{equation}
    \Lambda_{34}=\Lambda_{12}+\partial_x \Lambda_{12}\,,\quad \Lambda_{24}=\Lambda_{13}+\partial_y\Lambda_{13}\,
\end{equation}
where $\partial\Lambda$ and $\Lambda$ are both evaluated at the origin of $U_1.$ This in turn implies that the following gauge transformation is trivial
\begin{align}
    &Q_B(\partial_x\Lambda_{12}-\partial_y \Lambda_{13}) +Q_B[\Lambda_{12}\otimes \Lambda_{13}]+\sum_n [(\Psi_0)^n \otimes \Lambda_{12}\otimes \Lambda_{13}] +\mathcal{O}(\partial \Lambda^2)\equiv 0\,.
\end{align}
By treating $\Lambda$ as a small variation of order $\mathcal{O}(\epsilon)$, we can check that the above equation implies that gauge transformation by the following parameter 
\begin{equation}
    \partial_x \Lambda_{12} -\partial_y \Lambda_{13} +\sum_n\frac{1}{n!} [(\Psi_0)^n\otimes\Lambda_{12}\otimes\Lambda_{13}]
\end{equation}
should be trivial. Therefore, we shall impose that there is a zero ghost number gauge parameter $\Lambda_{123},$ with $\mathcal{O}(\Lambda_{123})=\mathcal{O}(\epsilon^2)$ such that
\begin{equation}
    \partial_x \Lambda_{12}-\partial_y \Lambda_{13} +\sum_n \frac{1}{n!} [(\Psi_0)^n \otimes \Lambda_{12}\otimes \Lambda_{13}]=Q_B \Lambda_{123} +\sum_{n}\frac{1}{n!}[(\Psi_0)^n \otimes \Lambda_{123}]\,.
\end{equation}
This integrability condition was first derived in \cite{Mazel:2025fxj} to construct the extended BV complex.

Once we confirm that the local background solutions are compatible with the diffeomorphism in the sense of \eqref{eqn:gluing top1}, we can define the global string field $\Psi$ and the global background solution $\Psi_0$ such that
\begin{equation}
    \Psi|_{U_i}:=\Psi^{(i)}\,,
\end{equation}
and
\begin{equation}
    \Psi_0|_{U_i}:=\Psi_0^{(i)}\,,
\end{equation}
for any $i.$ 

Because we are now going beyond the conventional construction of string field theory by patching up many copies of string field theories related by the gauge transformation, it is imperative to check if the consistency conditions required for the global formulation of string field theory are met. Crucially, one must check that the global form of the string field theory action constructed from the patch-by-patch description is gauge invariant and solves the quantum BV master equation. We will show in the later sections that indeed the string field theory action we construct using the patch-by-patch description meets the consistency conditions.

Given the globally well-defined background solution and the action, we can now formulate string field theory around the non-linear sigma model background. Let us define a few objects
\begin{equation}
    \{\Psi_1\dots\Psi_k\}':= \sum_{n=0}^\infty \frac{1}{n!}\{ \Psi_0^n \Psi_1\dots\Psi_k\}\,,\quad\text{for}\, k\geq2\,,
\end{equation}
\begin{equation}
    \{\Psi_1\}'=0\,,
\end{equation}
for arbitrary string fields $\Psi_i.$ We also define a new string bracket $[]'$ such that
\begin{equation}
    \langle \Psi_0|c_0^-|[\Psi_1\dots\Psi_n]'\rangle= \{\Psi_0\dots\Psi_n\}'\,.
\end{equation}
The BRST operator of the non-linear sigma model background is different from that of the free field CFT. We define $\hat{Q}_B$ and $\tilde{Q}_B$ that act on states in $\hat{\mathcal{H}}$ and $\tilde{\mathcal{H}},$ respectively, as
\begin{equation}
    \hat{Q}_B:=Q_B+\mathcal{G} K\,,
\end{equation}
and
\begin{equation}
    \tilde{Q}_B:=Q_B+K\mathcal{G}\,,
\end{equation}
where we define $K$ As
\begin{equation}
    K|\Psi\rangle:=\sum_{n=0}^\infty \frac{1}{n!}[\Psi_0^n\Psi]\,.
\end{equation}

By expanding the string field around the background solution
\begin{equation}
    |\Phi\rangle=|\Psi\rangle-|\Psi_0\rangle\,,\quad |\tilde{\Phi}\rangle=|\tilde{\Psi}\rangle-|\tilde{\Psi}_0\rangle\,,
\end{equation}
we can write down the 1PI action of string field theory at the non-linear sigma model background
\begin{equation}\label{eqn:1PI action NLSM}
    S_{1PI}=g_s^{-2}\left[ -\frac{1}{2}\langle\tilde{\Phi}|c_0^-Q_B\mathcal{G}|\tilde{\Phi}\rangle+\langle \tilde{\Phi}|c_0^-Q_B|\Phi\rangle+\frac{1}{2} \langle\Phi|c_0^-K|\Phi\rangle+\sum_{n=3}^{\infty}\frac{1}{n!}\{\Phi^n\}' \right]+S_0\,,
\end{equation}
where $S_0$ is the vev of the 1PI action evaluated at the background solution. It is important to note that \eqref{eqn:1PI action NLSM} is invariant under the following gauge transformation
\begin{equation}
    |\delta\Phi\rangle=\hat{Q}_B|\Lambda\rangle +\sum_{n=1}^\infty \frac{1}{n!} \mathcal{G}[\Phi^n\Lambda]'\,,\quad |\delta\tilde{\Phi}\rangle=Q_B|\tilde{\Lambda}\rangle+K|\Lambda\rangle+\sum_{n=1}^\infty \frac{1}{n!}[\Phi^n\Lambda]'\,.
\end{equation}

In order to compute on-shell amplitudes in the non-linear sigma model background, we need to first find the spectrum. To do so, we can solve linearized equations of motion
\begin{equation}
    \hat{Q}_B|\Phi_{linear}\rangle=0\,,
\end{equation}
and use $\Phi_{linear}$ and the string vertices $\{\}'$ to compute on-shell amplitudes. It is important to note that only the $\Phi_{linear}$ that are globally well-defined are admissible, where the gluing compatibility condition for the linearized states is given as
\begin{equation}
    |\Phi_{linear}^{(i)}\rangle= |\Phi_{linear}^{(j)}\rangle +\hat{Q}_B |\Lambda\rangle\,.
\end{equation}

\subsection{SFT In a Single Patch}

We first make the notion of string field theory more precise in a patch $U_i$ of the target space. The prescription we take is to associate to each patch a copy of the CFT Hilbert space, which we call $\hat{\mathcal{H}}_i$ and $\tilde{\mathcal{H}}_i$. The target space of this copy is the flat space $\mathbb{R}^{10}$, which may be thought of as the tangent space at the origin of $U_i$. $\hat{\mathcal{H}}_i$ and $\tilde{\mathcal{H}}_i$ are equipped with the field space of flat-space string field configurations. 

\begin{figure}[ht]
\centering
\includegraphics[width = 0.3\textwidth]{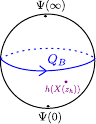}
\caption{A depiction of the kinetic term contributing to the action \eqref{eqn:h-act}. The crosses denote string field insertions. There is an additional insertion of $h(X)$, denoted by the dot, so we must additionally specify the choice of contour along which we insert $Q_B$.}\label{fig:kinetic-term}
\end{figure}

To each patch $U_i$ we associate a smeared-out characteristic function $h_i(X)$, which is $1$ nearly everywhere in the interior of $U_i$ and $0$ nearly everywhere in the exterior of $U_i$, except for the region along the boundary of width $\epsilon_i$ (see Fig. \ref{fig:smeared-boundaries}). We then define string vertices in the patch $U_i$ by taking the standard choice of string field, with extra data of a fixed but arbitrary point $z_h$ at which the $h_i(X)$ characteristic function is inserted:
\begin{equation}\label{eqn:hX-vertex}
\{\Psi^n\}_{h_i} := \int_{\mathcal{M}_n} \prod_{i=1}^{n-3} d t_i \left \langle e^{\mathcal{B}} h_i(X(z_{h})) \prod_{j=1}^n \Psi (z_j)  \right \rangle\,,
\end{equation}
where $\mathcal{M}_n$ parameterizes the fundamental domain of the moduli space of the $n$-punctured sphere, and the coordinates $t_i$ are some parameterizations of $\mathcal{M}_n$. $\mathcal{B}$ is the Beltrami differential. Because $h_i(X)$ comes with no $c$-ghost and $\eta$-ghost insertions, it commutes with $\mathcal{B}$.
It is important to emphasize that the kinetic term in \eqref{eqn:h-act} must also come with an $h_i(X)$ insertion (see Fig.\ref{fig:kinetic-term}).

There are a couple of points to note about \eqref{eqn:hX-vertex}. The first is that the string vertex appears to depend on the arbitrary coordinate $z_h$, and it may even have divergences when one of the $z_i$ hit $z_h$. In \S\ref{sec:patch-by-patch general}, we will show that these divergences drop out of the action once we sum over patches, and therefore any physical quantity we compute will not see the $z_h$ dependence or these divergences. In fact, for the purpose of physical observables we may compute \eqref{eqn:hX-vertex} by simply treating $h(X)$ as appearing in the integral over the $X$ zero modes. 

The second point is that if we use the objects defined in \eqref{eqn:hX-vertex} to define a target space action, as in
\begin{equation}\label{eqn:h-act}
S_{h}(\Psi):= g_s^{-2} \left[-\frac{1}{2} \langle \tilde{\Psi}|c_0^-Q_B\mathcal{G}|\tilde{\Psi}\rangle+\langle\tilde{\Psi}c_0^-Q_B|\Psi\rangle+ \sum_n \frac{1}{n!}\{\Psi^n\}_{h}\right]\,,
\end{equation}
the string vertices do not manifestly satisfy the appropriate boundary conditions. This is because the contribution to an $n$-particle scattering amplitude from a fundamental $n$-point vertex only comes with one insertion of $h_i(X)$, any Feynman diagram built from $m$-point vertices with $m < n$ will come with more than one insertion. The forms do not agree on the boundary of the moduli space, and generically, gauge invariance is broken.

The vertices we have defined are linear in $h_i(X)$, in the sense that
\begin{equation}
\{\Psi^n\}_{h_1} + \{\Psi^n\}_{h_2} = \{\Psi^n\}_{h_1 + h_2} \quad \leftrightarrow \quad S_{h_1}(\Psi) + S_{h_2}(\Psi) = S_{h_1 + h_2}(\Psi),
\end{equation}
In \S\ref{sec:patch-by-patch general} we demonstrate how the total action over a collection of patches $h_i$, which furnish a resolution of the identity of some target space without boundary, is gauge invariant and BV-quantizable.

\subsection{Building the Total Action}

Intuitively, the total action in the patch by patch description of a local theory is a sum over the individual actions evaluated on each patch. To write down a proposal for the full form of an action, we must first specify what our collection of patches is, and then take the appropriate sum. To do so, we must build up a few ingredients.

\begin{figure}[ht]
\centering
\includegraphics[width=0.65\textwidth]{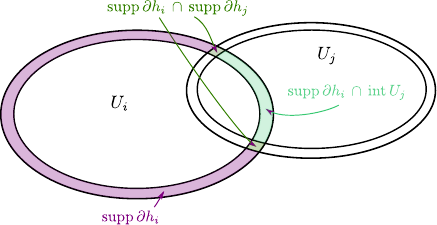}
\caption{Two overlapping patches, $U_i$ and $U_j$. The patches come with smeared boundaries, labeled $\text{supp}\, \partial h_i$ and of width $\epsilon_i$, corresponding to the region where the characteristic functions $h_i(X)$ and $h_j(X)$ have a nontrivial gradient. The smeared boundary of $U_i$ is shaded according to where it intersects the exterior of $U_j$, the interior of $U_j$, or the smeared boundaries. It is the region labeled $\text{supp}\, \partial h_i\, \cap \, \text{int}\, U_j$ that corresponds to a boundary condition \eqref{eqn:inter-constraints} for the interpolating field $\Psi^{(i \cap j)}$, equating it to $\Psi^{(i)}$. The regions $\text{supp}\, \partial h_i \, \cap \, \text{supp}\, \partial h_j$ are the `smeared corners', where there are no boundary conditions on $\Psi^{i \cap j}$. }\label{fig:smeared-boundaries}
\end{figure}

First, we define what properties we require from the regulated characteristic function $h_i(X)$ of the patch $U_i$. We allow $h_i(X)$ to differ from $0$ or $1$ only within a distance\footnote{We give $\epsilon_i$ an index as it is convenient to allow it to vary from region to region} $\epsilon_i$ from the boundary $\partial U_i$. Given this, it is clear that all points in the support of the gradient $\partial_{\mu} h(X)$ (for which we introduce the shorthand supp $\partial h_i$) are within a distance $\epsilon_i$ of supp $\partial h_i$. We refer to the region where $h_i(X) = 1$ as the `interior' of $U_i$ (denoted int $U_i$), the region where $h_i = 0$ the `exterior' (denoted ext $U_i$), and supp $\partial h_i$ as the `smeared out boundary'. (See Fig.\ref{fig:smeared-boundaries}).

Consider two such regulated characteristic functions, $h_i$ and $h_j$, for two overlapping patches $U_i$ and $U_j$. The new function
\begin{equation}\label{eqn:h-int-1}
h_{i \cap j}(X) := h_i(X)h_j(X)
\end{equation}
forms a natural regulated characteristic function for the overlap $U_{i \cap j}:= U_i \cap U_j$. First, note that $h_{i \cap j}(X) = 0$ implies $X$ is either in the exterior of $U_i$ or in the exterior of $U_j$, so $h_{i \cap j}(X)$ only has support on the interior of $U_{i \cap j}$ and some smearing of its boundary. The smeared boundary of $U_{i \cap j}$ is determined by the support of $\partial_{\mu} h_{i \cap j} = h_i \partial_{\mu} h_j + h_j \partial_{\mu} h_i$. Note that if $X$ is in supp $\partial h_{i \cap j}$, then it must either be in supp $\partial h_i$ or supp $\partial h_j$, and so it must either be within a distance $\epsilon_i$ of $\partial U_i$ or within a distance $\epsilon_j$ of $\partial U_j$. We therefore have
\begin{equation}
\epsilon_{i \cap j} \leq \epsilon_i + \epsilon_j.
\end{equation}

We may similarly define a smeared characteristic function for the union of $U_i$ and $U_j$ as
\begin{equation}
h_{i \cup j}(X) := h_i(X) + h_j(X) - h_{i \cap j}(X).
\end{equation}
We may then check
\begin{equation}
\partial^{\epsilon} U_{i \cup j} = \text{supp}\, \partial h_{i \cup j} = \partial^{\epsilon} U_i \cup \partial^{\epsilon} U_j - ((\partial^{\epsilon} U_i \cap \text{int}\,U_j) \cup (\partial^{\epsilon} U_j \cap \text{int}\,U_i)).
\end{equation}
In words, we have checked that the smeared boundary of $U_{i \cup j}$ is the union of the smeared boundaries of $U_i$ and $U_j$, with the bits contained in the interiors of $U_i$ and $U_j$ subtrated off. This is exactly what we expect for a notion of smeared boundary of the union of two patches. We may also directly see that
\begin{equation}
\epsilon_{i \cup j} \leq \epsilon_i + \epsilon_j,
\end{equation}
as to be in the smeared boundary of $U_{i \cup j}$, it is a necessary condition to be either near the smeared boundary of $U_i$ or the smeared boundary of $U_j$.

For a collection of more than two patches $U_{i_1}, \ldots U_{i_k}$, the functions $h_{i_1 \cap i_2 \cap \ldots \cap i_k}$ and $h_{i_1 \cup i_2 \cup \ldots \cup i_k}$ are defined recursively in the natural way. For three patches, for example, we have
\begin{equation}
\begin{split}
h_{i \cup j \cup k}(X) := &h_i(X) + h_j(X) + h_k(X) - h_i(X)h_j(X) - \\
&-h_i(X)h_k(X) - h_j(X)h_k(X) + h_i(X)h_j(X)h_k(X).
\end{split}
\end{equation}

Now, consider a family of patches $U_i$ that cover some target space $\mathcal{M}$ without boundary upon which we would like to define the SFT action. By this, we mean that (a) only a finite number of patches ever simultaneously overlap, and (b) that the smeared characteristic function of the total union, $h_{\bigcup_i}$, identically evalautes to 1 on $\mathcal{M}$. To each patch we associated a string field $\Psi^{(i)}$. Take two patches $U_i$ and $U_j$, and consider the overlap $U_i \cap U_j$. On this overlap, we introduce a new string field $\Psi^{(i \cap j)}(X)$, which we call an `interpolating string field'. We take this new string field to satisfy the following interpolation conditions:
\begin{equation}\label{eqn:inter-constraints}
\left.\Psi^{(i \cap j)}(X)\right|_{\partial^{\epsilon}U_i \cap\, \text{int}\, U_j} \equiv \left.\Psi^{(i)}(X)\right|_{\partial^{\epsilon}U_i \cap\, \text{int}\, U_j}, \quad i \leftrightarrow j.
\end{equation}
In words, $\Psi^{(i \cap j)}$ must exactly agree with $\Psi^{(i)}$ on the part of $\partial^{\epsilon}U_i$ that's contained entirely in the interior of $U_j$, and vice versa. In this way, it behaves as an interoplating functon between $\Psi^{(i)}$ on the boundary of $U_i$ and $\Psi^{(j)}$ on the boundary of $U_j$. Note that these boundary conditions do not fix what's going on in the intersection supp $\partial h_i$ $\cap$ supp $\partial h_j$, i.e. the (smeared) `corners' of $U^{(i \cap j)}$ (again see Fig. \ref{fig:smeared-boundaries}). The string fields on triple intersections $\Psi^{(i \cap j \cap k)}$, and larger numbers of intersections, are analogously defined in a recursive manner. 

\begin{figure}[ht]
\centering
\begin{subfigure}{.5\textwidth}
  \centering
  \includegraphics[width=\linewidth]{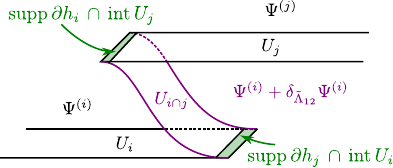}
  \caption{}
  \label{fig:interpolate-1}
\end{subfigure}%
\begin{subfigure}{.5\textwidth}
  \centering
  \includegraphics[width=\linewidth]{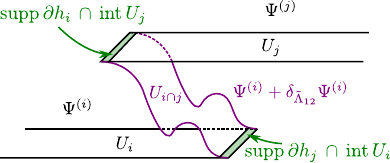}
  \caption{}
  \label{fig:interpolate-2}
\end{subfigure}

\caption{A depiction of the interpolating string field $\Psi^{(i \cap j)}$, which is taken to be related to $\Psi^{(i)}$ (or, equivalently, $\Psi^{(j)}$) by a gauge transformation. Figs. \ref{fig:interpolate-1} and \ref{fig:interpolate-2} represent two different interpolating string fields that solve the constraints \eqref{eqn:inter-constraints}. As can be directly seen from the figure, the full string field obtained from stitching together all the string fields is multivalued at each point in target space.}
\label{fig:interpoation-field}
\end{figure}

We may now write down the total action we consider as a sum over all patches and all possible intersections:
\begin{equation}\label{eqn:total-action}
\begin{split}
S_{tot}:= \sum_i S_{h_i}(\Psi^{(i)}) - \sum_{i \neq j} S_{h_{i \cap j}}(\Psi^{(i \cap j)}) + \sum_{i \neq j \neq k}S_{h_{i \cap j \cap j}}(\Psi^{(i \cap j \cap k)}) + \ldots
\end{split}
\end{equation}
Because only a finite number of patches may simultaneously intersect, the sum in \eqref{eqn:total-action} is finite. From now on, to compactify notation, we use $S_i$ as a shorthand for $S_{h_i}(\Psi^{(i)})$, leaving $h_i$ and $\Psi^{(i)}$ implicit.

Note that as we have defined it, the total string field may be multivalued (again, see Fig. \ref{fig:interpoation-field}). However, as we impose the constraint \eqref{eqn:gauge trans tor1}, we only ever consider configurations that are single valued up to gauge transformations. Furthermore, the arguments in the next sections will show that the action defined in \eqref{eqn:total-action} is gauge invariant and BV quantizable even if we do not impose any single-valuedness constraints such as \eqref{eqn:gauge trans tor1}.

\subsection{Gauge Transformations and the Antibracket}

In the patch by patch description of a gauge theory, gauge transformations are allowed to extend outside of any one given patch. Locally, when a single patch $U_i$ looks like a subset of $\mathbb{R}^D$, this means that the local string field $\Psi^{(i)}(X)$ are allowed to transform under gauge transformations that have nontrivial support on $\partial U_i$, and in general are allowed to have arbitrary boundary conditions. One way to ensure this is to consider $U_i$ as a subset of $\mathbb{R}^D$, and allow the gauge transformation $\Lambda^{(i)}(X)$ to arbitrarily extend outside of the boundary of the patch. This means that locally, a gauge transformation within a patch may be written in terms of the usual flat space string product with no $h_i(X)$ insertions:
\begin{equation}\label{eqn:local-gauge-transform}
\delta_{\Lambda^{(i)}} \Psi^{(i)} = Q_{B}\Lambda^{(i)} + \sum_{n}\frac{1}{n!}[\Psi^{(i)\,n},\Lambda^{(i)}].
\end{equation}

Once we include arbitrary intersections of patches, such as $U_{i \cap j}$, the constraints \eqref{eqn:inter-constraints} imply conditions on how the interpolating string fields transform. In particular, we may define `interpolating gauge transformations':
\begin{equation}\label{eqn:gauge-inter-constraint}
\left.\Lambda^{(i \cap j)}(X)\right|_{\text{supp}\,\partial h_i \cap \, \text{int}\, U_j} \equiv \left.\Lambda^{(i)}(X)\right|_{\text{supp}\,\partial h_i \cap \, \text{int}\,U_j}, \quad i \leftrightarrow j.
\end{equation}
The interpolating gauge transformation $\Lambda^{(i \cap j)}$ must agree identically with $\Lambda^{(i)}$ on the part of the smeared boundary of $U_i$ that is fully in the interior of $U_j$, and vice versa. Note that this places no constraints on how $\Lambda^{(i)}$ relates to $\Lambda^{(j)}$ -- because (perturbatively) the space of string fields is path connected, we may always find an interpolating string field of this form (although it will generically be smooth but non-analytic.)

It is now unambiguously defined how the action within a patch \eqref{eqn:h-act} transforms under the local gauge transformation \eqref{eqn:local-gauge-transform}:
\begin{equation}\label{eqn:local-act-transform}
\begin{split}
\delta_{\Lambda^{(i)}}S_{h_i} &= \sum_m \frac{1}{m!}\{Q_B \Psi, \Psi^m, \Lambda^{(i)}\}_{h_i} + \sum_n \frac{1}{n!}\{\Psi^n, Q_B \Lambda^{(i)}\}_{h_i} \\
&+\sum_{n,m} \frac{1}{n!m!}\{\Psi^n, [\Psi^m, \Lambda^{(i)}]\}_{h_i}.
\end{split}
\end{equation}
The important observation is that this transformation depends only on derivatives of $h_i(X)$ -- all dependence on the constant mode of $h_i(X)$ itself cancels out. To see this, it is convenient to introduce the local string product, denoted $[\, ,\, ]_{h_i}$ and defined through the relation
\begin{equation}\label{eqn:h-prod-def}
\bra{A}c_0^- \ket{[B_1, \ldots, B_k]_{h_i}} := \{A, B_1, \ldots, B_k\}_{h_i}.
\end{equation}
$h_i(X)$ can only contribute to worldsheet correlators through terms involving its derivatives, so we may observe
\begin{equation}\label{eqn:h-prod-exp}
[B_1, \ldots, B_k]_{h_i} \sim h_i(X)[B_1, \ldots, B_k] + (\text{terms involving derivatives of }h_i),
\end{equation}
where $\sim$ means equality up to terms proportional to derivatives of $h_i$. We now vary \eqref{eqn:local-act-transform} with respect to $\Lambda^{(i)}$ to find
\begin{equation}\label{eqn:del-del-Lambda}
\begin{split}
\frac{\delta \delta_{\Lambda^{(i)}} S_i}{\delta \Lambda^{(i)}} \sim h_i(X)\left(\sum_m \frac{1}{m!}[Q_B \Psi, \Psi^m] + \sum_{m,n}\frac{1}{m!n!}[\Psi^m,[\Psi^n]]\right) = 0.
\end{split}
\end{equation}
The RHS of \eqref{eqn:del-del-Lambda} vanishes due to the usual main identity of string field theory. As a simple corollary, $\delta_{\Lambda^{(i)}} S_{i}$ is only nonzero on supp $\partial h_i$.

We may likewise ask about the violation to the closure of the BV antibracket of $S_{i}$. This quantity is bilinear in $h_i(X)$, and is computed using moduli space integrals with two insertions of $h_i(X(z))$. The exact same computation again shows that the usual main identity implies $(S_{i}, S_{i}) \sim 0$. We shall demonstrate in the next sections how the interpolation conditions \eqref{eqn:inter-constraints} and \eqref{eqn:gauge-inter-constraint} ensure that the terms involving the derivatives of $h_i(X)$ also cancel in the gauge transformation and BV antibracket of \eqref{eqn:total-action}.

\subsection{A Clarifying Example: Two Patches}\label{sssec:two-patch-BV}

\begin{figure}[ht]
\centering
\includegraphics[width=0.4\textwidth]{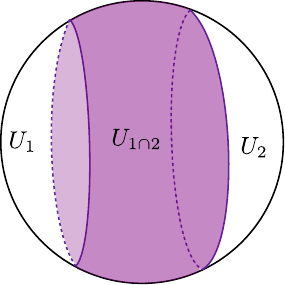}
\caption{An illustration of the target space with two patches. The two patches are $U_1$ and $U_2$, and their overlap is $U_{1 \cap 2}$ is shaded. Unlike the generic case depicted in Fig. \ref{fig:smeared-boundaries}, the boundaries $\partial U_1$ and $\partial U_2$ do not intersect, and the situation is simpler.}\label{eqn:fig-sphere-patches}
\end{figure}

Consider a target space manifold with topology that may just be foliated by two patches, $U_1$ and $U_2$ (see Fig.\ref{eqn:fig-sphere-patches}). We have chosen $U_1$ and $U_2$ such that the (smeared) boundaries of $U_1$ and $U_2$ do not intersect, so there is no overlap between supp $\partial h_i$ and supp $\partial h_j$. In particular, we have
\begin{equation}
\partial_{\mu} h_{i \cap j}(X) = \partial_{\mu}h_i(X) + \partial_{\mu}h_j(X), \quad \partial_{\mu}h_{i \cup j}(X) \equiv 0.
\end{equation}
We have three string fields, $\Psi^{(1)}$, $\Psi^{(2)}$, and $\Psi^{(1 \cap 2)}$ valued in the state spaces $\hat{\mathcal{H}}_1$, $\hat{\mathcal{H}}_2$, and $\hat{\mathcal{H}}_{1 \cap 2}$, and satisfying the interpolating constraints \eqref{eqn:inter-constraints}. There exists natural maps $\pi_{1 \rightarrow 2}$ and $\pi_{2 \rightarrow 1}$ from $\hat{\mathcal{H}}_{1}$ and $\hat{\mathcal{H}}_2$ to $\hat{\mathcal{H}}_{1 \cap 2}$, simply given by translation from the origin of $U_1$ (or $U_2$) to the origin of $U_{1 \cap 2}$. There is therefore a natural way to take overlaps between the different string fields, for example:
\begin{equation}\label{eqn:1-2-overlap}
\bra{\Psi^{(1)}} c_0^- \ket{\Psi^{(2)}} := \bra{\pi_{1 \rightarrow 2} \Psi^{(1)}}c_0^{-}\ket{\pi_{2\rightarrow 1} \Psi^{(2)}},
\end{equation}
where technically the computation of the overlap occurs in the $\hat{\mathcal{H}}_{1 \cap 2}$ state space.

The total action for this system is
\begin{equation}\label{eqn:2-patch-act}
S_{tot} = S_{h_1}(\Psi^{(1)}) + S_{h_2}(\Psi^{(2)}) - S_{h_{1\cap 2}}(\Psi^{(1 \cap 2)}).
\end{equation}
First, we demonstrate that this action is gauge invariant. Take gauge transformations $\Lambda^{(1)}$, $\Lambda^{(2)}$, and $\Lambda^{(1 \cap 2)}$ that satisfy the interpolating conditions \eqref{eqn:gauge-inter-constraint}. We now make the observation that the total system of constraints \eqref{eqn:inter-constraints} and \eqref{eqn:gauge-inter-constraint} imply
\begin{equation}\label{eqn:bdy-equiv}
\left.[\Psi^{(i)\,m},\Lambda^{(i)}]\right|_{\text{supp}\,\partial h_i} \equiv \left.[\Psi^{(i \cap j)\, m}, \Lambda^{(i \cap j)}]\right|_{\text{supp}\, \partial h_i},
\end{equation}
as to any order in $\alpha'$ the string products only depend on the local values of the field and their derivatives.

Any string vertex, global or local, involves a zero mode integral. Consider a set $U$ in the target space. For any (local) string vertex $\{\, , \,\}_{h_i}$, we denote by $\left.\{\}_{h_i}\right|_{U}$ the contribution to the zero mode integral from the region $U$. We may use similar notation for any quantity expressed as a sum of string vertices -- for example, $\left.S_{i}(\Psi^{(i)})\right|_{U}$ denotes the contribution to $S_{i}$ from the piece of the integral over $U$. Due to \eqref{eqn:h-prod-exp}, the gauge transformations of the local actions localize to integrals over the smeared boundaries, so we may write
\begin{equation}
\begin{split}
\delta_{\Lambda^{(1)}} S_{1} &= \left.\delta_{\Lambda^{(1)}} S_{1} \right|_{\text{supp}\, \partial h_1},\\
\delta_{\Lambda^{(2)}} S_{2} &= \left.\delta_{\Lambda^{(2)}} S_{2}\right|_{\text{supp}\, \partial h_2},\\
\delta_{\Lambda^{(1 \cap 2)}} S_{1 \cap 2} &= \left.\delta_{\Lambda^{(1 \cap 2)}} S_{1 \cap 2}\right|_{\text{supp}\, \partial h_1} + \left.\delta_{\Lambda^{(1 \cap 2)}} S_{1 \cap 2}\right|_{\text{supp}\, \partial h_2}.
\end{split}
\end{equation}
Because of \eqref{eqn:bdy-equiv}, we now have
\begin{equation}\label{eqn:bdy-equiv action}
\begin{split}
&\left.\delta_{\Lambda^{(1)}} S_{1}\right|_{\text{supp}\, \partial h_1} = \left.\delta_{\Lambda^{(1 \cap 2)}} S_{1 \cap 2}\right|_{\text{supp}\, \partial h_1},\\
&\left.\delta_{\Lambda^{(2)}} S_{2}\right|_{\text{supp}\, \partial h_2} = \left.\delta_{\Lambda^{(1 \cap 2)}} S_{1 \cap 2}\right|_{\text{supp}\, \partial h_2}.\\
\end{split}
\end{equation}
If we denote by $\Lambda$ the total gauge transformation due to the collection of local transformations $\Lambda^{(1)}$, $\Lambda^{(2)}$, and $\Lambda^{(1 \cap 2)}$, it is now easy to see from \eqref{eqn:2-patch-act} and \eqref{eqn:bdy-equiv} that $\delta_{\Lambda} S_{tot}$ vanishes, so the total two-patch action $S_{tot}$ is gauge invariant.

The same computation can be done for the BV antibracket, because the quantities $(S_{i}, S_{j})$ are also computed via string vertices and may be expressed as a local target space integrals at any finite order in $\alpha'$. The cancellation of the terms depending on the constant mode of $h_i(X)$ imply 
\begin{equation}\label{eqn:2-patch-BV-1}
\begin{split}
&(S_1, S_1) = \left.\left(S_1, S_1\right)\right|_{\text{supp}\, \partial h_1}, \quad \left(S_2, S_2\right) = \left.\left(S_2, S_2\right)\right|_{\text{supp}\, \partial h_2},\\
&(S_{1 \cap 2}, S_{1 \cap 2}) = \left.\left(S_{1 \cap 2}, S_{1 \cap 2}\right)\right|_{\text{supp}\, \partial h_1} + \left.\left(S_{1 \cap 2}, S_{1 \cap 2}\right)\right|_{\text{supp}\, \partial h_2},\\
&(S_1, S_{1 \cap 2}) = \left.\left(S_{1}, S_{1 \cap 2}\right)\right|_{\text{supp}\, \partial h_1} + \left.\left(S_{1}, S_{1 \cap 2}\right)\right|_{\text{supp}\, \partial h_2},\\
&(S_2, S_{1 \cap 2}) = \left.\left(S_{1}, S_{2 \cap 2}\right)\right|_{\text{supp}\, \partial h_1} + \left.\left(S_{2}, S_{1 \cap 2}\right)\right|_{\text{supp}\, \partial h_2},\\
&(S_1, S_2) = \left.(S_1, S_2)\right|_{\text{supp}\, \partial h_1} + \left.(S_1, S_2)\right|_{\text{supp}\, \partial h_2}.
\end{split}
\end{equation}
The interpolation conditions further imply
\begin{equation}\label{eqn:2-patch-BV-2}
\begin{split}
&\left.(S_1, S_1)\right|_{\text{supp}\, \partial h_1} = \left.(S_1, S_{1 \cap 2})\right|_{\text{supp}\,\partial h_1} = \left.(S_{1 \cap 2}, S_{1 \cap 2})\right|_{\text{supp}\,\partial h_1},\\
&\left.(S_2, S_2)\right|_{\text{supp}\, \partial h_2} = \left.(S_2, S_{1 \cap 2})\right|_{\text{supp}\, \partial h_2} = \left.(S_{1 \cap 2}, S_{1 \cap 2})\right|_{\text{supp} \, \partial h_2},\\
&\left.(S_1,S_2)\right|_{\text{supp}\, \partial h_1} = \left.(S_2, S_{1 \cap 2})\right|_{\text{supp}\, \partial h_1}, \quad \left.(S_1,S_2)\right|_{\text{supp}\, \partial h_2} = \left.(S_1, S_{1 \cap 2})\right|_{\text{supp}\, \partial h_2}.
\end{split}
\end{equation}
It is now an exercise in expansion and cancellation, using \eqref{eqn:2-patch-BV-1} and \eqref{eqn:2-patch-BV-2}, to show that
\begin{equation}
\begin{split}
(S_{tot}, S_{tot}) &= (S_1, S_1) + (S_2,S_2) + (S_{1 \cap 2}, S_{1 \cap 2})\\
&+ 2(S_1, S_2) -2(S_2,S_{1 \cap 2}) - 2(S_1,S_{1 \cap 2}) = 0.
\end{split}
\end{equation}

\subsection{Subtlety in the Corner}\label{sssec:general-BV}

The treatment of a general patch decomposition follows exactly as \S\ref{sssec:two-patch-BV}, with the additional subtlety that boundaries of patches may intersect, resulting in `corners' in the intersections of patches, as in Fig. \ref{fig:smeared-boundaries}. These corners are a priori a problem for the interpolation conditions \eqref{eqn:inter-constraints} and \eqref{eqn:gauge-inter-constraint}, as near the corners of $U_{i \cap j}$ the string field $\Psi^{(i \cap j)}$ must interpolate between $\Psi^{(i)}$ and $\Psi^{(j)}$ across an arbitrarily small length scale, which means terms in the action that depend on the derivatives of $\Psi^{(i \cap j)}$ might scale with negative powers of $\epsilon_{i \cap j}$.

The resolution to this issue becomes clear once we consider the case of three patches, say $U_1$, $U_2$, and $U_3$. Let us consider the problematic region near the intersection of $\partial U_1$ and $\partial U_2$, where the string field $\Psi^{(1 \cap 2)}$ must interpolate between $\left. \Psi^{(1)}\right|_{\partial U_1}$ and $\left. \Psi^{(2)}\right|_{\partial U_{2}}$ across an arbitrarily short separation. The important point to note is that $\Psi^{(1 \cap 2 \cap 3)}$ must satisfy the exact same interpolation conditions in this region of target space, but enters the action \eqref{eqn:total-action} with the opposite sign. A straightforward way to use this to cancel any divergences is to add the further interpolation conditions
\begin{equation}\label{eqn:corner-interpolation}
\left.\Psi^{(i \cap j \cap k)}\right|_{\text{supp}\,\partial h_i \cap\, \text{supp}\,\partial h_j} \equiv \left.\Psi^{(i \cap j)}\right|_{\text{supp}\, \partial h_i \cap \, \text{supp}\, \partial h_j}, \quad \forall i,j,k.
\end{equation}
This ensures that the corner contributions from the interpolation fields cancel from \eqref{eqn:total-action} (and therefore from the BV antibracket as well).

The only contribution to the action from the $\partial U_1 \cap \partial U_2$ corner region therefore comes directly from $S_1$ and $S_2$. However, this contribution scales with positive powers of $\epsilon_{1 \cap 2}$, as $\Psi^{(1)}$ and $\Psi^{(2)}$ are taken to have no divergent behavior at the corners. The violation to gauge invariance and the closure of the antibracket therefore goes to 0 as we take $\epsilon_i$ to 0 in all of the patches.

The case of higher codimension smeared corners, with more than two smeared boundaries intersecting, follows identically, with more indices dressing the patches and string fields.

\subsection{The General Case}\label{sec:patch-by-patch general}

We may now demonstrate the gauge invariance of \eqref{eqn:total-action} and the closure of its BV antibracket. The computation is no different from \S\ref{sssec:two-patch-BV}, there is simply more bookkeeping.

We order the intersections of the patches $U_i$ according to their degree of separation: $U_i \cap U_j$ is a double intersection, $U_i \cap U_j \cap U_k$ is a triple intersection, $U_{i_1} \cap U_{i_2} \cap \ldots \cap U_{i_n}$ is an $n$-fold intersection, and so on. We must specify the boundary conditions that the interpolating string fields $\Psi^{(i_1 \cap i_2 \cap \ldots \cap i_n)}$ satisfy on each $n$-fold intersection.

Let $\sigma := \{i\}$ be a collection of patch indices, with $k$ running form $1$ to $n$. We use the notation $\sigma/j$ to denote the set $\{i\}$ with the index $j$ excluded, or more generally $\sigma / \tilde{\sigma}$ to denote the set $\{i\}$ with some subset $\tilde{\sigma}$ of indices excluded. The smeared boundary of their intersection, $\partial^{\epsilon} \bigcap_{i \in \sigma} U_{i}$ may be generically decomposed into two parts -- the part where all $n$ smeared boundaries intersect (call it $\mathcal{B}_1$), and the part where no more than $n-1$ smeared boundaries intersect (call it $\mathcal{B}_2$). The volume of $\mathcal{B}_1$ generically scales as $\epsilon^{n}$, where we suppress the index $i$ on the smearing scales $\epsilon_i$, as we take them all to be of the same order of magnitude, and we will eventually take all to 0 simultaneously.

Every component of $\mathcal{B}_2$ is a component of the intersection of some subset of the boundaries. Suppose we consider the subset of $\mathcal{B}_2$ where the patch $U_j$ does not appear in the intersection. Call this subset $\mathcal{B}_{2;j}$. $\mathcal{B}_{2;j}$ subset has already defined boundary conditions for the interpolating field $\Psi^{(\bigcap_{i \in \sigma / j} i)}$. We may therefore impose the boundary conditions
\begin{equation}\label{eqn:generic-inter-conditions}
\left.\Psi^{(\bigcap_{i \in \sigma} i)}\right|_{\mathcal{B}_{2;j}} \equiv \left.\Psi^{(\bigcap_{i \in \sigma / j} i)}\right|_{\mathcal{B}_{2;j}}, \quad \forall j.
\end{equation}
Note that \eqref{eqn:generic-gauge-interpolation} simultaneously includes the generalization of conditions such as \eqref{eqn:inter-constraints} and \eqref{eqn:corner-interpolation}. We must check that these boundary conditions are single valued on the overlaps $\mathcal{B}_{2;j} \cap \mathcal{B}_{2;j'}$ for $j \neq j'$. Note that this intersection is part of the boundary of the intersection of the $n-2$ patches indexed by $\sigma / \{j,j'\}$, which comes with its own interpolating string field $\Psi^{(\bigcap_{i \in \sigma / {j,j'}} i)}$, and so we have already imposed the conditions
\begin{equation}\label{eqn:bdy-consistency}
\left. \Psi^{(\bigcap_{i \in \sigma / j} i)}\right|_{B_{2;j} \cap B_{2;j'}} \equiv \left. \Psi^{(\bigcap_{i \in \sigma / j'} i)}\right|_{B_{2;j} \cap B_{2;j'}} \equiv \left. \Psi^{(\bigcap_{i \in \sigma / \{j,j'\}} i)}\right|_{B_{2;j} \cap B_{2;j'}},
\end{equation}
in our definitions of $\Psi^{(\bigcap_{i \in \sigma / j} i )}$ and $\Psi^{(\bigcap_{i \in \sigma / j'} i )}$.

The interpolating gauge transformations satisfy the exact same conditions:
\begin{equation}\label{eqn:generic-gauge-interpolation}
\left.\Lambda^{(\bigcap_{i \in \sigma} i)}\right|_{\mathcal{B}_{2;j}} \equiv \left.\Lambda^{(\bigcap_{i \in \sigma / j} i)}\right|_{\mathcal{B}_{2;j}}, \quad \forall j. 
\end{equation}
The analog of \eqref{eqn:bdy-consistency} again guarantees that the boundary conditions \eqref{eqn:generic-gauge-interpolation} are well defined on the intersections of $\mathcal{B}_{2;j}$ and $\mathcal{B}_{2;j'}$. We may now prove that the total action \eqref{eqn:total-action} is gauge invariant up to terms that scale with positive powers of the boundary smearing $\epsilon$. The important identities are
\begin{equation}
\begin{split}
&\delta_{\Lambda} S_{\bigcap_{i \in \sigma}} = \left. \delta_{\Lambda} S_{\bigcap_{i \in \sigma}}\right|_{\mathcal{B}_{1}} + \left. \delta_{\Lambda} S_{\bigcap_{i \in \sigma}}\right|_{\mathcal{B}_{2}},\\
&\left. \delta_{\Lambda} S_{\bigcap_{i \in \sigma}}\right|_{\mathcal{B}_{2;j}} = \left. \delta_{\Lambda} S_{\bigcap_{i \in \sigma/j}}\right|_{\mathcal{B}_{2;j}}.
\end{split}
\end{equation}

Consider the first term in \eqref{eqn:total-action} as a starting point. The contribution to the violation of gauge invariance is from the collective boundaries of the sets:
\begin{equation}
\delta_{\Lambda} \sum_i S_i = \sum_i \left.\delta_{\Lambda} {S_i}\right|_{\text{supp}\, \partial h_i}.
\end{equation}
Each boundary $\partial U_i$ has two components: a component that intersects at most one other patch (call this component $\mathcal{C}_{1,i}$), and a component that intersects more than one patch\footnote{Every point on every boundary of a patch must be contained in at least one other patch, otherwise the entire collection of patches has nontrivial boundary and we have not adequately covered the target space.} (call this component $\mathcal{C}_{2,i}$). Each $\mathcal{C}_{1,i}$ may further be subdivided into components $\mathcal{C}_{1,i;j}$, with $j \neq i$, where $\partial^{\epsilon}U_i$ specifically intersects the interior of patch $U_j$. Note that by definition of $\mathcal{C}_{1;i}$, the intersection $\mathcal{C}_{1,i;j} \cap \mathcal{C}_{1,k;l}$ is only nontrivial if $i=k$ and $j=l$. Now, for each pair $\mathcal{C}_{1,i;j}$ and $\mathcal{C}_{j;i}$ we have
\begin{equation}
\left. \delta_{\Lambda} S_{i \cap j} \right|_{\mathcal{C}_{1,i;j} \cup \mathcal{C}_{1,j;i}} = \left. \delta_{\Lambda} S_i\right|_{\mathcal{C}_{1,i;j}} + \left. \delta_{\Lambda} S_j\right|_{\mathcal{C}_{1,j;i}}.
\end{equation}
Therefore, all contributions to the violation of gauge invariance from the $\mathcal{C}_{1,i;j}$ cancel when we combine the first two terms of \eqref{eqn:total-action}. The only violation to the gauge invariance of the first two terms therefore comes from $\bigcup_i \mathcal{C}_{2;i}$, where the patch boundaries intersect at least two other patches.

Let us now consider the components of $\partial U_i$ that intersect exactly $n$ other patches. Call such a component $\mathcal{C}_{2,i;\sigma}$, where $\sigma$ is a collection of $n$ indices that don't include $i$. No two $\mathcal{C}_{2,i;\sigma}$ and $\mathcal{C}_{2,i;\tau}$ have nontrivial overlap unless $\sigma = \tau$. By definition, $\mathcal{C}_{2,i;\sigma}$ forms part of the boundary of an $n+1$-fold intersection. Such an intersection will support interpolating string fields of the form $\Psi^{(\bigcap_{i' \in i \cup \tilde{\sigma}}i')}$ for any $\tilde{\sigma} \subset \sigma$. For a generic collection of indices $\tau$ (which has cardinality $|\tau|$), call $\Psi^{(\bigcap_{i \in \tau} i)}$ a $|\tau|$-fold interpolating string field. $\mathcal{C}_{2,i;\sigma}$ defines boundary conditions for exactly ${n \choose k}$ $k$-fold interpolating string fields (in addition to the original string field $\Psi^{(i)}$ associated to the patch $U_i$). All these string fields satisfy the same boundary conditions, so we have
\begin{equation}
\left. \delta_{\Lambda} S_{\bigcap_{i' \in i \cup \tilde{\sigma}}}\right|_{\mathcal{C}_{2,i;\sigma}} = \left. \delta_{\Lambda} S_i \right|_{C_{2,i;\sigma}}, \quad \forall \tilde{\sigma} \subset \sigma.
\end{equation}
The contribution to the gauge transformation of \eqref{eqn:total-action} due to $\mathcal{C}_{2,i;\sigma}$ may therefore be computed as
\begin{equation}\label{eqn:total-gauge-cancellation}
\left.\delta_{\Lambda} S_{tot}\right|_{\mathcal{C}_{2,i;\sigma}} = \sum_k (-1)^k {n \choose k} \delta_{\Lambda} \left. S_i \right|_{\mathcal{C}_{2,i;\sigma}} = 0. 
\end{equation}
We have therefore proven that $S_{tot}$ is gauge invariant under a generic collection of gauge transformations satisfying \eqref{eqn:generic-gauge-interpolation}.

The computation of the BV antibracket proceeds similarly. We again need to make sense of the antibracket between distinct string fields, which we may do by introducing the family of maps $\pi_{\tilde{\tau} \rightarrow \tau}$, which for any subset $\tilde{\tau} \subset \tau$ is a map from the Hilbert space $\mathcal{H}_{\bigcap_{i' \in \tilde{\tau}}i'}$ to the Hilbert space $\mathcal{H}_{\bigcap_{i' \in \tau}i'}$. We again pick a component of the boundary, $\mathcal{C}_{i;\sigma}$, where $\partial U_i$ intersects exactly $n$ other patches whose indices comprise $\sigma$ (we do not distinguish $\mathcal{C}_1$ and $\mathcal{C}_2$ here, as just as before the computation is identical).

For any subset $\tilde{\sigma} \subset \sigma$, there are two string fields in this problem: $\Psi^{(\bigcap_{i' \in \tilde{\sigma}}i')}$, whose indices don't include $i$, and $\Psi^{(\bigcap_{i' \in i \cup \tilde{\sigma}}i')}$, whose indices do include $i$. The latter string fields satisfy a boundary condition on $\partial U_i$, and so for any two subsets of $\sigma$ (call them $\tilde{\sigma}_1$ and $\tilde{\sigma}_2$) their contribution to the antibracket will satisfy
\begin{equation}
\left.(S_{\bigcap_{i' \in i \cup \tilde{\sigma}_1}},S_{\bigcap_{i' \in i \cup \tilde{\sigma}_2}})\right|_{\mathcal{C}_{i;\sigma}} = \left.(S_{i},S_{\bigcap_{i' \in i \cup \tilde{\sigma}_2}})\right|_{\mathcal{C}_{i;\sigma}}, \quad \forall\,\tilde{\sigma}_1,\tilde{\sigma}_2 \subset \sigma.
\end{equation}
We then have, just as in \eqref{eqn:total-gauge-cancellation},
\begin{equation}
\begin{split}
&\sum_{\tilde{\sigma}_1 \subset \sigma} (-1)^{|\tilde{\sigma}_1| + 1} \left.(S_{\bigcap_{i' \in i \cup \tilde{\sigma}_1}},S_{\bigcap_{i' \in i \cup \tilde{\sigma}_2}})\right|_{\mathcal{C}_{i;\sigma}} \\
&= -\sum_k (-1)^k{n \choose k}\left.(S_i, S_{\bigcap_{i' \in i \cup \tilde{\sigma}_2}})\right|_{\mathcal{C}_{i;\sigma}} = 0.
\end{split}
\end{equation}
The exact same cancellation holds for $S_{\bigcap_{i' \in \tilde{\sigma}_2}}$, components of the action whose indicies don't include $i$:
\begin{equation}
\begin{split}
&\sum_{\tilde{\sigma}_1 \subset \sigma} (-1)^{|\tilde{\sigma}_1| + 1} \left.(S_{\bigcap_{i' \in i \cup \tilde{\sigma}_1}},S_{\bigcap_{i' \in \tilde{\sigma}_2}})\right|_{\mathcal{C}_{i;\sigma}} \\
&= -\sum_k (-1)^k{n \choose k}\left.(S_i, S_{\bigcap_{i' \in  \tilde{\sigma}_2}})\right|_{\mathcal{C}_{i;\sigma}} = 0.
\end{split}
\end{equation}
Summing all of these zeros together with alternating signs, and summing over all such boundary components $\mathcal{C}_{i;\sigma}$, we find the equation we wished to show:
\begin{equation}
(S_{tot}, S_{tot}) = 0.
\end{equation}

\subsection{Quantum BV Equation}

Conveniently, the cancellation of boundary contributions in neither the gauge transformation nor the BV antibracket required either of the two to vanish -- the boundary conditions \eqref{eqn:generic-gauge-interpolation} are sufficient. This means it is straightforward to extend our proof of BV quantizability to arbitrary loop order. This is somewhat expected, as our proof of the BV quantizability can be directly applied to the 1PI action of string field theory \cite{Sen:2014dqa}, which contains the information about the loop diagrams. This is because the BV master equation of the quantum 1PI action takes the form of the classical BV master equation \cite{Sen:2014dqa}.

First, we introduce the higher loop order string products $\{\Psi^n\}_{h_i;g}$, which are exactly the same as \eqref{eqn:hX-vertex} with $\mathcal{M}_n$ replaced by $\mathcal{M}_{n;g}$, the moduli space of the $n$-punctured genus-$g$ Riemann surface. Again, the location $z_{h}$ of the $h_i(X)$ insertion does not matter, and only the zero-mode contribution where $h(X_i)$ deforms the target space integration measure matters. The action within a single patch is
\begin{equation}
S_{h_i}(\Psi^{(i)}) := \frac{1}{g_s^2}\sum_n \sum_g \frac{(g_s^2)^{g}}{n!}\{\Psi^{(i)n}\}_{h_i;g},
\end{equation}
and the total action is again \eqref{eqn:total-action}. Gauge transformations are also amended in the obvious way -- their expression again uses the usual genus-$g$ string product without any $h_i(X)$ insertions:
\begin{equation}
\delta_{\Lambda^{(i)}}\Psi^{(i)} = Q_{B}\Lambda^{(i)} + \sum_n \sum_g \frac{( g_s^2)^g}{n!}[\Psi^{(i)n}, \Lambda^{(i)}]_g,
\end{equation}
and they satisfy the same interpolation conditions \eqref{eqn:gauge-inter-constraint}.

The proof that the patch boundary term contributions to $\delta_{\Lambda}S_{tot}$ vanishes now proceeds identically as before, and the proof of gauge invariance again reduces to the familiar SFT main identity as we may again drop all terms that depend on derivatives of the $h_i(X)$:
\begin{equation}
    \frac{\delta (\delta_{\Lambda^{(i)}S_i})}{\delta \Lambda^{(i)}} \sim h_i(X)\left(\text{genus-corrected main identity} \right) = 0
\end{equation}
The same argument holds for the vanishing of the quantum BV master equation
\begin{equation}
(S_{tot},S_{tot}) - g_s^2 \Delta S_{tot} = 0,
\end{equation}
in the collective interior of the patches, as in this region all $h_i(X)$ are identically 1. The only thing we need to do is show that the interpolating conditions \eqref{eqn:generic-inter-conditions} are enough for the boundary terms due to $\Delta S_{tot}$ to collectively vanish.

\subsection{Non-perturbative $\alpha'$ effects}
The crucial assumption in the construction of the global SFT action that solves the BV master equation in the patch-by-patch description is that string fields and string field theory action exhibit the target space locality. However, the spacetime locality in string theory is only a perturbative mirage, as the theory is not local at finite $\alpha'$ due to non-perturbative effects in $\alpha',$ for example, worldsheet instanton effects \cite{Dine:1986zy,Dine:1987bq} famously introduce such a non-locality. One can therefore expect that at finite $\alpha',$ the gauge invariance and the BV solvability of the string field theory action we constructed will break down by $\mathcal{O}(e^{-l^2/\alpha'})$ effects.\footnote{We thank Ted Erler for emphasizing this point to us.} An important problem to understand is therefore how to restore the non-perturbative breakdown of the gauge invariance by including non-perturbative $\alpha'$ effects. We leave this problem for future investigation.

\section{Example 1: Toroidal compactifications}\label{sec:example 1}
In this section, we shall study toroidal compactifications in the non-linear sigma model description as a warm-up. It is well known that toroidal compactifications admit an exact worldsheet description, so adopting the non-linear sigma model approach does not add anything new. However, this case functions as a proof of concept that the NLSM approach can recover topologically non-trivial phenomena, such as winding modes.

We shall study a 26-dimensional flat spacetime, where one spatial direction is compactified on a circle $S^1.$ We shall parameterize a point $p$ in $S^1$ by $\theta\in\Bbb{R},$ such that 
\begin{equation}
    \theta\equiv \theta+2\pi,
\end{equation}
and
\begin{equation}
    S^1\equiv  \Bbb{R}/ (\theta\sim \theta+2\pi)\,.
\end{equation}
We shall denote the equivalence class under the identification $\theta\equiv\theta+2\pi$ by $[\theta]\in[0,2\pi].$ Hence, topologically, the target space is $\mathcal{M}=\Bbb{R}_{24,1}\times S^1.$ We shall construct local coordinate charts 
\begin{equation}
    U_i:= \Bbb{R}_{24,1}\times \left(\frac{2(i-1)}{N}\pi -\epsilon, \frac{2i}{N}\pi+\epsilon\right)/_{\sim}\,,
\end{equation}
for an arbitrary integer $i,$ a large integer $N,$ and a small real number $\epsilon.$ We shall also construct the coordinate maps $\varphi_i:U_i\rightarrow \Bbb{R}_{24,1}\times (-R (\pi/N+\epsilon), R(\pi/N+\epsilon))$
\begin{equation}
    \varphi_i(\theta)= x^{25}_i=R\left(\theta -\frac{(2i-1)\pi}{N}\right)\,,
\end{equation}
where $x_i^{25}$ is the local coordinate in $U_i.$ Consequently, for each $U_i\cap U_{i+1},$ we can define a homomorphism 
\begin{equation}
    \varphi_{i+1}\cdot \varphi_i^{-1} : \Bbb{R}_{24,1}\times R( -\pi/N-\epsilon,\pi/N+\epsilon)\,, \quad x^{25}_i\mapsto x^{25}_{i+1}=x_i^{25}+\frac{2\pi}{N}R\,. 
\end{equation}
Note that we chose the local coordinate charts such that intersections of local coordinate charts are connected and contractible. So far, we have only specified the topology of the target space. We shall endow a differential structure by declaring that the metric of $\varphi_i(U_i)$ is given by
\begin{equation}
    ds^2=dX^AdX_A+R^2(dX^{25})^2\,.
\end{equation}
Hence, $S^1$ has the volume $2\pi R.$

We shall now study the background $\Bbb{R}_{24,1}\times S^1$ using the patch-by-patch description in string field theory. Let us stress again that since we already have the exact worldsheet description for such a background, we are not gaining anything by adopting the patch-by-patch description. However, this patch-by-patch description in string field theory will be useful to study string backgrounds whose exact worldsheet formulations are unknown.

To each local coordinate patch $U_i,$ we shall append closed string field theory $\text{SFT}_i$ that is constructed from the free field CFT. To construct the string field theory in the patch-by-patch description, we need to determine the gauge transformation induced by the diffeomorphism. Before we write down the form of the gauge transformation, a few comments are in order. First, fields in the low-energy supergravity basis are related to the string fields by non-linear field redefinitions. Second, as we are formulating string field theory with the BRST quantization, the gauge transformation we are setting up is inevitably an infinitesimal one. Hence, in principle, we can only reliably describe the diffeomorphism that describes a small change of local coordinates in string field theory. However, this is not a serious problem, because we can always cover the target space with a large number $N$ of local charts. This also squares nicely with the goal of organizing the theory in $\alpha'$ expansion, which requires the size of local patches to be somewhat small.

We write the gauge transformation induced by the diffeomorphism $\varphi_{i+1}\cdot \varphi_i^{-1}$ as 
\begin{equation}
    \label{eqn:gauge trans tor1}\Psi^{(i)}|_{U_{i+1}}=\Psi^{(i+1)}|_{U_i}+\sum_n\frac{1}{n!}[(\Psi^{(i+1)})^n\Lambda]\,,
\end{equation}
because $Q_B\Lambda=0\,.$ We shall determine the gauge transformation up to the linear order in the field expansion. Note that for we wrote the string fields in the basis where the tree-level kinetic term comes with $g_c^{-2}$ factor. Had we rescaled the string field $\Psi\mapsto g_c\Psi,$ each string bracket $[\Psi^n \Lambda]$ is rescaled by a factor of $g_c^{n}.$ Hence, one can think of the expansion \eqref{eqn:gauge trans tor1} as $g_c$ expansion, agreeing with the graviton expansion of the low-energy supergravity which is an expansion in $1/M_{pl}$. 

To determine the gauge transformation, we shall first start with the generic form of the massless ghost number 1 fields of the form
\begin{equation}
    \Lambda= \frac{g_c}{2\pi\alpha'} (\lambda_A c\partial X^A-\bar{\lambda}_A \bar{c}\bar{\partial}X^A) +\frac{g_c}{2\pi}\mu (\partial c+\bar{\partial}\bar{c})\,.
\end{equation}
Similarly, we shall write $\Psi$ as 
\begin{equation}
    \Psi=-\frac{g_c}{2\pi\alpha'} \epsilon_{AB} c\bar{c}\partial X^A \bar{\partial}X^B+\frac{g_c}{2\pi} (ec\partial^2 c -\bar{e}\bar{c}\bar{\partial}^2\bar{c}) +\frac{g_c}{4\pi \alpha'} (\partial c+\bar{\partial} \bar{c}) (f_A c \partial X^A+\bar{f}_A \bar{c} \bar{\partial}X^A)\,. 
\end{equation}
We compute
\begin{align}
    Q_B\Lambda=&\frac{g_c}{2\pi\alpha'} \left( -\partial_B\lambda_A-\partial_A\bar{\lambda}_B \right) c\bar{c} \partial X^A \bar{\partial}X^B+\frac{g_c}{2\pi} (\partial_A \lambda^A+\mu)c\partial^2 c+\frac{g_c}{2\pi} (-\partial_A\bar{\lambda}^A+\mu)\bar{c}\bar{\partial}^2\bar{c}\nonumber\\
    &+\frac{g_c}{4\pi\alpha'}(\partial c+\bar{\partial}\bar{c})\left( -(\partial^2\lambda_A+\partial_A\mu/2) c\partial X^A+(\partial^2\bar{\lambda}_A-\partial_A \mu/2) \bar{c}\bar{\partial}X^A)\right)\,.
\end{align}
Hence, we find
\begin{equation}
    \delta \epsilon_{AB}= \partial_B\lambda_A+\partial_A\bar{\lambda}_B\,,\quad \delta e= \partial_A\lambda^A +\mu\,,\quad \delta\bar{e}=\partial_A\bar{\lambda}^A -\mu
\end{equation}
\begin{equation}
    \delta f_A= -\partial^2\lambda_A -\frac{1}{2}\partial_A\mu \,,\quad \delta \bar{f}_A =\partial^2\bar{\lambda}_A -\frac{1}{2}\partial_A \mu\,.
\end{equation}
We shall fix $e=\bar{e}=d/2$ by using $\mu.$ Note that $f_A$ and $\bar{f}_A$ are auxiliary fields, so we can focus on $\delta \epsilon_{AB}$ to deduce the relation between the diffemorphism and $\Lambda$ to the first order in $g_c$ expansion. As we can check, $\epsilon_{(AB)},$ $\epsilon_{[AB]},$ and $\epsilon_{AB}\eta^{AB}-d/2$ behave as a graviton, anti-symmetric two-tensor, and a scalar, respectively, under the diffeomorphism
\begin{equation}
    \partial_B \lambda^A=\partial_B \bar{\lambda}^A=\frac{\partial \delta X ^A}{\partial X^B}\,, 
\end{equation}
where the infinitesimal diffeomorphism acts on the local coordinates as
\begin{equation}
    X^A\mapsto X^A+\delta X^A\,.
\end{equation}
Hence, we shall fix the gauge transformation given the transition function $\varphi_{i+1}\cdot \varphi_i^{-1}$ as
\begin{equation}
    \partial_B\lambda^A_i=\partial_B \bar{\lambda}^A_i = \frac{\partial}{\partial X^B_i} \left(\varphi_{i+1}\cdot \varphi_i^{-1}(X_i^A)-X_{i+1}^A\right)\,.
\end{equation}
As we mentioned previously, the transition map $\varphi_{i+1}\cdot \varphi_i^{-1}$ is a linear function in the Atlas we used to define the cover of $S^1$, and hence the derivative of $\lambda$ vanishes. 

As we have identified the transition function, we can construct the background solution around which we can perturb. However, as $S^1$ has a trivial spacetime curvature, the background solution, at tree-level, is trivial to all order in $\alpha'$ in perturbation theory. Therefore, we can use the string field theory whose local data is obtained with the free field CFT to construct quantized states. 

Let us illustrate the BRST quantization of the patch-by-patch description with a tachyon state. Let us construct tachyon state $T_i$ in every $\text{SFT}_i$
\begin{equation}
    V_i=  T_i c\bar{c} e^{i (k_\mu X^\mu+k_{25}^L X^{25,L}_i+k_{25}^R X^{25,R}_i)}\,.
\end{equation}
It is important to note that we treated the left-moving momentum and the right-moving momentum independently. Since we are not studying the flat Minkowski space, $k_L$ need not be the same as $k_R.$ The BRST quantization condition in the local patch $U_i$ implies
\begin{equation}
    k_\mu k^\mu +(k_{25}^L)^2=k_\mu k^\mu +(k_{25}^R)^2=\frac{4}{\alpha}\,,
\end{equation}
and the level matching condition implies
\begin{equation}
    (k_{25}^L)^2=(k_{25}^R)^2\,,
\end{equation}
but they do not constrain the value of $k_{25}^L$ and $k_{25}^R.$ Furthermore, we shall require that the tachyonic states don't create a branch cut. Since the tachyonic states are in $(-1,-1)$ picture, we cannot cancel the phase ambiguity using the ghost sector. Therefore, the matter CFT component of the vertex operator must be local. This leads to the condition
\begin{equation}
    \pi \alpha' \left((k_{25}^L)^2-(k_{25}^R)^2\right)\in \Bbb{Z}\,.
\end{equation}
As $V_i$ is a linearized field, the gauge transformation relating $V_i$ to $V_{i+1}$ is given as
\begin{equation}
    V_{i+1}|_{U_i}=V_i|_{U_{i+1}} +Q_B\Lambda_i=V_i|_{U_{i+1}}\,,
\end{equation}
which is identical to
\begin{equation}
    T_{i+1} e^{i k_{25} X^{25}_{i+1}}= T_i e^{ik_{25} X^{25}_i}\,.
\end{equation}
In large $N$ limit, we can treat $T_{i+1}-T_i$ as 
\begin{equation}
    T_{i+1} -T_i \simeq \frac{\partial T}{\partial x}(x_i=0) \,,
\end{equation}
which leads to
\begin{equation}
    T_i=\exp\left( \frac{2\pi k_{25} (i-1)R}{N}\right)\,.
\end{equation}
As we identified $X_{N+i}^{25}$ with $X_i^{25},$ this implies that $k_{25}$ must be quantized
\begin{equation}
    2\pi R (k_{25}^L+k_{25}^R)\in \Bbb{Z}\,,
\end{equation}
which is the correct quantization condition. As a result, we have
\begin{equation}
    \frac{\alpha'}{2R}(k_{25}^L-k_{25}^R)\in \Bbb{Z}\,,
\end{equation}
and
\begin{equation}
    k_L=\frac{n}{2R}+\frac{m R}{4\alpha'}\,,\quad k_R=\frac{n}{2R}-\frac{m R}{4\alpha'}\,.
\end{equation}
For $m\neq0,$ we find winding states. Due to the level matching condition, the tachyon state can only either be a KK state or a winding state
\begin{equation}
    nm=0\,.
\end{equation}

A final remark is in order. As we have seen here, treating the left-moving and right-moving directions independently was crucial to get the correct spectrum. Had we chosen $k_{25}^L=k_{25}^R,$ the winding states wouldn't have been properly found. Since we are treating the left-moving and right-moving worldsheet boson to be independent fields, we must also integrate over zero modes of the left-moving and right-moving worldsheet bosons independently by the following replacement
\begin{equation}
    \int d X_{25}\mapsto\int d(X_{25}^L+X_{25}^R) d(X_{25}^L-X_{25}^R)\,.
\end{equation}
Note that this replacement was used in the derivation of the double field theory \cite{Hull:2009mi}.

\section{Example 2: Calabi-Yau compactification}\label{sec:example 2}
In this section, we shall study Calabi-Yau compactifications in type II string theories. There are many versions of the definition of Calabi-Yau manifolds. In this section, by Calabi-Yau manifolds, we shall mean smooth compact manifolds with $SU(n)$ holonomy. Often, we shall restrict our attention to $n=3.$ 

The goal of this section is manifold. First, to explicitly find the background solution perturbatively in $\alpha'$ expansion. This corresponds to constructing Calabi-Yau CFTs in $\alpha'$ expansion. Second, to construct vertex operators for moduli fields. Due to the extended supersymmetry of Calabi-Yau compactifications in target space, the kinetic action of moduli fields is very well understood. However, once the spacetime supersymmetry is broken to $\mathcal{N}=1$ or none, via orientifolding, non-trivial superpotential, explicit susy breaking sources, or a combination thereof, such exact results are not available. Furthermore, even with the understanding of the kinetic action, it is not possible to compute generic physical observables. Constructing the vertex operators for moduli fields is the first step towards having a more comprehensive understanding of the stringy physics in string compactifications. Third, to relate the extended supersymmetry of the worldsheet to the symmetries of the target space physics. 

\subsection{Perturbative background solution }\label{sec:back CY}
In this section, we shall study the background solution that corresponds to the Calabi-Yau CFT. As the Calabi-Yau metric and, consequently, the Riemann tensor admit no known analytic form, we will not dwell on the explicit form of the metric. Rather, we shall organize the background solution in terms of geometric quantities, such as Riemann tensors. This is not fully satisfactory, but at least, if one has access to such quantities via numerical techniques\footnote{For recent progress on numerical Calabi-Yau metric, see, for example, \cite{Larfors:2021pbb,Larfors:2022nep,Ahmed:2023cnw,Constantin:2024yxh,Fraser-Taliente:2024etl,Berglund:2024uqv,Butbaia:2024tje}.} one can use our results to calculate physical observables. Also, to determine the existence of the global solution, one needs to make sure that the set of local solutions found in each local coordinate patch is compatible with the transition map. Since we do not know the explicit form of the transition maps for Calabi-Yau manifolds, we shall not attempt to construct the global solution. However, provided that the transition maps are constructed via numerical techniques, extending the solution found in this section to that of a global one is straightforward.

As we illustrated before, to each local coordinate patch $U_i,$ we shall append string field theory $\text{SFT}_i$ constructed from the free field CFT. We shall then construct the background solution perturbatively in $\epsilon$
\begin{equation}
    \epsilon:=\sqrt{R_{abcd}R^{abcd}}\alpha'\,.
\end{equation}
Let us expand the background solution as
\begin{equation}
    \Psi_0^{(i)}=\sum_n\epsilon^n\Psi_{0,n}^{(i)}\,.
\end{equation}

We shall start with the first-order solution. To write the first-order solution, it is useful to recall that in the normal coordinates, the metric expands as
\begin{equation}
    g_{AB}=\delta_{AB}-\frac{1}{3} R_{ACBD}X^CX^D+\dots\,.
\end{equation}
For K\"ahler manifold, we have a more rigid structure
\begin{equation}
    g_{a\bar{b}}=\delta_{a\bar{b}}-R_{a\bar{b}c\bar{d}}Z^c\bar{Z}^d+\dots\,.
\end{equation}
Therefore, we can write
\begin{align}
    \Psi_{0,1}^{(i)}=& \frac{1}{4\pi} R_{ABc\bar{d}}^{(i)}Z^{c}Z^{\bar{d}} c\bar{c} e^{-\phi}\psi^A e^{-\bar{\phi}}\bar{\psi}^B\,,\\
    =&\frac{1}{4\pi} R_{a\bar{b}c\bar{d}}^{(i)}Z^cZ^{\bar{d}}c\bar{c} e^{-\phi}e^{-\bar{\phi}} \left(\psi^a\bar{\psi}^{\bar{b}}+\psi^{\bar{b}}\bar{\psi}^a\right)\,.
\end{align}
where $A$ and $B$ runs through both the holomorphic and the anti-holomorphic indices. As one can check, the first-order background solution satisfies the reality condition of string fields. 

Let us first check if $\Psi_{0,1}^{(i)}$ satisfies the perturbative equation of motion at the first order. We find
\begin{align}
    Q_B\Psi_{0,1}^{(i)}=& -\alpha' (\partial c+\bar{\partial}\bar{c}) R_{a\bar{b}c\bar{d}}^{(i)}\delta^{c\bar{d}} c\bar{c} e^{-\phi}e^{-\bar{\phi}}(\psi^a\bar{\psi}^{\bar{b}}+c.c)\nonumber\\
    &+i\sqrt{\frac{\alpha'}{2}} R_{a\bar{b}c\bar{d}}^{(i)} Z^{\bar{d}} c\bar{c} \eta e^{-\bar{\phi}} \bar{\psi}^a \delta^{c\bar{b}}+c.c.\,.
\end{align}
Note that we used $v^A v_A=2 v_a v_{\bar{b}}g^{a\bar{b}}$ and $\eta^{a\bar{b}}=2\delta^{ab},$ $\eta_{a\bar{b}}=\delta_{ab}/2.$ As the curvature satisfies the Ricci-flatness condition, we conclude 
\begin{equation}
    Q_B\Psi_{0,1}^{(i)}=0\,.
\end{equation}

Now, let us study the second-order equation of motion
\begin{equation}
    Q_B\Psi_{0,2}^{(i)}=-\frac{1}{2} \mathcal{G}[(\Psi_{0,1}^{(i)})^2]\,.
\end{equation}
To solve the above equation, we shall split the equation of motion into the $L_0^+$ nilpotent components and the rest. We shall denote the projection to the $L_0^+$ nilpotent sector by $\Bbb{P}.$ As was studied in \cite{deLacroix:2017lif}, $(1-\Bbb{P})$ component of the equation of motion can be easily solved by using
\begin{equation}
    \{Q_B,b_0^+\}=L_0^+\,,
\end{equation}
that leads to
\begin{equation}
    (1-\Bbb{P})\Psi_{0,2}^{(i)}=-\frac{1}{2}\frac{b_0}{L_0^+}(1-\Bbb{P})\mathcal{G} [(\Psi_{0,1}^{(i)})^2]\,. 
\end{equation}
On the other hand, $\Bbb{P}$ projected component is non-trivial
\begin{equation}
    Q_B\Bbb{P}\Psi_{0,2}^{(i)}=-\frac{1}{2}\Bbb{P}\mathcal{G} [(\Psi_{0,1}^{(i)})^2]\,,
\end{equation}
and we need to explicitly solve it. We find
\begin{align}
    Q_B\Bbb{P}\Psi_{0,2}^{(i)}=\frac{\alpha'C_{S^2}}{2} S_{AB} (\partial c+\bar{\partial }\bar{c}) c\bar{c} e^{-\phi}e^{-\bar{\phi}}\psi^A\bar{\psi}^B
\end{align}
where 
\begin{align}
    32\pi^2\mathcal{S}_{AB}=&-\partial_C\delta g_{DB}\partial_A\delta g^{CD}-\partial_D\delta g_{AC}\partial_B\delta g^{CD}+2\partial_C\delta g_{AD}\partial^D\delta g^C_{~B}+\frac{1}{2}\partial_A\delta g_{CD}\partial_B\delta g^{CD}\nonumber\\
    &-\frac{1}{2}\delta g_{CD}\partial_A\partial_B\delta g^{CD}+\delta g_{CD}\partial_A\partial^D\delta g^{C}_{~B}+\delta g_{CD}\partial^C\partial_B\delta g^{D}_{~A}-2\delta g_{CD}\partial^C\partial^D\delta g_{AB}
\end{align}
where
\begin{equation}
    \delta g_{AB}^{(i)}=-\frac{1}{3}R_{ACBD}X^CX^D\,.
\end{equation}
Note that we placed the test field at $\infty$ to compute the square bracket. We averaged over PCO locations.

To solve the second-order background equation, we shall use the following ansatz
\begin{align}
    \Bbb{P}\Psi_{0,2}^{-1,-1}=&\mathcal{G}_{AB} c\bar{c}e^{-\phi}\psi^Ae^{-\bar{\phi}}\bar{\psi}^B+\mathcal{D} c\bar{c}(\eta \bar{\partial}\bar{\xi}e^{-2\bar{\phi}}-\partial\xi e^{-2\phi}\bar{\eta})\nonumber\\
    &+ \frac{i}{2}\mathcal{F}_A (\partial c+\bar{\partial}\bar{c}) c\bar{c} (e^{-\phi}\psi^A e^{-2\bar{\phi}}\bar{\partial}\bar{\xi}+e^{-2\phi}\partial\xi e^{-\bar{\phi}}\bar{\psi}^A)\,.
\end{align}
We find \cite{Kim:2024dnw}
\begin{align}
Q_B \Bbb{P}\Psi_{0,2}^{-1,-1}=&\mathcal{A}_{AB} (\partial c+\bar{\partial }\bar{c})c\bar{c} e^{-\phi}\psi^Ae^{-\bar{\phi}}\bar{\psi}^B +\mathcal{B}_A c\bar{c} (\eta e^{-\bar{\phi}}\bar{\psi}^A+c.c)\nonumber\\
&+ \mathcal{C} (\partial c+\bar{\partial}\bar{c}) c\bar{c}(\eta e^{-2\bar{\phi}}\bar{\partial}\bar{\xi}-e^{-2\phi}\partial\xi\bar{\eta})\,,
\end{align}
where
\begin{equation}
    \mathcal{A}_{AB}=-\frac{1}{2}\partial^2\mathcal{G}_{AB}-\frac{1}{2}(\partial_B\mathcal{F}_A+\partial_A\mathcal{F}_B)\,,
\end{equation}
\begin{equation}
    \mathcal{B}_A=i\partial^B\mathcal{G}_{AB} +i\mathcal{F}_A+i\partial_A\mathcal{D}\,,
\end{equation}
\begin{equation}
    \mathcal{C}=\frac{1}{2}\partial^A\mathcal{F}_A-\frac{1}{2}\partial^2\mathcal{D}\,.
\end{equation}
The background equation is then translated to
\begin{equation}
    \mathcal{A}_{AB}=\frac{\alpha' C_{S^2}}{2} S_{AB}\,,\quad \mathcal{B}_A=0\,,\quad \mathcal{C}=0\,.
\end{equation}
We shall choose the following ansatz as we studied in \S\ref{app:diff2}
\begin{equation}
    \mathcal{G}_{AB}=W_{AB}+ \delta \mathcal{G}_{AB} +a\frac{ C_{S^2}}{2\pi^2} \delta g_A^{~C} \delta g_{CB}\,,\quad  \mathcal{D}=W+\delta\mathcal{D}-b\frac{ C_{S^2}}{8\pi^2} \delta g_{AB}\delta g^{AB}\,,
\end{equation}
where $(a-b/2)=-1/32,$ and $W_{AB}$ and $W$ correspond to the zero modes. 

We can further simplify the equations of motion by adopting the complex coordinates
\begin{equation}
    32\pi^2\mathcal{S}_{ab}=\partial_a\delta g_{c\bar{d}}\partial_b\delta g^{c\bar{d}}\,,
\end{equation}
\begin{equation}
    32\pi^2\mathcal{S}_{\bar{a}\bar{b}}=\partial_{\bar{a}}\delta g_{c\bar{d}}\partial_{\bar{b}}\delta g^{c\bar{d}}\,,
\end{equation}
\begin{equation}
32\pi^2 S_{a\bar{b}}= \partial_a\delta g_{c\bar{d}}\partial_{\bar{b}}\delta g^{c\bar{d}}-3\delta g_{c\bar{d}}\partial^c\partial^{\bar{d}}\delta g_{a\bar{b}}\,.
\end{equation}

Assuming $\delta \mathcal{G}_{ab}=\delta\mathcal{G}_{\bar{a}\bar{b}}=0\,,$ we find
\begin{align}
    \mathcal{A}_{AB}=&-\frac{1}{2} \partial^2\delta \mathcal{G}_{AB}+\frac{1}{2} \partial_A\partial^C\delta \mathcal{G}_{CB}+\frac{1}{2}\partial_B\partial^C\delta \mathcal{G}_{CA}+\partial_A\partial_B\mathcal{D}\nonumber\\&-\frac{C_{S^2}}{4\pi^2} [a\partial^2 (\delta g_A^{~C}\delta g_{CB})-a \partial_A \partial^C(\delta g_{BD}\delta g^D_{~C})-a\partial_B \partial^C (\delta g_{AD}\delta g^{D}_{~C})]\nonumber\\
    &-b\frac{ C_{S^2}}{8\pi^2} \partial_A\partial_B (\delta g_{CD}\delta g^{CD})\,,
\end{align}
\begin{equation}
    \partial^2 W_{AB}-\partial_A \partial^CW_{CB}-\partial_B\partial^C W_{CA}-2\partial_A\partial_BW=0\,,
\end{equation}
\begin{equation}
    \partial^A W_{AB}+W_B+\partial_BW=0\,,
\end{equation}
and
\begin{equation}
    \partial^AW_A=\partial^2W\,.
\end{equation}
We shall furthermore require that
\begin{equation}
    W_{AB}\eta^{AB}+2W=-\frac{1}{16\pi} \log\lambda \partial_C\delta g_{AB}\partial^C\delta g^{AB}-\frac{1}{16\pi}\partial_C\partial_D\delta g_{AB}\partial^C\partial^D\delta g^{AB}(1-\log^2\lambda)\,.
\end{equation}

Let us first solve for the zero modes. Let us write $W$ and $W_{AB}$ as
\begin{equation}
    W=\alpha \log\lambda \partial_C\delta g_{AB}\partial^C\delta g^{AB}+\beta \partial_C\partial_D\delta g_{AB}\partial^C\partial^D\delta g^{AB}\,,
\end{equation}
\begin{equation}
    W_{AB}=\gamma \log\lambda \partial_C\delta g_{AD}\partial^C \delta g_{B}^{~D}+\delta \log\lambda \delta g_{CD}\partial^C\partial^D  \delta g_{AB}+\epsilon \partial_A \partial_C\delta g_{DE}\partial_B \partial^C \delta g^{DE}\,.
\end{equation}
Then, we find
\begin{equation}
    \partial^2 W_{AB}-\partial_A\partial^CW_{CB}-\partial_B \partial^{C}W_{CA}=4\alpha \log\partial_A\partial_C\delta g_{DE}\partial_B\partial^C\delta g^{DE}\,.
\end{equation}
We shall require $W_{ab}=W_{\bar{a}\bar{b}}=0\,.$ Then, we find
\begin{equation}
    \partial_a\partial^{\bar{c}}W_{b\bar{c}}+\partial_b\partial^{\bar{c}}W_{a\bar{c}}=0\,,
\end{equation}
\begin{equation}
    2\gamma \log\lambda \partial_c\partial_{\bar{d}}\delta g_{a\bar{e}}\partial^c\partial^{\bar{d}}\delta g_{\bar{b}}^{~\bar{c}} -4\delta \log\lambda \partial_a \partial^{c}\delta g_{d\bar{e}} \partial^d\partial^{\bar{e}} \delta g_{c\bar{b}} =16\alpha\log\lambda\partial_a\partial_{\bar{c}}\delta g_{d\bar{e}}\partial_{\bar{b}}\partial^{\bar{c}}\delta g^{d\bar{e}}\,,
\end{equation}
and
\begin{equation}
    2\alpha+\gamma=-\frac{1}{16\pi}\,,\quad  2\beta+\epsilon =-\frac{1}{16}(1-\log^2\lambda)\,.
\end{equation}
In total, there are five parameters and we only have three non-trivial constraints. So, the system is not fully constrained, which indicates that there exists moduli. We shall not attempt to fix the moduli completely in this work.

We write the non-zero mode parts in holomorphic coordinates
\begin{align}
    \mathcal{A}_{ab}=&\frac{1}{2}\partial_a\partial^{\bar{c}}\delta \mathcal{G}_{b\bar{c}}+\frac{1}{2}\partial_b\partial^{\bar{c}}\delta\mathcal{G}_{a\bar{c}}-\partial_a\partial_b \mathcal{G}_{c\bar{d}}\eta^{c\bar{d}}\nonumber\\&+\frac{aC_{S^2}}{4\pi^2} (\partial_a\partial^{\bar{c}}( \delta g_{b\bar{d}}\delta g^{\bar{d}}_{~\bar{c}})+\partial_b\partial^{\bar{c}}( \delta g_{a\bar{d}}\delta g^{\bar{d}}_{~\bar{c}}))-b\frac{ C_{S^2}}{4\pi^2} \partial_a\partial_b (\delta g_{c\bar{d}}\delta g^{c\bar{d}})\nonumber\\
    =&\frac{1}{2}\partial_a\partial^{\bar{c}}\delta \mathcal{G}_{b\bar{c}}+\frac{1}{2}\partial_b\partial^{\bar{c}}\delta\mathcal{G}_{a\bar{c}}-\partial_a\partial_b \mathcal{G}_{c\bar{d}}\eta^{c\bar{d}}+(a-b)\frac{ C_{S^2}}{2\pi^2} \biggr[R_{b\bar{d}c \bar{e}} R_{a\bar{c} d \bar{f}}\eta^{c\bar{c}}\eta^{d\bar{d}} x^{\bar{f}} x^{\bar{e}}\biggr]\,,
\end{align}
\begin{align}
    \mathcal{A}_{a\bar{b}}=& - \partial^c\partial_c\delta\mathcal{G}_{a\bar{b}}+\frac{1}{2}\partial_a\partial^{c}\delta\mathcal{G}_{c\bar{b}}+\frac{1}{2}\partial_{\bar{b}}\partial^{\bar{c}}\delta \mathcal{G}_{a\bar{c}}-\partial_a\partial_{\bar{b}} \delta\mathcal{G}_{c\bar{d}}\eta^{c\bar{d}}\nonumber\\&-b\frac{ C_{S^2}}{4\pi^2} \partial_a\partial_{\bar{b}} (\delta g_{c\bar{d}}\delta g^{c\bar{d}})-a\frac{ C_{S^2}}{4\pi^2}\left[2\partial_d\partial^d (\delta g_a^{~c}\delta g_{c\bar{b}})-\partial_a\partial^c (\delta g_{d\bar{b}}\delta g^{d}_{~c})-\partial_{\bar{b}}\partial^{\bar{c}}(\delta g_{a\bar{d}}\delta g^{\bar{d}}_{~\bar{c}}) \right] \\
    =&- \partial^c\partial_c\delta\mathcal{G}_{a\bar{b}}+\frac{1}{2}\partial_a\partial^{c}\delta\mathcal{G}_{c\bar{b}}+\frac{1}{2}\partial_{\bar{b}}\partial^{\bar{c}}\delta \mathcal{G}_{a\bar{c}}-\partial_a\partial_{\bar{b}} \delta\mathcal{G}_{c\bar{d}}\eta^{c\bar{d}}\nonumber\\
    &-b\frac{C_{S^2}}{2\pi^2}(\delta g^{c\bar{d}}\partial_a\partial_{\bar{b}}\delta g_{c\bar{d}}+\partial_a\delta g_{c\bar{d}}\partial_{\bar{b}}\delta g^{c\bar{d}})-a\frac{C_{S^2}}{2\pi^2} \biggr( \partial_d\delta g_{a\bar{c}}\partial^d\delta g^{\bar{c}}_{~\bar{b}}- \delta g^{c\bar{d}}\partial_a\partial_{\bar{b}}\delta g_{c\bar{d}} \biggr)\,\\
    =&\frac{C_{S^2}}{32\pi^2} (\partial_a\delta g_{c\bar{d}}\partial_{\bar{b}}\delta g^{c\bar{d}}-3\delta g_{c\bar{d}}\partial^c\partial^{\bar{d}}\delta g_{a\bar{b}})\,.
\end{align}
Note that we determined in \S\ref{app:diff2} and \S\ref{app:diff3}
\begin{equation}
    a-\frac{1}{2}b=-\frac{1}{32}\,,
\end{equation}
and
\begin{equation}
    a=-\frac{1}{8}\,.
\end{equation}
Note that requiring the metric is K\"ahler, meaning
\begin{equation}
    \delta\mathcal{G}_{a\bar{b}}:=\partial_a\partial_{\bar{b}}\mathcal{K}\,,
\end{equation}
also leads to $a=-1/8$ and $b=-3/16.$

We finally find
\begin{equation}
    \partial^c\partial_c \delta\mathcal{G}_{a\bar{b}}=-\frac{C_{S^2}}{8\pi^2} \partial_a\delta g_{c\bar{d}}\partial_{\bar{b}}\delta g^{c\bar{d}}-\frac{C_{S^2}}{8\pi^2}\delta g_{c\bar{d}}\partial^c\partial^{\bar{d}}\delta g_{a\bar{b}}=-\frac{C_{S^2}}{8\pi^2} \partial_c\delta g_{a\bar{d}}\partial^c \delta g^{\bar{d}}_{~\bar{b}}-\frac{C_{S^2}}{8\pi^2}\delta g_{c\bar{d}}\partial^c\partial^{\bar{d}}\delta g_{a\bar{b}}\,.
\end{equation}
Note that the last equality holds as we are using the K\"ahler normal coordinates. 

We shall use the ansatz
\begin{equation}
    \mathcal{G}_{a\bar{b}}=\partial_a\bar{\partial_b }\mathcal{K}\,,
\end{equation}
with
\begin{equation}
    \mathcal{K}=\frac{1}{(3!)^2} K_{a\bar{b}c\bar{d}e\bar{f}} x^ax^{\bar{b}}x^cx^{\bar{d}}x^ex^{\bar{f}}\,,
\end{equation}
where $K$ is symmetric under the exchange of holomorphic indices and anti-holomorphic indices, respectively. Then we need to solve, 
\begin{equation}
    K_{a\bar{b}c\bar{d}e\bar{f}} x^ex^{\bar{f}}\eta^{c\bar{d}} =-\frac{C_{S^2}}{8\pi^2} R_{a\bar{d}c\bar{f}} R^{\bar{d}c}_{~~\bar{b}e}x^ex^{\bar{f}}-\frac{C_{S^2}}{8\pi^2} R_{c\bar{d}e\bar{f}} R^{c\bar{d}}_{~~a\bar{b}}x^ex^{\bar{f}}\,.
\end{equation}
We can write the right-hand side as
\begin{equation}
    -3\frac{C_{S^2}}{8\pi^2} \eta^{g\bar{h}}\eta^{c\bar{d}}x^e x^{\bar{f} }(R_{g(\bar{b}a\bar{d}}R_{c\bar{f}e)_h\bar{h}})\,,
\end{equation}
and consequently, we find
\begin{equation}
    K_{a\bar{b}c\bar{d}e\bar{f}}=-3\frac{C_{S^2}}{8\pi^2} \eta^{g\bar{h}}(R_{g(\bar{b}a\bar{d}}R_{c\bar{f}e)_h\bar{h}})\,.
\end{equation}

\subsection{Shifted BRST charge}
In this section, we shall study the $L_0^+$ nilpotent sector of the BRST charge of the Calabi-Yau background. One of the main purposes is to read off the shifted energy-momentum tensor of the Calabi-Yau background. 

Given the perturbative background solution
\begin{equation}
    \Psi_0^{(i)}=\sum_n\epsilon^n \Psi_{0,n}^{(i)}\,,
\end{equation}
the shifted BRST operators are defined as
\begin{equation}
    \hat{Q}_B:=Q_B+\mathcal{G}K\,,
\end{equation}
and
\begin{equation}
    \tilde{Q}_B:=Q_B+K\mathcal{G}\,,
\end{equation}
where
\begin{equation}
    K|\psi\rangle:= \sum_n \frac{1}{n!}\left[\left( \Psi_0^{(i)}\right)^n \otimes \psi\right]\,.
\end{equation}
For the NSNS sector, we have $\tilde{Q}_B=\hat{Q}_B,$ because $\mathcal{G}=1.$ 

Once we have found $\tilde{Q}_B,$ we can equate $\tilde{Q}_B$ with
\begin{equation}
    \tilde{Q}_B= c\left(\tilde{T}_m-\frac{1}{2}(\partial\phi)^2-\partial^2\phi-\eta\partial\xi\right)+\eta e^\phi \tilde{T}_F +bc\partial c -\eta\partial\eta be^{2\phi}\,,
\end{equation}
where $\tilde{T}_m$ is the shifted energy-momentum tensor, and $\tilde{T}_F$ is the shifted worldsheet supercurrent. 
\subsection{Spacetime supersymmetry}
Spacetime supersymmetry is phrased as a fermionic gauge symmetry of superstring field theory under which the background solution does not transform such that
\begin{equation}
    Q_B|\Lambda\rangle+\sum_n \frac{1}{n!} \mathcal{G}[\Lambda\otimes\Psi_0^n]=0\,,
\end{equation}
where $\Lambda$ is in (NS,R) and (R,NS) sectors. In this section, we shall first explicitly study the spacetime supersymmetry to the first order in $\epsilon$ expansion, and argue that the spacetime supersymmetry persists to all orders in $\epsilon$ expansion. We will then use the spacetime supersymmetry to argue that the background solution exists to all orders in $\epsilon$ expansion, which was already expected from the exactness of (2,2) SCFT. 

Before studying the spacetime supersymmetry, we shall first collect conventions for spacetime spinors and extended supersymmetry of the worldsheet. 

For spin fields of the worldsheet, we shall use 16 component Weyl spinors. Let us define the 10d chirality operator $\Gamma_{10},$ the 4d chirality operator $\Gamma_4,$ and the 6d chirality operator $\Gamma_6$ as 
\begin{equation}
    \Gamma_{10}:=\Gamma^0\Gamma^1\dots\Gamma^9\,,
\end{equation}
\begin{equation}
    \Gamma_4:= -i \Gamma^0\Gamma^1\Gamma^2\Gamma^3\,,
\end{equation}
\begin{equation}
    \Gamma_6:=i\Gamma^4\Gamma^5\Gamma^6\Gamma^7\Gamma^8\Gamma^9\,.
\end{equation}
We shall use Greek letters $\alpha,\beta,\dots$ for 16 component indices. Chiral spinor index will be denoted by a lower Greek index. Similarly, anti-chiral spinor index will be denoted by an upper Greek index. Four-dimensional chiral spinor index will be denoted by a lower index $\alpha_{(4)},$ and four-dimensional anti-chiral spinor index will be denoted by an upper index $\alpha_{(4)}.$ Similarly, six-dimensional chiral and anti-chiral indices will be denoted by lower and upper index $\alpha_{(6)},$ respectively. We shall denote the ten-dimensional spinor fields by either $\Sigma_\alpha$ or $\Sigma^\alpha.$ Four-dimensional spinor fields will be denoted by $S_{\alpha_{(4)}}$ and $S^{\alpha_{(4)}}.$ Lastly, six-dimensional spinor fields will be denoted by $\mathcal{S}_{\alpha_{(6)}}$ and $\mathcal{S}^{\alpha_{(6)}}.$ It is useful to note that
\begin{equation}
    16=(2,4)\oplus(\bar{2},\bar{4})\,,\quad \bar{16}=(2,\bar{4})\oplus (\bar{2},4)\,.
\end{equation}

We shall choose the GSO projection such that 
\begin{equation}
    e^{-\phi/2}\Sigma_\alpha\,, e^{-3\phi/2} \Sigma^\alpha\,, e^{-\bar{\phi}/2} \overline{\Sigma}_\alpha\,,e^{-3\bar{\phi}/2}\overline{\Sigma}^\alpha\,,
\end{equation}
are projected in. Hence, we shall work with type IIB string. To work with type IIA string, one can change the GSO projection so that the GSO even states are not symmetric under the exchange of holomorphic and anti-holomorphic sectors. 

In a Calabi-Yau compactification, not all internal spinors are globally well-defined due to a topological obstruction. Globally well-defined internal spinors are naturally described with complex geometry, as are many other properties of Calabi-Yau manifolds. For this reason, we shall summarize the index convention for tangent and cotangent vectors. We shall denote the ten-dimensional indices by capital Latin letters, 4 dimensional indices by lower Greek letters, and 6 dimensional indices by last half of lower Latin letters. For holomorphic indices for the internal directions, we shall use first half of lower Latin letters, and for anti-holomorphic indices for the internal directions, we shall use barred first half of lower Latin letters. Also, for free fields along the internal direction, we may often use tilded fields to avoid confusion. 

Also, Calabi-Yau manifolds are complex K\"ahler. Hence, the target space metric and, correspondingly, the background solution should satisfy the K\"ahlerity. We shall assume that background solution at any given order then takes the following form
\begin{align}
    \Psi_0=& -\frac{g_c}{4\pi} \mathcal{G}_{a\bar{b}} c\bar{c} e^{-\phi}\psi^a e^{-\bar{\phi}}\bar{\psi}^{\bar{b}}+c.c +\frac{g_c}{4\pi}\mathcal{D}c\bar{c} (\eta e^{-2\bar{\phi}}\bar{\partial}\bar{\xi}-e^{-2\phi}\partial\xi \bar{\eta})\nonumber\\&+i\frac{g_c}{4\pi\sqrt{2\alpha'}}  \mathcal{F}_A(\partial c+\bar{\partial}\bar{c})  c\bar{c} (e^{-\phi}\psi^A e^{-2\bar{\phi}}\bar{\partial}\bar{\xi} +e^{-2\phi}\partial\xi e^{-\bar{\phi}}\bar{\psi}^B)\,.
\end{align}
In general, there is a non-trivial field redefinition between string fields and the supergravity field. Hence, it is not straightforward to show that the background solution for a K\"ahler manifold should take the above form, although we have checked in the first two orders our ansatz is correct. Nevertheless, we will show that given the ansatz for the background solution, we can show the existence of the massless graviton. Also, we shall assume that the physical dilaton stays constant, and we can solve SFT eom order by order in both $\alpha'$ and $g_s$ expansions.

With the K\"ahlerity of the internal manifold, we can now define the operators that determine the extended $\mathcal{N}=2$ supersymmetry of the worldsheet. Let us first define
\begin{equation}
    I:= \psi^a\psi_a\,,\quad X:= \tilde{\psi}^1\tilde{\psi}^2\tilde{\psi}^3\,,\quad \tilde{X}:=\tilde{\psi}^{\bar{1}}\tilde{\psi}^{\bar{2}}\tilde{\psi}^{\bar{3}}\,.
\end{equation}
We can bosonize $I,$ $X,$ and $\tilde{X}$ as 
\begin{equation}
    I=i\sqrt{3}\partial H\,, X=e^{i\sqrt{3}H}\,, \tilde{X} =e^{-i\sqrt{3}H}\,,
\end{equation}
where 
\begin{equation}
    H(z) H(0)\sim -\log|z|^2\,.
\end{equation}
A state $V_q$ with $U(1)_R$ charge $q$ can be written as
\begin{equation}
    V_q= e^{i\frac{q}{\sqrt{3}}H} \tilde{V}_q\,.
\end{equation}
It is important to note $\tilde{V}_q$ does not have a non-trivial OPE with respect to $e^{i\alpha H}$ and $\partial H.$ We can now define spinor fields
\begin{equation}
    \eta:= e^{i\frac{\sqrt{3}}{2}H}\,, \bar{\eta}:= e^{-i\frac{\sqrt{3}}{2}H}\,.
\end{equation}
The fields $\eta$ and $\bar{\eta}$ correspond to spinors that are invariant under $SU(3)$ holonomy of the Calabi-Yau. 

We shall first establish that $I$ is a well-defined current of the worldsheet theory. It is straightforward to check that $(0,0)$ picture integrated form of the background solution has trivial $U(1)_R$ charge both in the holomorphic and the anti-holomorphic sector. Hence, in the conformal perturbation theory in $\alpha'$ expansion, the weight of the $U(1)_R$ current is not changed. As a result, despite the $U(1)_R$ current being constructed in the large volume approximation, we conclude that the $U(1)_R$ current remains exact to all orders in $\alpha'$ perturbation theory. Note that this agrees with the well-known results of $\mathcal{N}=2$ non-linear sigma models \cite{Sen:1986mg}.

The existence of the $U(1)_R$ current imposes a strong restriction on the correlators. In particular, only the conformal correlators that have the vanishing $U(1)_R$ charge can be non-trivial. This can be understood as follows. For a conformal correlator, we can insert a loop encircling no vertex operator on which the $U(1)_R$ current is integrated. Now, to evaluate the effect of the $U(1)_R$ current insertion, we need to make a choice on how to interpret the contour. The first choice is to pick up the poles within the circle, and the second choice is to pick up poles from the outside of the circle, and the two must agree modulo the sign. Because the first option gives zero, and the second option produces the original correlator multiplied by the sum of the $U(1)_R$ charge, unless the total $U(1)_R$ charge vanishes, the conformal correlator should vanish.

We shall now study the condition for unbroken spacetime supersymmetry. Due to some technicalities, it is easier to study the existence of spacetime supercharge indirectly. Rather than showing that there is a global symmetry of SFT in the Ramond sector, we shall show that the four dimensional gravitino state remains massless to an arbitrary order in $\alpha'.$ The gravitino state takes the following form
\begin{equation}
    \Lambda_1:=\epsilon^{\alpha \mu}c\bar{c} e^{-\phi/2}S_\alpha \eta_s e^{-\bar{\phi}}\bar{\psi}_\mu \,,\quad \Lambda_2:=\epsilon_\alpha^\mu c\bar{c} e^{-\phi/2}S^\alpha \tilde{\eta}_s e^{-\bar{\phi}}\bar{\psi}_\mu\,, 
\end{equation}
and similar forms for the anti-holomorphic gravitino states.

To show that the gravitino remains massless to all orders in $\alpha',$ we can show that the following string vertices vanish for an arbitrary $n$
\begin{equation}
    \{ \Psi_0^n \otimes \Lambda\otimes \mathcal{V}_t\}\,,
\end{equation}
where $\mathcal{V}_t$ is a test gravitino field
\begin{equation}
    \mathcal{V}_t= f c\bar{c} e^{-\phi/2} S^\alpha\tilde{\eta}_s e^{-\bar{\phi}}\bar{\psi}^\nu\,.
\end{equation}
and the similar fields for the complex conjugate and the opposite 4d chirality. Note that in the free theory, we can in fact write more gravitino states. However, in Calabi-Yau manifolds, due to the topological obstruction, $\Lambda$ and $\mathcal{V}_t$ and their opposite chirality are the only choices for the four-dimensional gravitino operators.

Now, let us show that the following string vertex vanishes for an arbitrary $n$
\begin{equation}
\mathcal{A}_{\Lambda}=\{    \Psi_0^n\otimes \Lambda_1\otimes  \mathcal{V}_{t}\}. 
\end{equation}
For the evaluation of the above string vertex, we shall use a different gauge in which the background solution takes the following form
\begin{equation}
    \Psi_0=\delta g_{a\bar{b}} c\bar{c}e^{-\phi}\psi^a e^{-\bar{\phi}}\bar{\psi}^{\bar{b}}+c.c +e c\bar{c} e^{-2\phi}\partial\xi \bar{\eta} +\bar{e} c\bar{c} \eta e^{-2\bar{\phi}}\bar{\partial}\bar{\xi}\,.
\end{equation}
We shall focus on the contraction of the CFT correlator where $n_{g,h},$ $n_{g,a},$ $n_e,$ and $n_{\bar{e}}$ of the first term, second term, third term, and the last terms in $\Psi_0$ are used. To evaluate $\mathcal{A}_{\Lambda,1},$ we need to insert $n-1$ holomorphic PCOs, and $n$ anti-holomorphic PCOs. By $n_c,~n_{F,h},~n_{F,a}~n_{\eta}$ we shall denote the numbers of $c\partial \xi,$ $e^\phi T_F^+,$ $e^\phi T_F^-,$ and $-\partial \eta b e^{2\phi}-\partial(\eta be^{2\phi})$ of PCOs that are used in the contraction. We define $\bar{n}_c,~\bar{n}_{F,h},~\bar{n}_{F,a},$ and $\bar{n}_\eta$ similarly. Then we find that the net number of $b,c$ ghost, $U(1)_R$ charge, background $\phi$ charge, and  the net number of $\xi,~\eta$ ghosts are
\begin{center}
    \begin{tabular}{|c|c|}\hline
      $n_{c-b}$   & $3-n_\eta+n_c$  \\
      $Q_{U(1)_R}$   & $n_{g,h}-n_{g,a}+n_{F,h}-n_{F,a}$\\
      $Q_\phi$&$-1-n_{g,h}-n_{g,a}-2n_e+n_{F,h}+n_{F,a}+2n_\eta$\\ $n_{\xi-\eta}$& $n_c-n_\eta+n_e-n_{\bar{e}}$   \\
      $\bar{n}_{c-b}$& $3-\bar{n}_\eta +\bar{n}_c$\\
      $\bar{Q}_{U(1)_R}$&$-n_{g,h}+n_{g,a}+\bar{n}_{F,h}-\bar{n}_{F,a}$\\
      $\bar{Q}_\phi$& $-2-n_{g,h}-n_{g,a}-2n_{\bar{e}}+\bar{n}_{F,h}+\bar{N}_{F,a}+2\bar{n}_\eta$\\
      $\bar{n}_{\xi-\eta}$& $\bar{n}_c-\bar{n}_\eta +n_{\bar{e}}-n_e$\\\hline
    \end{tabular}
\end{center}

For the correlation function to be non-vanishing, we need
\begin{equation}
    n_{c-b}=3\,,\quad Q_{U(1)_R}=0\,,\quad Q_\phi=-2\,,\quad n_{\xi-\eta}=0\,,
\end{equation}
with the constraints 
\begin{equation}
    n_c+n_{F,h}+n_{F,a}+n_\eta=n-1\,,\quad \bar{n}_c +\bar{n}_{F,h}+\bar{n}_{F,a}+\bar{n}_\eta=n\,.
\end{equation}
The only non-trivial equations are
\begin{equation}
    n_{g,h}-n_{g,a}+n_{F,h}-n_{F,a}=0\,,\quad n_{g,h}-n_{g,a}+\bar{n}_{F,a}-\bar{n}_{F,h}=0\,.
\end{equation}
Let us assume $n$ is odd. Since $n_c=n_\eta,$ $n_{F,h}+n_{F,a}$ must be even. On the other hand, $\bar{n}_{F,h}+\bar{n}_{F,a}$ must be odd. Because 
\begin{equation}
    n_{F,h}-n_{F,a}\equiv n_{F,h}+n_{F,a}\equiv 0 \mod 2\,,
\end{equation}
and 
\begin{equation}
    \bar{n}_{F,h}-\bar{n}_{F,a}\equiv 1\mod 2\,,
\end{equation}
we conclude that there is no solution because $n_{g,h}-n_{g,a}$ cannot be both odd and even. Similarly, when $n$ is even, we conclude that there is no solution. Therefore, 
\begin{equation}
    \{\Psi_0^n\otimes\Lambda_1\otimes \mathcal{V}_{t,1}\}=0\,,
\end{equation}
for an arbitrary $n.$

As a result, we conclude that the four-dimensional gravitino remains massless and henceforth spacetime supersymmetry is preserved to all orders in $\alpha'$ expansion. 

The existence of the spacetime supersymmetry to all orders in $\alpha'$ in turn implies that the perturbative background solution exists to all orders in $\alpha'$ \cite{Sen:2015uoa}.

\subsection{Vertex operators for moduli fields}
Finally, we shall find the local forms of the vertex operator for moduli fields of Calabi-Yau compactifications. Of the hyper multiplets, we will study K\"ahler moduli. And among the vector multiples, we will study complex structure moduli. At the leading order in $\alpha',$ vertex operators for both moduli take the following form \cite{Alexandrov:2021shf,Alexandrov:2021dyl}
\begin{equation}
    V_{m,0}^{-1,-1}=\frac{1}{4\pi} \delta g_{AB,m} c\bar{c}\left(e^{-\phi}\psi^A e^{-\bar{\phi}}\bar{\psi}^B -\frac{\eta^{AB}}{2}(\eta\bar{\partial}\bar{\xi}e^{-2\bar{\phi}}-e^{-2\phi}\partial\xi\bar{\eta})\right)e^{ik\cdot x}\,,
\end{equation}
where $k$ is the four-momentum. The index $m$ can either take the value $z$ or $t$, depending on whether we are considering the complex structure moduli or the K\"alher moduli. The BRST condition imposes
\begin{equation}
    k^2=0\,.
\end{equation}
For complex structure moduli, $\delta g_{AB,z}$ is given by
\begin{equation}
    \delta g_{a\bar{b},z}=0\,, 
\end{equation}
\begin{equation}
    \delta g_{ab,z}=\delta \bar{z}^i \frac{\Omega_{acd} \eta^{c\bar{c}}\eta^{d\bar{d}}}{|\Omega|^2} (\overline{\chi}_i)_{b\bar{c}\bar{d}}\,,
\end{equation}
with the similar form for $\delta g_{\bar{a}\bar{b}}.$ For K\"ahler moduli, $\delta g_{AB,t}$ is given by
\begin{equation}
    \delta g_{ab,t}=\delta g_{\bar{a}\bar{b},t}=0\,,
\end{equation}
\begin{equation}
    \delta g_{a\bar{b},t}=\delta t^i (\omega_i)_{a\bar{b}}\,,
\end{equation}
where $t^i\omega_i$ is the K\"ahler form. 

The higher order terms in $V_{m,n}^{-1,-1}$ can be found perturbatively. We shall illustrate how to find the first order correction to the moduli vertex operator. At the first order, we have
\begin{equation}
    Q_B |V_{m,1}^{-1,-1}\rangle= -\Bbb{P} \mathcal{G} [ \Psi_{0,1}^{(i)}\otimes V_{m,0}^{-1,-1}]\,,
\end{equation}
whose explicit form for the complex structure moduli and K\"ahler moduli are given as
\begin{equation}
    Q_B\Bbb{P}|V_{z,1}^{-1,-1}\rangle=\mathcal{S}^{z}_{AB} \frac{\alpha' C_{S^2}}{64\pi^2} (\partial c+\bar{\partial}\bar{c}) c\bar{c}e^{-\phi}\psi^A e^{-\bar{\phi}}\bar{\psi}^B e^{ik\cdot x}\,,
\end{equation}
\begin{align}
    \mathcal{S}^z_{ab}=&-\delta g_{cd,z} \partial_a \partial^d\delta g^{(i)c}_{~~~~b}-\delta g_{cd,z} \partial^c\partial_b \delta g^{(i)d}_{~~~~a}\,,\\
    =&2\delta g_{cd,z} R_{a~b}^{~d~c}\,,
\end{align}
and
\begin{equation}
    \mathcal{S}^z_{\bar{a}\bar{b}}=\overline{\mathcal{S}_{ab}^z}\,,
\end{equation}
\begin{equation}
\begin{split}
    Q_B\Bbb{P}|V_{t,1}^{-1,-1}\rangle&=\frac{3\alpha' C_{S^2}}{64\pi^2} (\delta g_{t}^{c\bar{d}}) \partial_c\partial_{\bar{d}} \delta g_{a\bar{b}}^{(i)}(\partial c+\bar{\partial}\bar{c})c\bar{c} e^{-\phi}\psi^a e^{-\bar{\phi}}\bar{\psi}^{\bar{b}} e^{ik\cdot x}+(a\leftrightarrow \bar{b}) \\
    &=-\frac{3\alpha' C_{S^2}}{64\pi^2} (\delta g_t^{c\bar{d}}) R^{(i)}_{a\bar{b}c\bar{d}}(\partial c+\bar{\partial}\bar{c})c\bar{c} e^{-\phi}\psi^a e^{-\bar{\phi}}\bar{\psi}^{\bar{b}} e^{ik\cdot x}+(a\leftrightarrow \bar{b}) \,.
    \end{split}
\end{equation}
As one can check, the form of the first-order equation of motion for the moduli sates is analogous to the background equation of motion. We can therefore write the ansatz for the moduli field at the first order as
\begin{align}
    V_{m,1}^{-1,-1}=&\mathcal{G}_{AB}^m c\bar{c} e^{-\phi}\psi^A e^{-\bar{\phi}}\bar{\psi}^B e^{ik\cdot x} +\mathcal{D}^m c\bar{c}(\eta\bar{\partial}\bar{\xi} e^{-2\bar{\phi}}-\partial\xi e^{-2\phi}\bar{\eta})e^{ik\cdot x}\nonumber\\
    &+ \frac{i}{2}\mathcal{F}_A^m (\partial c+\bar{\partial}\bar{c}) c\bar{c}(e^{-\phi}\psi^A e^{-2\bar{\phi}}\bar{\partial}\bar{\xi} +e^{-2\phi} \partial \xi e^{-\bar{\phi}}\bar{\psi}^A) e^{ik\cdot x}\,,
\end{align}
with
\begin{equation}
    -\frac{1}{2} \partial^2 \mathcal{G}_{ab}^z-\frac{1}{2}(\partial_b\mathcal{F}_a^z+\partial_a\mathcal{F}_b^z)= \frac{\alpha'C_{S^2}}{64\pi^2} S_{ab}^z\,,\quad \partial^B\mathcal{G}^z_{aB}+\mathcal{F}_a^z+\partial_a\mathcal{D}^z=0\,,
\end{equation}
\begin{equation}
    \partial^A \mathcal{F}_A^z-\partial^2\mathcal{D}^z=0\,,
\end{equation}
and
\begin{equation}
    -\frac{1}{2}\partial^2 \mathcal{G}_{a\bar{b}}^t -\frac{1}{2}(\partial_{\bar{b}}\mathcal{F}_a^t +\partial_a\mathcal{F}_{\bar{b}}^t)=\frac{3\alpha'C_{S^2}}{64\pi^2} \delta g_t^{c\bar{d}} \partial_c \partial_{\bar{d}} \delta g_{a\bar{b}}^{(i)}\,,\quad \partial^B\mathcal{G}_{a B}^t+\mathcal{F}_a^t+\partial_a\mathcal{D}^t=0\,,
\end{equation}
\begin{equation}
    \partial^A\mathcal{F}_A^t-\partial^2\mathcal{D}^t=0\,.
\end{equation}
Let us first solve for the complex structure moduli. We shall assume $\mathcal{D}^z=0$ and $\mathcal{G}_{a\bar{b}}^z=0.$ After imposing the harmonic gauge, we find
\begin{equation}
    \mathcal{G}_{ab}^z=-\frac{\alpha'C_{S^2}}{32\pi^2}\delta g_{cd,z} R_{a~b}^{~d~e}x^c x_e\,.
\end{equation}
Similarly, we shall assume
\begin{equation}
    \mathcal{D}^t=-\eta^{a\bar{b}}\mathcal{G}_{a\bar{b}}^t\,,
\end{equation}
and impose the harmonic gauge to find
\begin{equation}
    \mathcal{G}_{a\bar{b}}^t=-\frac{3\alpha'C_{S^2}}{128\pi^2} (\delta g_{e\bar{f},t}) R_{a\bar{b}c\bar{d}}^{(i)} (\eta^{e\bar{d}}x^c x^{\bar{f}}+\eta^{c\bar{f}}x^{\bar{d}}x^e)+(a\leftrightarrow \bar{b})\,.
\end{equation}
Note that one can add $Q_B$ exact terms to both $\Bbb{P} V_{z,1}^{-1,-1}$ and $\Bbb{P}V_{t,1}^{-1,-1}.$ An addition of such a $Q_B$ exact term won't affect a generic on-shell amplitudes. However, for the more subtle calculation, such as calculations involving the moduli states with zero momentum, the correction identification of the $Q_B$ exact term is necessary, which we leave for future work.

\section{Weyl frame and local coordinates}\label{sec:Wey and local coords}
In this section, we shall review the construction of $SL(2;C)$ string vertices and the prescription of Tseytlin for the non-linear sigma model approach, and compare and contrast the two approaches.

The prescriptions developed by Tseytlin and collaborators in \cite{Tseytlin:1988rr,Tseytlin:1988tv,Tseytlin:2000mt} for the treatment of NLSMs in string theory, recently reformulated into the modern language of conformal perturbation theory in \cite{Ahmadain:2022tew,Ahmadain:2022eso,Ahmadain:2024hdp}, are built on the relationship between worldsheet RG and target space physics. Fundamental to the prescription is the introduction of a worldsheet UV cutoff, $\epsilon$, to render the sphere partition function $Z_{S^2}$ well defined. The target space effective action $S$ is then computed by taking derivatives with respect to $\log \epsilon$:
\begin{equation}\label{eqn:tseyt-acts}
S^{\textbf{T1}} := -\frac{\partial}{\partial \log \epsilon} Z_{S^2}, \quad S^{\textbf{T2}}:=  -\left(1 + \frac{1}{2}\frac{\partial}{\partial \log \epsilon}\right)\frac{\partial}{\partial \log \epsilon} Z_{S^2}.
\end{equation}
The \textbf{T1} or \textbf{T2} superscripts in \eqref{eqn:tseyt-acts} denote two different prescriptions associated with two different choices differential operator with which to act on $Z_{S^2}$.

In \cite{Sen:2019jpm}, it was argued that replacing the contributions from the Feynman region of the moduli space with the Feynman diagrams involving ``simpler" string vertices in the large-stub limit may be thought of as a worldsheet UV cutoff. It is interesting to ask whether there is a relationship between this limit of tree-level SFT and the derivative prescriptions \eqref{eqn:tseyt-acts}. What we shall find is that a simple choice of string vertices in string field theory reduces to that of Tseytlin. We will also explain why it is fine to use such a vertex for restricted purposes, such as finding the background solution. We review the relevant aspects of the large stub limit in \S\ref{ssec:local-coords-review}, Tseytlin's prescriptions in \S\ref{ssec:Tseytlin-Review}, and compare and contrast the two in \S\ref{ssec:NLSM-SFT}.

\subsection{Local coordinates in string field theory}\label{ssec:local-coords-review}
Off-shell amplitudes of string theory are computed with off-shell vertex operators. Notably, because the off-shell vertex operators are not invariant under conformal transformations, insertions of such operators do depend on the local data such as the metric of the worldsheet. As we only need to know the worldsheet metric around the punctures to evaluate the off-shell amplitudes, an efficient way to encode the metric is to use a local coordinate. We will be mostly concerned with Riemann sphere with punctures. However, generalizations to Riemann surfaces with and without boundaries of higher genus is straightforward. 

To see how the local coordinates encode the metric around the punctures, we shall assume that there is a puncture on a Riemann sphere whose global coordinate is denoted by $z.$ We shall now then denote a coordinate of a unit disk by $w$ such that $|w|\leq1.$ We shall map the unit disk to the Riemann sphere such that $w$ maps to a puncture located at $z=0.$ We declare that the local coordinate $w$ is related to the global coordinate by
\begin{equation}
    w= h_1(z)\,,
\end{equation}
such that $h(0)=0.$ We shall assume that $h(z)$ is invertible. Then, we find that the following map maps from the unit disk to the punctured Riemann sphere
\begin{equation}
    h_1^{-1}: D^2\rightarrow S^2\,.
\end{equation}
We shall now map $C^2$ to $S^2/\{pt\}$ via a map
\begin{equation}
    h_2: C^2\rightarrow S^2/\{pt\}\,.
\end{equation}

We shall use the local coordinate maps as follows. First, the local metric of $S^2$ around the puncture shall be determined by $h_1,$ or equivalently, by the metric we endow on the unit disk. As is common in the CFT calculations, computing CFT correlators is often much easier to do in flat space than on the sphere. Therefore, we shall use $h_2^{-1}$ to map the operator insertion from the sphere to the flat space. In practice, therefore, we can construct a composite local coordinate map
\begin{equation}
    h^{-1}:= h^{-1}_2\cdot h_1^{-1} : D^2 \rightarrow C^2\,.
\end{equation}

We shall now endow a natural differential structure on the unit disk. We shall start with the metric on the flat $C^2$
\begin{equation}
    ds^2=|dz|^2\,.
\end{equation}
Then the metric on the unit disk is given as
\begin{equation}
    ds^2= |\partial h^{-1}/\partial dw|^2 |dw|^2\,.
\end{equation}

Suppose that we insert a conformal primary 
\begin{equation}
    c\bar{c} V\,,
\end{equation}
at the puncture $w=0.$ Then, in terms of the global coordinate, the very same insertion is written as
\begin{equation}
    |dh^{-1}/dw|^{2h} c\bar{c} V(0,0)\,,
\end{equation}
where $h$ is the conformal dimension of $c\bar{c} V(0,0).$ To make the dimension of the operator to vanish, we can accordingly multiply an arbitrary length scale $l_0$ 
\begin{equation}
    l_0^{2h}  |dh^{-1}/dw|^{-2h} c\bar{c} V(0,0)\,.
\end{equation}
Note that $l_0$ is not a priory related to the string length scale $l_s,$ and $l_0$ should be thought of as a book-keeping device.

We shall now comment on the stub parameter. We are allowed to rescale each local coordinate by a parameter $\lambda$
\begin{equation}
    w=\lambda h(z)\,.
\end{equation}
A utility of the stub is that taking large stub limit effectively acts as an on-shell limit. This statement can be understood as follows. With the inclusion of the stub parameter $\lambda,$ now each insertion of a conformal primary into a puncture on a Riemann surface is given as
\begin{equation}
\left| \frac{dh^{-1}}{dw} \frac{1}{l_0\lambda}\right|^{2h} c\bar{c} V(0,0)\,.
\end{equation}
Hence, in the large stub limit $\lambda\rightarrow\infty,$ the contributions to the off-shell amplitudes by off-shell states with $h>0$ are exponentially damped. 

Now we shall review how the local coordinates are related to the string vertices. String vertices are defined to be a region of moduli space that cannot be covered by Feynman diagrams made by joining simpler string vertices with propagators. The latter region of the moduli space is often called the Feynman region. The Feynman region of the moduli space can be constructed by gluing Riemann surfaces with punctures by plumbing fixtures. Such plumbing fixtures provide the boundary data of string vertices. For details, see \S\ref{sec:vertices}. 

\begin{figure}[ht]
\centering
\begin{subfigure}{.5\textwidth}
  \centering
  \includegraphics[width=0.5\linewidth]{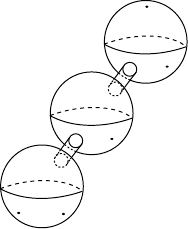}
  \caption{}
  \label{fig:3-spheres}
\end{subfigure}%
\begin{subfigure}{.5\textwidth}
  \centering
  \includegraphics[width=0.5\linewidth]{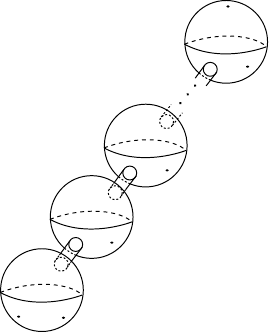}
  \caption{}
  \label{fig:n-spheres}
\end{subfigure}

\caption{A depiction of sphere diagrams that set boundary conditions on local coordinates for degenerating punctures. Dots represent the insertions of string fields. Fig. \ref{fig:3-spheres} sets the boundary condition for three degenerating punctures -- in the coordinates of the top right sphere, the two bottom left dots are a distance $\mathcal{O}(\lambda^{-4})$ apart. The generalization to $n$ degenerating punctures is drawn in Fig. \ref{fig:n-spheres}, wherein the two bottom left punctures are $\mathcal{O}(\lambda^{2-2n})$. These minimal separations may be interpreted as UV regulators for integrals of CFT correlators over moduli space, as in \cite{Sen:2019jpm}.}
\label{fig:degenerating-spheres}
\end{figure}

A few important comments are in order. For simplicity, we shall focus on the moduli space of the four-punctured sphere $\mathcal{M}_{0,4}$ and its vertex region $\mathcal{V}_{0,4}.$ In the large stub limit, string vertices $\mathcal{V}_{0,4}$ cover most of the moduli space $\mathcal{M}_{0,4}.$ The boundaries of $\mathcal{V}_{0,n}$ are given by the hard disk cutoff, whose size in general depends on the points in the moduli space. In the simplest case, $\mathcal{V}_{0,4}$ has three boundaries where the modulus $z$ is around $0,~1,~\infty.$ The boundary of the string vertex $\mathcal{V}_{0,4}$ around $z=0,$ for example, is given by, in the $SL(2;C)$ vertex,
\begin{equation}\label{eqn:vertex-boundary}
    z= \frac{16e^{i\theta} \lambda^{-2}}{(4+e^{i\theta}\lambda^{-2})^2}\,.
\end{equation}
Hence, effectively, in the large stub limit, we can see that the integral over $\mathcal{V}_{0,4}$ can be understood as integral over $\mathcal{M}_{0,4}$ with the hard-disk UV cutoff whose radius is given by $\lambda^{-2}+\dots.$ We stress again that the radius of the hard-disk UV cutoff, in general, depends on moduli. Also, in the case of a four-punctured sphere, we found that the size of the hard cut-off is $\lambda^{-2}.$ In higher punctures, the size of the hard cut-off won't be in general of order $\mathcal{O}(\lambda^{-2}).$ For example, let us study a five-punctured sphere, and send two movable punctures to a puncture fixed at $0.$ We shall force the Riemann sphere with five punctures to degenerate into three Riemann spheres with three punctures each (see Fig. \ref{fig:degenerating-spheres}). In such a case, the hard cut-off scales as $\mathcal{O}(\lambda^{-4})$, as can be noted by mapping all points of the plumbing fixture in Fig. \ref{fig:3-spheres} into the local coordinate system of just one of the spheres.

\subsection{Weyl frames and UV cutoff in the non-linear sigma model}\label{ssec:Tseytlin-Review}

In Tseytlin's prescription of off-shell formulation of string theory based on non-linear sigma model perturbation theory \cite{Tseytlin:1988rr}, one proceeds somewhat similarly. We will mostly follow the reformulation of Tseytlin's formulation in modern language studied by \cite{Ahmadain:2022tew,Ahmadain:2022eso}.

In contrast to string field theory, the space of field configurations in Tseytlin's approach to the target space effective action is not the worldsheet CFT Hilbert space, but the space of 2d worldsheet QFTs\footnote{The proof of background independence \cite{Sen:1993kb} demonstrates how to move between different worldsheet CFTs (and, more speculatively, QFTs \cite{Zwiebach:1996ph}) in the formalism of SFT. However, demonstrating that SFT built on different worldsheet theories is equivalent is conceptually distinct from treating the space of worldsheet theories itself as field configuration space.}, $\mathcal{Q}$. From this point of view, $c=0$ worldsheet theories (equivalently, a $c=26$ matter sector) correspond to on-shell backgrounds solving the equations of motion, and all other points in theory space are off shell configurations. $\mathcal{Q}$ is difficult to define (see \cite{Gaiotto:2024gii} for recent progress), but it may in part be parameterized by the space of coupling constants in the action.

The target space action is a function on field configuration space, and so in Tseytlin's approach is a function on $\mathcal{Q}$. A natural starting point for building functions on $\mathcal{Q}$ is the sphere\footnote{We are free to choose other 2d topologies, but as we are interested in tree level closed string theory, we focus on the sphere.} partition function. Away from the space of $c=0$ CFTs, the sphere partition function depends on the choice of sphere Weyl frame $\omega(z)$, so we denote it $Z[\omega]$ (we leave the fact that $Z$ is computed on $S^2$ implicit). In particular, $Z$ may depend on the overall scale of the geometry, denoted as $r$. For QFTs near a particular CFT, the partition function (and similarly any correlation functions) may be computed in terms of correlation functions in the CFT using conformal perturbation theory. For example, if the worldsheet QFT is defined by perturbing the worldsheet CFT action by some operator $\mathcal{A}(z)$ of conformal dimension $\Delta$ as
\begin{equation}\label{eqn:S-QFT-pert}
S_{\text{QFT}} = S_{\text{CFT}} + \kappa \int e^{(2-\Delta) \omega(z)}d^2z\, \mathcal{A}(z),
\end{equation}
the partition function may be expanded in terms of $\mathcal{A}(z)$ $n$-point functions as
\begin{equation}\label{eqn:Z-QFT}
\begin{split}
Z_{\text{QFT}} &= \left \langle \left\langle \exp(- \kappa \int e^{(2-\Delta) \omega(z)}d^2z\, \mathcal{A}(z)) \right \rangle \right\rangle_{\text{CFT}} \\
& = Z_{\text{CFT}} - \kappa \int e^{(2 - \Delta) \omega(z)}d^2z \langle \langle \mathcal{A}(z) \rangle \rangle_{\text{CFT}} \\
&+\frac{\kappa^2}{2}\int e^{(2-\Delta)\omega(z_1)}e^{(2-\Delta)\omega(z_2)}d^2z_1d^2 z_2 \langle \langle \mathcal{A}(z_1) \mathcal{A}(z_2) \rangle \rangle_{\text{CFT}} + \mathcal{O}(\lambda^3).
\end{split}
\end{equation}
Because $\kappa$ here functions as a coordinate on $\mathcal{Q}$, we interpret it as one mode of a target space field. 

If $\Delta \geq 1$, the integrated $n$-point function in \eqref{eqn:Z-QFT} is generically UV divergent and requires the introduction of a UV cutoff $\epsilon$ in order to be well defined. To make contact with SFT, we take $\epsilon$ to be a hard disc regulator, which prevents worldsheet operator insertions from coming within a distance $\epsilon$ as measured according to the Weyl frame $\omega$.

The correlators involved in the computation of $Z_{\text{QFT}}$ \eqref{eqn:Z-QFT} integrate over the location of all resulting operator insertions on the worldsheet. However, the sphere diagram in string theory has an additional non-compact SL$(2,\mathbb{C})$ conformal killing group that remains unfixed by a choice of Weyl frame. The standard procedure in worldsheet string theory (or SFT) is to fix this additional gauge invariance by fixing the location of three punctures. This procedure is more subtle for the linear or quadratic in $\lambda$ terms in \eqref{eqn:Z-QFT}, as there are fewer than three insertions.\footnote{For a careful treatment of a two-point amplitude, see \cite{Erbin:2019uiz}.} The observation of Tseytlin in \cite{Tseytlin:1988tv} is that $\epsilon$ also serves as a regulator on the divergent volume of non-compact SL$(2,\mathbb{C})$ orbits. We therefore arrive at the prescriptions \eqref{eqn:tseyt-acts} -- as reviewed in \cite{Ahmadain:2022tew}, these prescriptions are designed to remove a factor of this divergent volume from $Z_{\text{QFT}}$.

The introduction of a UV cutoff generically introduces scheme-dependent behavior into $Z_{\text{QFT}}$ (and, therefore, derivatives of $Z_{\text{QFT}}$), which appear as power law divergences in $\epsilon$. In particular, these power law terms depend on the choice of Weyl frame. The approach taken in \cite{Tseytlin:1988rr,Tseytlin:1988tv,Ahmadain:2022tew,Ahmadain:2022eso} is simply to drop these power law divergences, arguing that these are gauge-dependent (or scheme-dependent, from the RG perspective) terms that disappear from any physical amplitude. The NLSM approach to computing the target space effective action (specifically in the Euclidean S-matrix regime, to which we are making the most direct comparison) may therefore be summarized as:
\begin{enumerate}
    \item Compute the target space effective action by taking derivatives of $Z_{\text{QFT}}$.
    \item Drop all power law divergences in $\epsilon$.
    \item Take $\epsilon \rightarrow 0$.
\end{enumerate}
There are a few comments to make about the regime of validity of this approach, as laid out in \cite{Ahmadain:2022tew}. The techniques used are understood to apply to compact target spaces normalizable deformations of non-compact target spaces (excluding e.g. exact plane wave modes), bosonic string theory (the appropriate treatment of PCO operators or some alternative prescription is not yet understood), and Euclidean target spaces (imposing unitarity on the worldsheet -- some comments will be made about this in \S\ref{ssec:NLSM-SFT}). We will therefore focus our comparison of the two approaches to this overlapping regime of validity, keeping in mind that the SFT approach appears to apply far more generally.

\subsection{Relation between NLSM and SFT}\label{ssec:NLSM-SFT}
One of the biggest advantages of the Tseytlin's approach (and its generalizations) over the current formulation of closed string field theory is its computational efficiency. In particular, the NLSM approach does not require finding consistent choices of local coordinates. For example, SL$(2,\mathbb{C})$ invariant choices of local coordinates are difficult to find \cite{Erbin:2023rsq} and generically might not exist, forcing a sum over permutations of vertex operator insertions that may grow exponentially in the number of terms. It is therefore instructive to ask under what conditions the NLSM approach agrees with string field theory, which techniques from the NLSM approach may be ported into SFT as computational shortcuts, and how lessons from SFT may help extend the validity of Tseytlin's formulation.

It is important to note that at least perturbatively, there is a natural identification between the field configuration space of the NLSM approach and SFT. Perturbatively, small deformations to the worldsheet theory may be identified with operators in the worldsheet CFT we perturb around, as in \eqref{eqn:S-QFT-pert}. Take $\lambda\mathcal{A}(z)$ to be such an operator, dressed with its coupling $\kappa$. It is natural to identify this operator with a string field $\Psi$ that satisfies
\begin{equation}
\mathcal{B} \mathcal{B} \Psi = \kappa \mathcal{A},
\end{equation}
where $\mathcal{B}$ is the Beltrami differential. As an example, if $\mathcal{A}(z)$ is a minimally coupled and purely in the matter sector this implies
\begin{equation}
\Psi = \kappa c \bar{c} \mathcal{A}.
\end{equation}

Given the behavior of the stub parameter $\lambda$ as a UV cutoff on the integrated worldsheet correlators in SFT (see Fig. \ref{fig:degenerating-spheres} and the discussion below \eqref{eqn:vertex-boundary}), $\lambda$ on the SFT side and $\epsilon^{-1}$ on the NLSM side appear to play equivalent roles, at least up to fourth order in perturbation theory for on-shell vertex operators. More specifically, we identify the scheme-dependent $\epsilon$ dependence on the NLSM side with the non-gauge-invariant $\lambda$-dependent terms on the SFT side. Whereas the approach in SFT is to choose appropriate boundary conditions to ensure cancellation of $\lambda$-dependence (and, more generally, independence of all local coordinate ambiguities), the approach on the NLSM side is to just drop all offending terms.

\subsubsection{The Quadratic Term}

The term in the target space action quadratic in the fields (which we may call $K$, standing for kinetic) determines the propagator. Suppose we take two worldsheet operators $\mathcal{A}_1$ and $\mathcal{A}_2$, with corresponding couplings $\kappa_1$ and $\kappa_2$. The two terms to compare are
\begin{equation}\label{eqn:K-T2}
\begin{split}
K_{\text{SFT}} &= \kappa_1 \kappa_2 \bra{c \bar{c} \mathcal{A}_1}c_0^{-}Q_B \ket{c \bar{c}\mathcal{A}_2},\\
\quad K^{\textbf{T2}} &= -\frac{\partial}{\partial \log \epsilon}\left(1 + \frac{1}{2}\frac{\partial}{\partial \log \epsilon}\right)\int_{|z| = \epsilon}^{\infty} e^{(2-\Delta_1) \omega(z)}d^2z\, \langle \langle \mathcal{A}_1(z) \mathcal{A}_2(0) \rangle \rangle.
\end{split}
\end{equation}
Suppose that the $\mathcal{A}_i$ are primaries with conformal dimension $\Delta_i$. We have
\begin{equation}\label{eqn:kinetic-action}
K_{\text{SFT}} = K^\textbf{T2} = \mathfrak{g}^{ij}\kappa_i \kappa_j(2 - \Delta_i),
\end{equation}
where $\mathfrak{g}^{ij}$ is the Zamolodchikov metric\footnote{$\mathfrak{g}$ an unusual choice of symbol for the Zamolodchikov metric ($\kappa^{ij}$ is more common), but we have exhausted the more traditional choices elsewhere.}. For this class of string fields, the NLSM and SFT approaches agree. In particular, we may identify string fields built on primaries with $\Delta = 2$ as solutions to the linear equations of motion.

The agreement (and the efficacy of the \textbf{T2} prescription) has yet to be fully understood for Lorentzian target spaces, which introduce non-unitary deformations to the worldsheet theory. A specific example is the case of a dilaton vertex operator $\mathcal{V}_1 = \Phi R(z)e^{i k^{\mu}X_{\mu}}$ and vector field vertex operator $\mathcal{V}_2 = F_{\nu} \partial X^{\nu} \bar{\partial} g e^{-i k_{\mu} X^{\mu}}$. These two operators have a kinetic mixing in the target space SFT action, of the form $\Phi k_{\mu}F^{\mu}$, that is responsible for lifting the spurious dilaton with non-trivial momentum. Let us compute explicitly what happens when they are inserted into the \textbf{T2} prescription as in \eqref{eqn:K-T2} directly following \cite{Ahmadain:2022tew}, in particular working on the round sphere metric and dropping all even integer power law behavior in $\epsilon$. The appropriate integrated two point function is
\begin{equation}\label{eqn:dil-vec-2pt}
\begin{split}
&K_{0,2} \supset \Phi F_{\nu}k^{\nu} \epsilon^{k^2} \int_{|z| > \epsilon}\sqrt{g}d^2z\, |z|^{-k^2}(1 + |z|^2)^{\frac{k^2}{2} + 1} \frac{\bar{\partial} g}{z}\\
&=- 16\Phi F_{\mu}k^{\mu} \epsilon^{k^2}\int_{|z| > \epsilon} d^2z\, |z|^{-k^2}(1 + |z|^2)^{\frac{k^2}{2}-6} = \\
& = -32 \pi F_{\mu}k^{\mu} \epsilon^{k^2}\left(\frac{1 - \epsilon^{2 - k^2}}{2 - k^2} + \sum_{p \geq 0} a_{p} \epsilon^{2 + 2p - k^2}\right). 
\end{split}
\end{equation}
where we have inserted the explicit form of the round sphere metric $g_{z\bar{z}} = 4 (1 + |z|^2)^{-2}$. We are now instructed to drop the even integer power laws in $\epsilon$, as they may be absorbed into tadpoles and are pure scheme. An application of the \textbf{T2} prescription now results in
\begin{equation}
\left(1 + \frac{1}{2}\frac{\partial}{\partial \log \epsilon}\right)\frac{\partial}{\partial \log \epsilon} K_{0,2} \supset -16 \pi F_{\mu}k^{\mu}k^2\frac{2 + k^2}{2 - k^2}\epsilon^{k^2},
\end{equation}
which has an unwanted zero at $k^2 = 0$.

Another way to see this result is to compute \eqref{eqn:dil-vec-2pt} directly for $k^2 = 0$:
\begin{equation}\label{eqn:dil-vec-2pt-2}
K_{0,2} \supset -16\Phi F_{\nu}k^{\nu}\int_{|z| > \epsilon} \frac{1}{(1 + |z|^2)^6} \propto \frac{\Phi F_{\nu}k^{\nu}}{(1 + \epsilon^2)^5}. 
\end{equation}
Ostensibly, applying \textbf{T2} to \eqref{eqn:dil-vec-2pt-2} now produces the correct mixing term. However, the RHS of \eqref{eqn:dil-vec-2pt-2} is pure scheme per Tseytlin's prescription, as it is an analytic function of $\epsilon^2$ that may be expanded purely in terms of non-negative integer powers of $\epsilon^2$, so it cannot contribute in a universal, gauge-invariant way to the action according to the prescription of \cite{Ahmadain:2022tew}.

This appears to be a failure mode for Tseytlin's prescriptions in Lorentzian target spaces, in the same way that they are incomplete for irrelevant worldsheet operators. However, there may be a non-trivial off-shell field redefinition between the \textbf{T2} action and SFT that gives another non-minimally coupled and/or descendant operator the role that the vector plays in SFT. It should be noted that in the prescription of  \cite{Ahmadain:2024hdp} this issue is avoided, as there is no need to drop power law divergences in $\epsilon$ by hand.

In contrast, the SFT linear equation of motion is $Q_B \Psi = 0$. The corresponding implication for Tseytlin's approach is that the matter operator must be annihilated by all Virasoro generators $L_{n}$, $\bar{L}_n$ with $n \geq 1$, and must have total conformal dimension $\Delta = 2$. So even at quadratic order in the target space action, there is an important mismatch between Tseytlin's approach and SFT which becomes relevant as soon as we wish to consider Lorentzian target spaces.

\subsubsection{Interaction Vertices}

SFT is built on the principle that all dependence on local coordinates, including the stub parameter $\lambda$, must vanish from the on-shell quantities, including on-shell target space action and on-shell amplitudes. The analogous issue in the conformal perturbation theory approach is RG scheme dependence, including dependence on the worldsheet Weyl factor $\omega(z)$ and the UV cutoff $\epsilon$. The key difference in the treatment of interaction vertices in the two approaches is how the dependence on these gauge ambiguities is handled.

In SFT, we resolve this ambiguity by ensuring that the fundamental vertex and Feynman regions cover moduli space exactly once, and that the local coordinates on the fundamental region connect continuously with local coordinates constructed from the Feynman regions. This ensures that $\lambda$ dependence in observables drops out automatically.

In conformal perturbation theory, we note that for generic conformal dimensions of worldsheet insertions (roughly corresponding to generic target space momenta) divergences in the integrated correlators only ever come with one power of $\log \epsilon$, perhaps dressed with a power law. This factor of $\log \epsilon$ is then stripped off when we apply one of the derivative prescriptions \eqref{eqn:tseyt-acts}, leaving behind only power law divergences in $\epsilon$. The prescription given by \cite{Ahmadain:2022tew} is then to identify these power law divergences as pure scheme, and drop them, leaving behind terms that are entirely independent of $\epsilon$. For worldsheet operators with $\Delta \leq 2$ (note that there is no requirement for these operators to be nearly marginal), this procedure is argued in \cite{Ahmadain:2022tew}\footnote{The proof by induction therein is for the case of bosonic strings, compact target spaces, and generic external momenta and normalizable target space deformations, but the on-shell external momenta that interest us here result in issues related to the higher powers of $\log \lambda$ that can arise but are not directly addressed in \cite{Ahmadain:2022tew}.} to return the correct equations of motion (and correspondingly, the correct on-shell tree level amplitudes).

The analogous prescription in SFT would be to simply compute observables by only including the vertex region contribution as a function of $\lambda$, and then to drop all $\lambda$-dependent terms, trusting that they will cancel after the inclusion of the Feynman regions. It is important to remark that this prescription relies on a crucial assumption that the Feynman region contributions do not contain a finite piece that neither diverges nor vanishes in the large stub limit. However, this assumption is not true in general, as was illustrated in the calculation of spectrum under the vacuum shift \cite{Sen:2019jpm}. 

Following \cite{Sen:2019jpm}, let us give a rough sketch of why this works in some special cases in bosonic SFT. Consider the Virasoro-Shapiro amplitude. The vertex operator for an on-shell tachyon state is given as
\begin{equation}
    V_T(z,k):= c\bar{c} e^{ik\cdot x}(z)\,,
\end{equation}
where $k^2=-m^2=4$ in $\alpha'=1.$ The Virasoro-Shapiro amplitude, modulo the overall normalization, takes the following form
\begin{align}
    A_{0,4}\propto &\int d^2 z \langle V_T(0,k_1)\otimes V_T(1,k_2)\otimes V_T(\infty,k_3) \otimes V_T(z,k_4) \rangle\,,\\
    =&\int d^2 z |z|^{-u/2+m^2} |z-1|^{-s/2+m^2}\,,
\end{align}
where the domain of the integral is formally defined as $\Bbb{C}.$ The Mandelstam variables are defined as
\begin{equation}
    s:=-(k_3+k_4)^2\,,\quad t=-(k_2+k_4)^2\,,\quad u=-(k_1+k_4)^2\,.
\end{equation}
The moduli integral is not convergent at every point in the kinematically allowed region. For example, if $u\geq m^2,$ the integral around $z=0$ diverges due to the following contribution
\begin{equation}
    \int_{|z|\leq \lambda^{-2}} d^2z |z|^{-2-(u-m^2)/2}+\dots\,.
\end{equation}
Similarly, if $t\geq m^2$ and and $s\geq m^2$ the integral diverges at $z=\infty$ and $z=1,$ respectively.

To evaluate the moduli integral in SFT, we divide the moduli space into the vertex region and the Feynman regions around which the worldsheet degenerates. The vertex contribution shall be denoted by $\mathcal{V}_{0,4},$ which can be computed by
\begin{equation}
    \mathcal{V}_4(k_1,k_2,k_3,k_4)=\frac{\partial^4 S}{\partial T_1(k_1)\partial T_2(k_2)\partial T_3(k_3)\partial T_4(k_4)}\biggr|_{\Psi=0}\,,
\end{equation}
where $T_i$ are tachyon string fields. To write down the Feynman region contributions, in the s-channel, we note that locally the moduli integral is represented as
\begin{equation}
    \propto \mathcal{V}_{0,3}(k_1,k_2,-k_1-k_2)\times \mathcal{V}_{0,3}(k_1+k_2,k_3,k_4)\times\lambda^{-L_0^+} \int_{|q|\leq1} d^2q |q|^{L_0^+-2}\,,
\end{equation}
where $L_0^+$ refers to the eigenvalue of $L_0^+$ of the propagating state. Note that we normalized the string vertices such that $\mathcal{V}_{0,3}$ is independent of $\lambda.$ The insight of SFT is that while the Schwinger parameterization above is ill-defined for $L_0^+\leq0,$ the representation of the same integral in terms of the propagator
\begin{equation}
    \biggr\langle\Psi\biggr|\frac{b_0^+}{L_0^+}\biggr|\Psi\biggr\rangle\,,
\end{equation}
is perfectly well defined for any value of $L_0^+.$ Hence, using the physical insight, we shall replace the Feynman region contributions with
\begin{equation}
    \sum_{\Psi}\mathcal{V}_{0,3}\times \mathcal{V}_{0,3} \times \biggr\langle\Psi\biggr|\frac{b_0^+}{L_0^+}\lambda^{-L_0^+} \biggr|\Psi\biggr\rangle\,,
\end{equation}
schematically. This replacement can be understood as a surgical analytic continuation of the problematic region of the moduli integral. We then have
\begin{equation}
    \mathcal{A}_{0,4}=\mathcal{V}_{0,4}+\sum_{\Psi}\mathcal{V}_{0,3}\times \mathcal{V}_{0,3} \times \biggr\langle\Psi\biggr|\frac{b_0^+}{L_0^+}\lambda^{-L_0^+}\biggr|\Psi\biggr\rangle\,,
\end{equation}
again, schematically.

By construction, $\mathcal{A}_{0,4}$ is independent of $\lambda,$ despite that both the Feynman and vertex contributions depend on $\lambda.$ A simplification in the computation of $\mathcal{A}_{0,4}$ can occur. If $L_0^+=\Delta-2$ of every propagating state is not $0,$ then under the large stub limit $\lambda\rightarrow\infty,$ the diverging contributions from the Feynman region must cancel, as well as the rest of the terms that vanish independently in the large stub limit. Therefore, in this case, we can effectively write the Virasoro-Shapiro amplitude as
\begin{equation}
    \mathcal{A}_{0,4}\equiv \text{Reg} (\mathcal{V}_{0,4})\,,
\end{equation}
where
\begin{equation}
    \text{Reg}(\mathcal{V}_{0,4})
\end{equation}
is obtained by dropping all the diverging terms in $\mathcal{V}_{0,4}$ in $\lambda\rightarrow \infty$ limit. 


\begin{figure}[ht]
\centering
\includegraphics[width=0.9\textwidth]{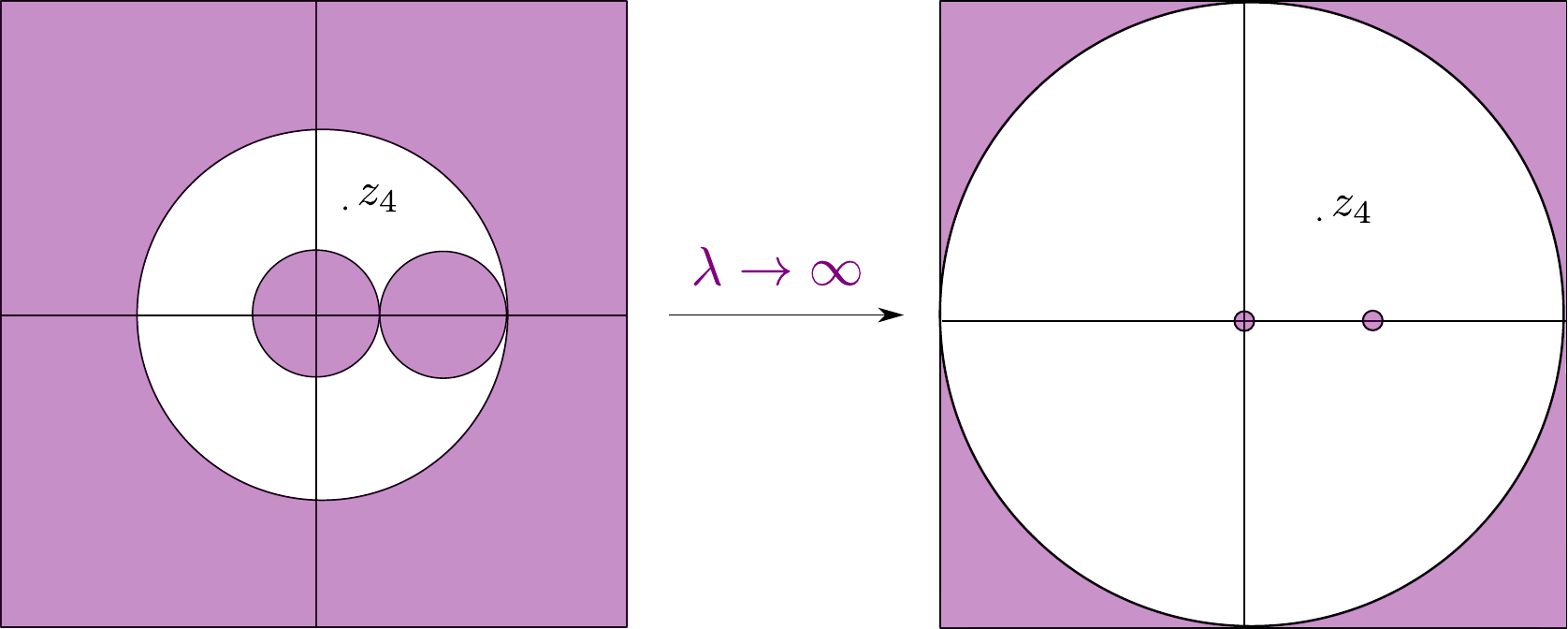}
\caption{We depict the moduli space of four-punctured spheres, with the stub parameter $\lambda$ getting large on the right-hand diagram. The fixed punctures are at $z = 1,$ $0$, $\infty$, and the movable puncture is labeled as $z_4$. The shaded regions are the Feynman regions of the vertex, and the unshaded region is the fundamental region. As we take $\lambda$ large (dropping divergent terms, which are all power-law in $\lambda$ for generic conformal dimension $\Delta$) the four-point amplitude is dominated by the fundamental vertex region, and we recover the effective action regime of \cite{Ahmadain:2022tew}. }\label{fig:large-stub-moduli-space}
\end{figure}

More generally, by the nature of the operator product expansion, we expect an $n$-point correlator to be expanded as a series of power laws when one puncture at $z_i$ approaches another at $z_j$. Either in SFT (or in Tseytlin's prescriptions), these correlators may contribute power law or logarithmic divergences in $\lambda$ (or $\epsilon^{-1}$). In supersymmetric SFT, where we must also insert PCOs, there may be $1/z_{ij}$ behavior in the correlators which contributes a universal non-cutoff-dependent term once integrated around a boundary of moduli space, but in bosonic SFT we expect the UV behavior of a correlator near any vertex operator insertion to be purely real as all operators satisfy the level matching condition. Therefore, at least in bosonic SFT, we identify non-trivial power laws in the OPE in the region near a puncture (a regime that the large stub limit generically brings about near the boundary of the fundamental vertex and Feynman region) with non-universal $\lambda$-dependent terms in the eventual amplitude.

In summary, it appears that the procedure of dropping power-law-in-$\epsilon$ contributions to the target space effective action in Tseytlin's prescriptions relies on these same terms exactly canceling out in string field theory, thereby leaving no constant term behind. This is a fairly reliable assumption in bosonic SFT, so it is not surprising that Tseytlin's prescriptions have passed the basic checks for agreement with string-theory amplitudes. However, this assumption is far less reliable once we shift the background, include PCOs, include boundaries or crosscaps, or wish to study worldsheet operators that may introduce branch cuts. This makes manifest a concrete challenge that must be resolved in order to extend Tseytlin's prescriptions to the supersymmetric case or a vacuum shift.

\subsection{Summary}

The relationship between Tseytlin's approach to off-shell string theory as formulated in \eqref{eqn:tseyt-acts} and SFT is sharpest in the large-stub limit of SFT, and is limited to unitary worldsheet theories, normalizable target space deformations, and primary vertex operators with total conformal dimension $\Delta \in [0,2]$. In this regime, we may draw the relationship
\begin{enumerate}
    \item UV cutoff $\epsilon$ $\leftrightarrow$ stub parameter $\lambda^{-1}$,

    \item Derivative prescription \eqref{eqn:tseyt-acts} and dropping of power-law divergences $\leftrightarrow$ cancellation of $\lambda$-dependence with Feynman regions.
    
\end{enumerate}

Despite the lack of careful cancellation of gauge-dependent data, to the extent that it has been directly tested it appears that Tseytlin's prescriptions correctly recover target space equations of motion and certain tree-level (on-shell) scattering amplitudes. At least in closed bosonic string theory, this suggests that for many tree-level questions in target space some computational short-cuts may be taken, such as not being careful with a choice of local coordinates around worldsheet punctures. The identification of $\epsilon$ with the stub parameter $\lambda$ as in Fig. \ref{fig:degenerating-spheres} suggests that Tseytlin's prescriptions may need a refinement once five punctures are introduced on the worldsheet -- the local coordinate boundary conditions require moduli-dependent stub parameters for the punctures, something which a fixed UV cutoff does not allow for. It would therefore be interesting to check this explicitly by going to fifth order in perturbation theory both in SFT and in Tseytlin's approach, which we leave for future work. It would likewise be interesting to learn what techniques from SFT may be used to consistently extend Tseytlin's formalisms to the supersymmetric or higher-genus cases, in the hope that similar shortcuts may be found to simplify and expedite arduous SFT calculations for the same cases.

\section{Conclusions}\label{sec:conclusions}
In this paper, we revisited the non-linear sigma model in the framework of string field theory. Our result provides a systematic treatment of the perturbation theory both in $\alpha'$ and $g_s$ expansion. Furthermore, our result opens up a possibility to investigate target spaces with non-trivial metric and curvature directly within string theory that goes beyond the supergravity approximation. For example, we can compute $\alpha'$ and $g_s$ corrections both perturbatively and non-perturbatively to the spacetime physics. Lastly, we have found the relations between the string field theoretic formulation of the non-linear sigma model and the approach based on the conformal perturbation theory. 

There are a number of directions that deserve future investigation.
\begin{itemize}
    \item Our work is limited by the fact that we only considered trivial worldsheet saddle contributions. It is, however, well understood that the contributions from the non-trivial worldsheet saddles, e.g., worldsheet instantons \cite{Dine:1986zy,Dine:1987bq}, play crucial roles. For example, to understand topology change, such as flops and conifold transitions \cite{Gendler:2022ztv}, and the emergence of stringy geometry \cite{Witten:1996qb}, it is imperative to understand the roles of the worldsheet instantons. It would be, therefore, important to understand how to incorporate the non-trivial worldsheet saddles into our description.
    \item The formulation of the non-linear sigma model through the patch-by-patch description obscures the relationships between physical observables and the global topologies. It is not clear how the perturbative approach will teach us more global questions, e.g., formulation of physical observables in cosmological backgrounds in string theory. It would be interesting to study how non-trivial worldsheet saddles and non-perturbative $g_s$ effects intertwine the global topology with the question of physical observables in string theory in curved backgrounds.
    \item In particular, it would be fascinating to apply this technology to the study of black hole backgrounds in string field theory. In \cite{Horowitz:1992jp}, Horowitz notes some supersymmetric black hole solutions that satisfy the string equations of motion (at least, away from potential singularities). In some simple cases, the exact CFT is known, but not in general. The framework we build here might be useful in studying new stringy phenomena, and their quantum corrections, across the black hole horizon and into the interior.
    
    \item Our work provides a framework to study off-shell fluctuations near an on-shell gravitational background directly in string theory, albeit perturbatively. An interesting application of our result would be, therefore, to study entanglement entropy of gravitational backgrounds directly in string theory \cite{Susskind:1994sm} (and perhaps an extension of the results of \cite{Balasubramanian:2018axm} to closed string theory). Furthermore, it would be interesting to study time-dependent phenomena, such as the decay of a non-supersymmetric black hole.
    \item One of the main motivations behind this work is to develop a tool to consistently compute $\alpha'$ and $g_s$ corrections, both perturbatively and non-perturbatively, in cosmological backgrounds in string theory in the context of flux compactifications. The long-term goal is to compute $\alpha'$ and $g_s$ corrections to the effective potential in the Calabi-Yau orientifold compactifications with fluxes, with and without spacetime supersymmetry. An immediate short-term goal, as a warm-up, is to compute the $g_s$ corrections to the K\"ahler potential and determine the overall normalization of the non-perturbative superpotential in Calabi-Yau orientifold compactifications in which the Ramond-Ramond tadpole is canceled with spacetime filling D-branes \cite{Kim:2023cbh,Kim:2023sfs,Kim:2023eut}. 
    \item One important lesson we learned by comparing the string field theoretic approach to the non-linear sigma model to that of Tsetlyn is that there might be a way to construct a rather na\"ive version of the string vertices, for a restricted set of questions, to simplify the complexity of constructing the full-fledged string vertices \cite{Zwiebach:1992ie,Zwiebach:1990nh,Headrick:2018ncs,Costello:2019fuh,Cho:2019anu,Firat:2021ukc,Firat:2023suh,Firat:2024ajp,Erbin:2022rgx,Erbin:2023rsq}.\footnote{Recently the flat vertices were proposed to simplify the construction of string vertices \cite{Mazel:2024alu}.} It would be worthwhile to investigate along this line.
\end{itemize}

\section*{Acknowledgments}
We thank Yuji Okawa, Michael Haack, Xi Yin, Ashoke Sen, Atakan Hilmi Fırat, Ted Erler, Carlo Maccaferri, Alessandro Tomasiello for interesting discussions. AF would like to in particular thank Aron C. Wall for lengthy and insightful discussions from which significant points of intuition for \S\ref{sec:Wey and local coords} emerged. We thank Aron Wall and Ted Erler for comments on the draft. The work of MK is supported in part by the Leinweber Institute for Theoretical Physics, and Simons Investigator Award (MPS-SIP-00507021).

\appendix
\section{Diffeomorphism and background solution}\label{app:diff}
In this section, we shall study the action of the diffeomorphism on the background solution of the form
\begin{align}
    \Psi_0=& -\frac{g_c}{4\pi} \mathcal{G}_{AB} c\bar{c} e^{-\phi}\psi^A e^{-\bar{\phi}}\bar{\psi}^B +\frac{g_c}{4\pi}\mathcal{D}c\bar{c} (\eta e^{-2\bar{\phi}}\bar{\partial}\bar{\xi}-e^{-2\phi}\partial\xi \bar{\eta})\nonumber\\&+i\frac{g_c}{4\pi\sqrt{2\alpha'}}  \mathcal{F}_A(\partial c+\bar{\partial}\bar{c})  c\bar{c} (e^{-\phi}\psi^A e^{-2\bar{\phi}}\bar{\partial}\bar{\xi} +e^{-2\phi}\partial\xi e^{-\bar{\phi}}\bar{\psi}^A)\,.
\end{align}
Understanding how the diffeomorphism acts on the background solution will determine how to interpret each term in the background solution. Note that the background solution written above should be thought of as an n-th order term in the perturbative expansion in $\alpha'.$ However, we shall omit the index $n$ to simplify the expression. 

The gauge generator $\Lambda$, we take has the form
\begin{equation}
\Lambda = \lambda_A c \bar{c}\left(e^{-\phi}\psi^Ae^{- 2 \bar{\phi}}\bar{\partial}\bar{\xi} + e^{-2\phi}\partial \xi e^{-\bar{\phi}}\bar{\psi}^A\right).
\end{equation}
To determine the form of the gauge transformation, we must compute the string product $[\Psi_0, \Lambda]$, which first requires the computation of the string product $\{,\}$ with all possible background operators. Using $G, \Lambda,$ and $D$ as shorthand for the graviton, gauge, and dilaton vertex operators, we find
\begin{equation}
\{G, \Lambda, D^c\} = 0
\end{equation}
For the graviton, we use $\tilde{G}$ as the coefficent for the conjugate vertex operator to find
\begin{equation}
\begin{split}
\{G,\Lambda, G^c\} = 9i\sqrt{2\alpha'}\left(2\tilde{G}_{ab}G^{bc}\partial_c\lambda^a + \tilde{G}_{ab}G^{ab}\partial_c\lambda^c + 2 \lambda^cG^{ab}\partial_c\tilde{G}_{ab} + \lambda^cG^a_c\partial^b\tilde{G}_{ab}\right)
\end{split}
\end{equation}

We may therefore read off that (upon dropping total derivative terms)
\begin{equation}
\begin{split}
[\Lambda,G] &= \mathcal{G}_{ab}c\bar{c}e^{-\phi}e^{-\bar{\phi}}\psi^a \bar{\psi}^b,\\
\mathcal{G}_{ab} &= 9i\sqrt{2\alpha'}\left( G_b^c\partial_c\lambda_a - G_{ab}\partial_c \lambda^a -2\lambda^c\partial_cG_{ab} - \lambda^c\partial_b G_{ac} \right)
\end{split}
\end{equation}
\section{Disk amplitudes}
In this appendix, we will evaluate a disk one-point function and a two-point function $\{\Psi_{0,1}\}_{D^2}. $ Note that we shall use $a,~b,\dots$ to denote real indices in this appendix.

For a generic on-shell string field 
\begin{equation}
    V^{-1,-1}=G_{AB} c\bar{c} e^{-\phi}\psi^A e^{-\bar{\phi}}\bar{\psi}^B +D (\eta \bar{\partial}\bar{\xi}e^{-2\bar{\phi}}-\partial\xi e^{-2\phi} \bar{\eta}) \,,
\end{equation}
where $G_{AB}$ and $D$ are polynomials in $X,$ the disk one-point function is given by
\begin{equation}
    \{V\}_{D^2}=\frac{1}{2\pi}\langle c_0^- V\rangle=\frac{1}{2\pi} C_{D^2} 2^{\partial\cdot D\cdot\partial}  (-G_{AB}D^{AB}+2D) \,.
\end{equation}
Note that the self-contraction of the worldsheet bosons without the derivatives would lead to additional contributions, e.g.,
\begin{equation}\label{eqn:ambiguous term}
    -\frac{1}{2\pi} C_{D^2}  \partial\cdot D\cdot\partial (-G_{AB}D^{AB}+2D)\log 2\,.
\end{equation}
However, for an on-shell field, we have
\begin{equation}
    \partial^2 G_{AB}=\partial^2 D=0\,.
\end{equation}
Furthermore, 
\begin{equation}
    \partial \cdot D\cdot \partial (a G_{AB}+b D)=-\partial^2 (a G_{AB}+b D)=0\,.
\end{equation}
Therefore, the term \eqref{eqn:ambiguous term} vanishes.

If we evaluate the closed string one-point function with the properly normalized first-order background solution
\begin{equation}
    \Psi_{0,1}=\frac{1}{12\pi} R_{ACBD}x^Cx^Dc\bar{c}e^{-\phi}\psi^A e^{-\bar{\phi}}\bar{\psi}^B\,,
\end{equation}
we are supposed to reproduce the first-order term in the $\alpha'$ expansion of the worldvolume 
\begin{equation}
    -T_p \int  d^{p+1}xe^{-\Phi} \sqrt{-g}=-T_p\int d^{p+1}x e^{-\Phi} \left(1+\frac{1}{2}h^\mu_\mu+\dots \right)\,,
\end{equation}
where
\begin{equation}
    h_{\mu\nu}= -\frac{1}{3} R_{\mu\rho\nu\sigma}x^\rho s^\sigma\,.
\end{equation}
The disk one-point function, while treating $X$ as the zero modes, is given as
\begin{equation}
    -T_p\int d^{p+1}x e^{-\Phi} \frac{1}{6} R_{\mu c\nu d} x^c x^d \eta^{\mu\nu}\,, 
\end{equation}
where $\mu,~\nu$ are worldvolume indices. As one can check, this precisely reproduces the answer one expects from the DBI action. Note that we used $R_{ab}=0.$ 

Let us write $(0,-1)$ picture and $(-1,0)$ picture first order background solutions
\begin{equation}
    \Psi_{0,1}^{-1,0}=-i\sqrt{\frac{2}{\alpha'}} \frac{1}{4\pi}\delta g_{AB} c\bar{c} \partial X^A e^{-\bar{\phi}}\bar{\psi}^B+i\sqrt{\frac{\alpha'}{2}}\frac{1}{4\pi} \partial_C \delta g_{AB} c\bar{c} \psi^C\psi^Ae^{-\bar{\phi}}\bar{\psi}^B+\dots \,,
\end{equation}
\begin{equation}
    \Psi_{0,1}^{0,-1}=-i\sqrt{\frac{2}{\alpha'}} \frac{1}{4\pi} \delta g_{AB}c\bar{c} e^{-\phi}\psi^A \bar{\partial}X^B+i\sqrt{\frac{\alpha'}{2}} \frac{1}{4\pi} \partial_C \delta g_{AB}e^{-\phi}\psi^A \bar{\psi}^C\bar{\psi}^B+\dots\,.
\end{equation}

We shall first compute 
\begin{equation}
    \frac{1}{2}\{\Psi_{0,1}^2\}_{D^2}\,.
\end{equation}
Let us first evaluate the disk two-point in the interior of the moduli space. We shall place a picture $(-1,0)$ vertex at $i$ and a picture $(0,-1)$ vertex at $iy.$ In the interior of the moduli space
\begin{equation}
    \mu^{-2}\leq y\leq 1-2\lambda^{-2}+\dots\,,
\end{equation}
we find
\begin{align}
    \frac{1}{2}\{\Psi_{0,1}^2\}_{D^2}=& \frac{C_{D^2}}{8\pi^2}\int_{\mu^{-2}}^{1-2\lambda^{-2}} dy (1-y^2) 2^{-\partial_1\cdot D_1\partial_1} (2y)^{-\partial_2\cdot D\cdot \partial_2} (1-y)^{-2\partial_1\cdot \partial_2}(1+y)^{-2\partial_1\cdot D\partial_2}\nonumber\\& \biggr[ \frac{\delta g_{ab} D^{aa'}D^{bb'}\delta g_{a'b'}}{(1+y)^4}-\frac{4 D^{bc}(\partial_{||}^a+y\partial_\perp^a)\delta g_{cd} (\partial_\perp^d+y\partial_{||}^d)\delta g_{ab}}{(1-y^2)^2(1+y)^2} \nonumber\\
    &+\partial_e\delta g_{ab}\partial_f\delta g_{cd}\biggr(\frac{D^{af}D^{be}D^{cd}}{4y(1+y)^2} +\frac{D^{af}\eta^{bd}\eta^{ce}}{(1-y^2)^2}-\frac{D^{ad}\eta^{bf}\eta^{ce}}{(1-y^2)^2}-\frac{D^{ad}D^{be}D^{cf}}{4y(1+y)^2}\nonumber\\
    &-\frac{D^{af}D^{bc}D^{de}}{(1+y)^4}+\frac{\eta^{ac}\eta^{bf}D^{de}}{(1-y^2)^2}+\frac{D^{ab}D^{cf}D^{de}}{4y(1+y)^2}+\frac{D^{ad}D^{bc}D^{ef}}{(1+y)^4}\nonumber\\
    &-\frac{\eta^{ac}\eta^{bd}D^{ef}}{(1-y^2)^2}-\frac{D^{ab}D^{cd}D^{ef}}{4y(1+y)^2}\biggr)\biggr]\,.
\end{align}
The vertical integration is computed to be
\begin{equation}
    \mathcal{B}=-\frac{C_{D^2}}{64\pi^2}   \delta g_{ab}\delta g^{cd}D^{ac}D^{bd}\,.
\end{equation}
It is important to note that the above expression is well defined modulo terms that are proportional to
\begin{equation}
    \partial_1\cdot \partial_2+\partial_1\cdot D\cdot\partial_1+\partial_1\cdot D\cdot \partial_2\,,
\end{equation}
and
\begin{equation}
    \partial_1\cdot \partial_2+\partial_1\cdot D\cdot\partial_2+\partial_2\cdot D\cdot\partial_2\,.
\end{equation}

Note that in the weak field expansion, the Riemann tensor is written as
\begin{equation}
R_{\alpha\beta\mu\nu}=\frac{1}{2}\left(\partial_\beta\partial_\mu \delta g_{\alpha\nu}+\partial_\alpha\partial_\nu\delta g_{\beta\mu}-\partial_\alpha\partial_\mu\delta g_{\beta\nu}-\partial_\beta\partial_\nu\delta g_{\alpha\mu}\right)\,,
\end{equation}
and the Riemann square is given as
\begin{equation}
    R_{\alpha\beta\gamma\delta}R^{\alpha\beta\gamma\delta}= \partial_\alpha \partial_\beta \delta g_{\gamma\delta}\partial^\alpha\partial^\beta\delta g^{\gamma\delta} -2\partial_\beta\partial_\gamma \delta g_{\alpha\delta} \partial^\alpha \partial^\gamma \delta g^{\beta\delta}+\partial_\alpha\partial_\gamma \delta g_{\beta\delta} \partial^\beta\partial^\delta\delta g^{\alpha\gamma}\,.
\end{equation}
Note that for a complex K\"ahler manifold, with treating $\delta g_{ab}$ as the first order term of the metric in the normal coordinate expansion, we have
\begin{equation}
    2\partial_\beta\partial_\gamma \delta g_{\alpha\delta}\partial^\alpha\partial^\gamma\delta g^{\beta\delta}-\partial_\alpha\partial_\gamma\delta g_{\beta\delta}\partial^\beta\partial^\delta \delta g^{\alpha\gamma}=0\,.
\end{equation}

\section{Background solution and constant dilaton}\label{app:diff2}
In the background solution obtained in \S\ref{sec:back CY}, the spacetime covariance is not explicit as the supergravity is related to the string field via non-trivial field redefinition. In particular, the background solution we aim to find is a constant dilaton background that preserves supersymmetry. In this section, we shall find the constraint on the background solution that leads to the constant physical dilaton by using a D(-1) instanton probe. Note that we shall use $a,~b,\dots$ to denote real indices in this appendix.

We shall evaluate the tension of the D(-1)-instanton to the second order in the large radius expansion, and the goal is to show that the D(-1) instanton action stays constant. 

Let us write $(0,-1)$ picture and $(-1,0)$ picture first order background solutions
\begin{equation}
    \Psi_{0,1}^{-1,0}=-i\sqrt{\frac{2}{\alpha'}} \frac{1}{4\pi}\delta g_{AB} c\bar{c} \partial X^A e^{-\bar{\phi}}\bar{\psi}^B+i\sqrt{\frac{\alpha'}{2}}\frac{1}{4\pi} \partial_C \delta g_{AB} c\bar{c} \psi^C\psi^Ae^{-\bar{\phi}}\bar{\psi}^B+\dots \,,
\end{equation}
\begin{equation}
    \Psi_{0,1}^{0,-1}=-i\sqrt{\frac{2}{\alpha'}} \frac{1}{4\pi} \delta g_{AB}c\bar{c} e^{-\phi}\psi^A \bar{\partial}X^B+i\sqrt{\frac{\alpha'}{2}} \frac{1}{4\pi} \partial_C \delta g_{AB}e^{-\phi}\psi^A \bar{\psi}^C\bar{\psi}^B+\dots\,.
\end{equation}

We shall first compute 
\begin{equation}
    \frac{1}{2}\{\Psi_{0,1}^2\}_{D^2}\,.
\end{equation}
An efficient way to evaluate the above string vertex is to treat $\delta g_{AB}$ as
\begin{equation}
\delta g_{AB}=\epsilon_{AB} e^{ik\cdot X}\,,
\end{equation}
compute the CFT correlator, and perform the expansion in momenta to the second order. We shall insert the $(-1,0)$ picture vertex at $z=i,$ and the $(0,-1)$ picture vertex at $z=iy.$ In the large stub limit, the integration domain is given by, as we studied in \S\ref{app:disk CC},
\begin{equation}
    \mu^{-2}\leq y\leq 1-2\lambda^{-2}+\dots\,.
\end{equation}
We shall, for simplicity, assume that metric and derivatives are all along the parallel directions.

In $\alpha'=2$ unit, before expanding in momentum, $\{\Psi_{0,1}^2/2\}_{D^2}$ modulo the vertical integration reads 
\begin{align}
\frac{1}{2}\{\Psi_{0,1}^2\}_{D^2}=&\frac{C_{D^2}}{8\pi^2} \int_{\mu^{-2}}^{1-2(\lambda\lambda_o)^{-1}+\dots} dy (1-y^2) \left(\frac{1-y}{1+y}\right)^{-2\partial_1\cdot \partial_2}  \biggr[ \frac{\delta g_{ab}\delta g^{ab}}{(1+y)^4}+ \frac{4y\partial^d \delta g_{ab} \partial^a\delta g^b_{~d}}{(1-y)^2(1+y)^4}\nonumber\\
&-\frac{4y\partial^d \delta g_{ab} \partial^a\delta g^b_{~d}}{(1-y)^2(1+y)^4}+ \frac{4y\partial^c\delta g_{ab}\partial_c\delta g^{ab}}{(1-y)^2(1+y)^4}\biggr]\,,\\
=&\frac{C_{D^2}}{8\pi^2} \int_{\mu^{-2}}^{1-2(\lambda\lambda_o)^{-1}+\dots} dy (1-y^2) \left(\frac{1-y}{1+y}\right)^{-2\partial_1\cdot \partial_2}  \biggr[ \frac{\delta g_{ab}\delta g^{ab}}{(1+y)^4}+ \frac{4y\partial^c\delta g_{ab}\partial_c\delta g^{ab}}{(1-y)^2(1+y)^4}\biggr]\,,\\
=& \frac{C_{D^2}}{32\pi^2} \left( \delta g_{ab}\delta g^{ab} +2\log( \lambda\lambda_0)\partial_c\delta g_{ab}\partial^c\delta g_{ab}+2\partial_c\partial_d\delta g_{ab}\partial^c\partial^d\delta g^{ab}\log^2\lambda\lambda_0\right)\,.
\end{align}

Let us now compute the vertical integration. Despite the fact that picture numbers are correctly distributed the precise location of the PCOs is off by $\mathcal{O}(\lambda^{-2}).$ Combined with the fact that the leading singularity due to the collision of $(-1,0)$ and $(-1,-1)$ vertices scales as $\lambda^2,$ the vertical integration can produce a non-trivial result. Following \cite{Sen:2015hia}, the vertical integration can be obtained by replacing the factor 
\begin{equation}
    \mathcal{X}(i) \mathcal{X}(-iy) idy \left( -\oint b(w)dw -\oint \bar{b}(\bar{w})d\bar{w}\right)\,, 
\end{equation}
with
\begin{equation}
    (\xi(i)-\xi(W_1))\mathcal{X}(-iy_0)+\mathcal{X}(W_1)(\xi(-iy_0)-\xi(W_2))\,,
\end{equation}
where 
\begin{equation}
    y_0=1-2(\lambda\lambda_0)^{-1}+\dots\,,
\end{equation}
\begin{equation}
    W_1=i-i (3\pm 2\sqrt{3})(\lambda\lambda_0)^{-1}+\dots\,,
\end{equation}
\begin{equation}
    W_2= -i+i (3\pm2\sqrt{3})(\lambda\lambda_0)^{-1}+\dots\,.
\end{equation}
After permuting over different PCO locations and vertical integration paths, we find
\begin{equation}
\mathcal{B}= -\frac{C_{D^2}}{64\pi^2} \left(\lambda\lambda_0+\dots\right)^{2\partial_1\cdot \partial_2} \delta g_{ab}\delta g^{ab}\,.
\end{equation}
As a result, we find that the full disk two-point function is given as
\begin{equation}\label{eqn:disk two d(-1)}
    \frac{1}{2}\{\Psi_{0,1}^2\}_{D^2}=  \frac{C_{D^2}}{64\pi^2} \left( \delta g_{ab}\delta g^{ab} +2\log (\lambda\lambda_0)\partial_c\delta g_{ab}\partial^c\delta g_{ab}+2\partial_c\partial_d\delta g_{ab}\partial^c\partial^d\delta g^{ab}\log^2(\lambda\lambda_0)
    \right)\,.
\end{equation}
Had we treated $X$ as the classical field, we would've instead obtained
\begin{equation}
    \frac{1}{2}\{\Psi_{0,1}^2\}_{D^2}=\frac{C_{D^2}}{64\pi^2} (\delta g_{ab}\delta g^{ab} -2\partial_c\delta g_{ab}\partial^c\delta g^{ab} (1-
    2\log(\lambda \lambda_o))\,.
\end{equation}

We then now compute $\{\Psi_{0,2}\}_{D^2}\,,$
\begin{align}
    \{\Psi_{0,2}\}_{D^2}=&\frac{1}{2\pi} \left\langle c_0^-\left( \frac{1}{2}\lambda_o\right)^{\partial^2}\Psi_{0,2}\right\rangle\,,\\
    =&\frac{C_{D^2}}{2\pi} \lambda_o^{\partial^2} \left(\mathcal{G}_{AB}\eta^{AB}+2\mathcal{D} \right)\,,\\
    =&\frac{C_{D^2}}{2\pi } (1+\log \lambda_o \partial^2+\log^2\lambda_o (\partial^2)^2/2) \left(\mathcal{G}_{AB}\eta^{AB}+2\mathcal{D}\right)\,.
\end{align}

We shall now solve for $\mathcal{G}_{AB}$ and $\mathcal{D}$ such that the on-shell action of the D(-1)-instanton vanishes at the second order. We shall decompose $\mathcal{G}_{AB}$ and $\mathcal{D}$ into a $Q_B$ closed component and the rest
\begin{equation}
    \mathcal{G}_{AB}= W_{AB}+\delta \mathcal{G}_{AB} +a\frac{C_{S^2}}{2\pi^2} \delta g_A^{~C}\delta g_{CB}\,,
\end{equation}
\begin{equation}
    \mathcal{D} =W-\frac{1}{2} \delta \mathcal{G}_{AB}\eta^{AB}-b\frac{C_{S^2}}{8\pi^2}\delta g_{AB}\delta g^{AB}\,,
\end{equation}
where $W_{AB}$ and $W$ depend on the closed string off-shell parameters, whereas the rest of the terms do not.

Then $\{\Psi_{0,2}\}_{D^2}$ is re-written as
\begin{equation}
    \{\Psi_{0,2}\}_{D^2}= \frac{C_{D^2}}{2\pi} \left(1+\log \lambda_o \partial^2 +\log^2 \lambda_o (\partial^2)^2/2\right)\left(W_{AB}\eta^{AB}+2W+\frac{1}{\pi} \left(a-\frac{1}{2}b \right) \delta g_{AB}\delta g^{AB}\right)\,. 
\end{equation}
Note that we used $C_{S^2}=2\pi.$ To cancel the quartic terms in \eqref{eqn:disk two d(-1)}, we shall require 
\begin{equation}
    a-\frac{1}{2} b= -\frac{1}{32}\,.
\end{equation} 
Then we find
\begin{align}
    \frac{1}{2}\{\Psi_{0,1}^2\}_{D^2}+\{\Psi_{0,2}\}_{D^2}=&\frac{C_{D^2}}{64\pi^2} \biggr[2\log \lambda \partial_c\delta g_{ab}\partial^c\delta g_{ab} +2\partial_c\partial_d\delta g_{ab}\partial^c\partial^d\delta g^{ab}(\log^2\lambda+2\log\lambda\log \lambda_o)\nonumber\\
    &+32\pi (1+\log\lambda_o\partial^2+\log^2\lambda_o(\partial^2)^2/2)(W_{AB}\eta^{AB}+2W)\biggr]\,.
\end{align}

We shall require
\begin{equation}
    W_{AB}\eta^{AB}+2W =-\frac{1}{16\pi} \log\lambda \partial_c\delta g_{ab}\partial^c\delta g_{ab} -\frac{1}{16\pi}\partial_c\partial_d\delta g_{ab}\partial^c\partial^d\delta g^{ab}\log^2\lambda\,.
\end{equation}
Then, we conclude
\begin{equation}
    \frac{1}{2}\{\Psi_{0,1}^2\}_{D^2}+\{\Psi_{0,2}\}_{D^2}=0\,.
\end{equation}

We close this appendix with an important remark. If we had treated $X$ as the zero modes, then due to the fact that $\{\Psi_{0,2}\}_{D^2}$ contains $\log^2\lambda_o$ contribution as $\Psi_{0,2}$ is an off-shell field, we cannot find the second order background solution in which the dilaton variation is zero unless $\Psi_{0,2}$ also depends on the off-shell parameter of open string diagrams. If this were to be true, it would imply that the dilaton felt by different D-branes would be different. On the other hand, as we showed in this appendix, if we treat $X$ as the quantum field, then one can find the second-order closed string background with the trivial dilaton variation that is insensitive to the open string off-shell parameters. As a result, we conclude that the right prescription for $X$ is to treat $X$ as the quantum field.
\section{D-particle and metric}\label{app:diff3}
In this section, extending the idea of the previous appendix, we will determine the proper identification of the spacetime metric by using a giant graviton obtained by a Dp-brane wrapped on a p-dimensional cycle with the nontrivial background. We should point out one crucial subtlety, however. In the computation of the on-shell value of the D-brane action in the non-trivial background, the total derivative terms along the worldvolume directions are ambiguous, unlike the derivatives along the transverse directions. Although it is tempting to completely fix the total derivative ambiguity to compute $\alpha'^2$ corrected on-shell value of the worldvolume action, we do not know how to do so. So we shall only fix the terms without the derivatives. We hope to come back to this issue in the future.

We first compute $\{\Psi_{0,1}^2/2\}_{D^2}$
\begin{align}
    \frac{1}{2} \{\Psi_{0,2}^2\}_{D^2}=&\frac{C_{D^2}}{8\pi^2} \int_{\mu^{-2}}^{1-2(\lambda\lambda_o)^{-1}+\dots} dy(1-y^2)  \left[ \frac{\delta g_{ab}\delta g^{ab}}{(1+y)^4}\right]+\mathcal{B}\,,\\
    =&\frac{C_{D^2}}{64\pi^2}\delta g_{ab}\delta g^{ab}
\end{align}
with
\begin{equation}
    \mathcal{B}=-\frac{C_{D^2}}{64\pi^2} \delta g_{ab}\delta g^{ab}\,.
\end{equation}
Note that we dropped the terms that contain derivatives along the worldvolume directions.
\begin{align}
    \{\Psi_{0,2}\}_{D^2}=&\frac{C_{D^2}}{2\pi}\left(-\mathcal{G}_{AB}D^{AB}+2\mathcal{D}\right)\,,\\
    =&\frac{C_{D^2}}{2\pi}\left(-2\mathcal{G}_{\mu\nu}\eta^{\mu\nu}+\mathcal{G}_{AB}\eta^{AB}+2\mathcal{D}\right)\,.
\end{align}

By combining $\{\Psi_{0,1}^2/2\}_{D^2}+\{\Psi_{0,2}\}_{D^2},$ we find
\begin{align}
    \frac{1}{2}\{\Psi_{0,1}^2\}_{D^2}+\{\Psi_{0,2}\}_{D^2}=&\frac{C_{D^2}}{2\pi} (-2\mathcal{G}_{\mu\nu}\eta^{\mu\nu})\,.
\end{align}
To find the second-order background solution that is compatible with the desired non-linear sigma model background, we need the on-shell value of the action at the second order to be
\begin{equation}
    -T_p \int d x^{p+1} e^{-\Phi}\left(1+\frac{1}{2} (h^{(2)})_\mu^\mu-\frac{1}{4} (h^{(1)})^\mu_{~\nu} (h^{(1)})^\nu_{~\mu}\right)\,.
\end{equation}
Note that we write $\mathcal{G}_{\mu\nu}$ as
\begin{equation}
    \mathcal{G}_{\mu\nu}=W_{\mu\nu}+\delta \mathcal{G}_{\mu\nu} +a\frac{1}{\pi} \delta g_{\mu\rho}\delta g^\rho_{~\nu}\,,
\end{equation}
where $W_{\mu\nu}$ contains derivative terms, and $C_{S^2}=2\pi.$ We shall identify
\begin{equation}
    \mathcal{G}_{\mu\nu}=\frac{1}{4\pi} h^{(2)}_{\mu\nu}\,,
\end{equation}
and
\begin{equation}
    \delta g_{\mu\nu}=h^{(1)}_{\mu\nu}\,.
\end{equation}
Then we conclude $a$ shall be normalized as 
\begin{equation}
    a =-\frac{1}{8}\,. 
\end{equation}

\section{Boundary terms and the failure of the BRST decoupling argument}
One subtlety that arises in the development of the patch-by-patch description is that the closure of the local coordinate patches, in general, has a boundary. The existence of the boundary, in the intermediate step of the constructions, combined with the polynomial profiles of various vertex operators that show up in the calculation leads to the failure of the well-known BRST decoupling arguments. Here we show two simple instances of such failures. First, non-trivial dependence on the PCO location of on-shell amplitudes. Second, the failure of the na\"ive gauge invariance. 

\subsection{Dependence on the PCO location}
In this section, we shall take the large stub limit so that an n-point string vertex
\begin{equation}
    \frac{1}{n!}\{\Psi^n\}_{S^2}\,,
\end{equation}
recovers the n-point on-shell amplitudes for 
\begin{equation}
Q_B\Psi=0\,,
\end{equation}
provided that the PCOs are correctly distributed in the degeneration limit. For simplicity, we shall assume that $\Psi$ only contains conformal primaries. Then, the n-point on-shell amplitude is given as
\begin{equation}
    \int_{\mathcal{M}}d^{2n-6}z \frac{1}{n!}\biggr\langle \Psi(0)\otimes \Psi(1)\otimes \Psi(\infty) \otimes \prod_{i=1}^{n-3} \Psi(z_i) \otimes\prod_i^{n_{NS}+n_R/2-2} \mathcal{X}(p_i) \otimes\prod_i^{n_{NS}+n_R/2-2}\overline{\mathcal{X}}(\bar{p}_i)\biggr\rangle_{S^2},,
\end{equation}
where we omitted an appropriate overall numerical factor, and the zero mode integral. A few important remarks are in order. The moduli space $\mathcal{M}$ has a non-trivial boundary component, that corresponds to degeneration limits. To make the on-shell amplitude well defined, we shall require that the PCO locations are non-trivial piece-wise functions of moduli that are correctly distributed on the boundary of the moduli space. 

Now, we shall move a PCO location in the interior of the moduli space while fixing the location of the boundary. The effect of this change from $p_j\mapsto q_j$ is encoded in the following amplitude
\begin{align}\label{eqn:PCO change}
    \int_{\mathcal{M}}d^{2n-6}z \frac{1}{n!}\biggr\langle \Psi(0)\otimes \Psi(1)\otimes \Psi(\infty) \otimes \prod_{i=1}^{n-3} \Psi(z_i) &\otimes\prod_{i\neq j }^{n_{NS}+n_R/2-2} \mathcal{X}(p_i) \otimes\prod_i^{n_{NS}+n_R/2-2}\overline{\mathcal{X}}(\bar{p}_i)\nonumber\\
    &\{Q_B,\xi(p_j)-\xi(q_j)\}\biggr\rangle_{S^2}\,.
\end{align}
Assuming that there is no apparent pole in the BRST current, we can freely deform the BRST contour to encircle the rest of the PCOs and the string field insertions. As $\Psi$ is on-shell, and PCO is defined as
\begin{equation}
    \mathcal{X}:=\{Q_B,\xi\}\,,
\end{equation}
one can conclude that \eqref{eqn:PCO change} vanishes.

However, the BRST contour deformation argument we used is incomplete for the class of problems we are interested in. We will illustrate with a simple example that the argument must break down. We will also present a general analysis of how the BRST current can pick up an additional pole. 

Let us compute a three-point function of three BRST closed operators
\begin{equation}
\mathcal{A}:=    \{V_1\otimes V_2\otimes V_3\}_{S^2}\,,
\end{equation}
where
\begin{equation}
    V_1=\epsilon_{AB}(X_1) c\bar{c} e^{-\phi}\psi^A e^{-\bar{\phi}}\bar{\psi}^B\,,
\end{equation}
\begin{equation}
    V_2=\frac{1}{4\pi}\delta g_{ab}(X_2) c\bar{c} e^{-\phi}\psi^ae^{-\bar{\phi}}\bar{\psi}^b\,,
\end{equation}
and
\begin{equation}
    V_3= \frac{1}{4\pi}\delta g_{ab}(X_3)c\bar{c} e^{-\phi}\psi^a e^{-\bar{\phi}}\bar{\psi}^b\,,
\end{equation}
where we shall require that $V_1,~V_2,$ and $V_3$ are BRST closed. $V_1$ will play the role of the test field, whereas $V_2$ and $V_3$ will play the role of the first order background solution. Hence, we shall require
\begin{equation}
    \partial^2 \epsilon_{AB}=\partial^A \epsilon_{AB}=\partial^B\epsilon_{AB}=0\,,
\end{equation}
and
\begin{equation}
    \delta g_{ab}(X)= -\frac{1}{3} R_{acbd}X^cX^d\,,
\end{equation}
where $R_{acbd}$ is Weyl tensor. To evaluate $\mathcal{A},$ we shall insert one holomorphic PCO and one anti-holomorphic PCO. We shall insert the holomorphic PCO at $p_1,$ and the anti-holomorphic PCO at $p_2.$ We find
\begin{align}
    \mathcal{A}=&\frac{\alpha' C_{S^2}}{32\pi^2}\int_{\bar{U}_i} d^{10}X\prod_i^3 (z_i-p_1)(\bar{z}_i-p_2)\left(\frac{\eta^{ai}\eta^{k\alpha}}{z_{12}z_{3p}}-\frac{\eta^{ak}\eta^{i\alpha}}{z_{13}z_{2p}}+\frac{\eta^{a\alpha}\eta^{ik}}{z_{1p}z_{23}}\right)\nonumber\\
    &\times \left(\frac{\eta^{bj}\eta^{l\beta}}{\bar{z}_{12}\bar{z}_{3p}}-\frac{\eta^{bl}\eta^{j\beta}}{\bar{z}_{13}\bar{z}_{2p}}+\frac{\eta^{b\beta}\eta^{jl}}{\bar{z}_{1p}\bar{z}_{23}}\right)\left(\frac{\partial_{1\alpha}}{z_{1p}}+\frac{\partial_{2\alpha}}{z_{2p}}+\frac{\partial_{3\alpha}}{z_{3p}}\right)\left(\frac{\partial_{1\beta}}{\bar{z}_{1p}}+\frac{\partial_{2\beta}}{\bar{z}_{2p}}+\frac{\partial_{3\beta}}{\bar{z}_{3p}}\right)\nonumber\\
    &\times \epsilon_{ab}(z_1)\delta g_{ij}(z_2)\delta g_{kl}(z_3)\,,
\end{align}
where $\partial_1,~\partial_2,$ and $\partial_3$ are short-hand notations for spacetime derivatives acting on $\epsilon_{ab},~\delta g_{ij},$ and $\delta g_{kl},$ and $z_{ij}:=z_i-z_j,~z_{ip}:=z_i-p_1,~\bar{z}_{ip}:=\bar{z}_i-p_2.$ Note that the contractions between the zero mode of the worldsheet bosons $X$ are treated as total derivative terms. This is because 
\begin{equation}
    \partial_{1a}+\partial_{2a}+\partial_{3a}
\end{equation}
generates a total derivative term. Therefore, we have
\begin{equation}
0=\partial_{3a}\partial_3^a\equiv (\partial_{1a}+\partial_{2a})(\partial^{a}_1+\partial_2^a) =2\partial_{1a}\partial^{a}_2\,,
\end{equation}
modulo total derivative terms.

As one can check, $\mathcal{A}$ depends explicitly on the PCO locations $p_1$ and $p_2.$ We can further simplify the expression given above to identify the source of the PCO dependence. As an example, let us focus on $\eta^{ai}\eta^{k\alpha}$ term that contains the following factors in the holomorphic sector
\begin{equation}\label{eqn:app hf1}
    \prod_i (z_i-p_1) \frac{\eta^{ai}\eta^{k\alpha}}{z_{12}z_{3p}} \left(\frac{\partial_{1\alpha}}{z_{1p}}+\frac{\partial_{2\alpha}}{z_{2p}}+\frac{\partial_{3\alpha}}{z_{3p}}\right) \,.
\end{equation}
We can then use 
\begin{equation}
    \partial_{1\alpha}=-\partial_{2\alpha}+(\partial_{1\alpha}+\partial_{2\alpha}+\partial_{3\alpha})\,,
\end{equation}
and
\begin{equation}
    \partial^k\delta g_{kl}=0\,,
\end{equation}
to rewrite \eqref{eqn:app hf1} as
\begin{equation}
    \eta^{ai}\eta^{k\alpha} \partial_{2\alpha}+\eta^{ai}\eta^{k\alpha} (\partial_1+\partial_2+\partial_3)_\alpha \frac{z_{2p}}{z_{12}}\,.
\end{equation}
As a result, we found that the term that depends on the PCO position can be re-written as the total derivative term. Using this observation, we rewrite $\mathcal{A}$ as
\begin{align}
    \mathcal{A}=&\frac{\alpha' C_{S^2}}{32\pi^2} \int_{\bar{U}_i}d^{10}X \left(\eta^{ai}\partial_2^k-\eta^{ak}\partial_3^i+\eta^{ik}\frac{-\partial_2^a+\partial_3^a}{2}\right)\left(\eta^{bj}\partial_2^l-\eta^{bl}\partial_3^j+\eta^{jl}\frac{-\partial_2^b+\partial_3^b}{2}\right)\nonumber\\
    &\times \epsilon_{ab}\delta g_{ij}\delta g_{kl}+\text{Total derivatives}\,.
\end{align}
Note that only the total derivative terms contain the explicit PCO location dependence.

Breakdown of the $SL(2;C)$ invariance and the BRST decoupling argument presents a serious obstacle towards formulating string field theory with a non-trivial target space. Hence, we should understand the source of the breakdown of both of the properties. 
\section{String vertices and the moduli space}\label{sec:vertices}
In this appendix, we write down the details of a few string vertices.  We shall use the $SL(2;R)$ vertices following \cite{Sen:2019jpm,Sen:2020eck}. Following \cite{Sen:2020eck}, we will denote a closed string puncture by C and an open string puncture by O. 

\subsection{Sphere with C-C-C}
The moduli space for a there-punctured sphere is a point. By using the $SL(2;C),$ we shall place the punctures at $z_1=0,$ $z_2=1,$ and $z_3=\infty.$ To each puncture around $z_i,$ we attach a unit disk parametrized by a coordinate $w_i$ such that
\begin{equation}
w_i=\lambda_c f_i(z)\,,
\end{equation}
where $\lambda_c$ is the closed string stub parameter that will be taken to infinity at the end of the computations. We choose the local coordinates as
\begin{equation}
f_1(z) =\frac{2z}{z-2} \,,\quad f_2(z)=-2\frac{1-z}{1+z} \,,\quad f_3(z)=\frac{2}{1-2z}\,.
\end{equation}
Following \cite{Sen:2019jpm}, we shall place PCOs at symmetric locations that are invariant under the $SL(2;C)$ maps that permute over the closed string punctures
\begin{equation}
p_\pm=\frac{1}{2}\pm i\frac{\sqrt{3}}{2}\,,
\end{equation}
and average over the two choices.

\subsection{Disk with C}
The moduli space of a disk with one closed string puncture is zero-dimensional. We shall define the global coordinate on a unit disk to be $y,$ such that $|y|\leq1,$ and the global coordinate on the upper-half-plane to be $z.$ We shall place the closed string puncture at $y=0.$ The global coordinates $z$ and $y$ are related by
\begin{equation}
    z=i\frac{1-y}{1+y}\,.
\end{equation}
The closed puncture at $y=0$ is mapped to $z=i.$

We shall define the local coordinate around the closed string puncture as
\begin{equation}
    w=\lambda_o y=\lambda_o \frac{i-z}{i+z}\,.
\end{equation}

\subsection{Disk with C-O}
The moduli space of a disk with one closed string puncture and one open string puncture is zero-dimensional. 

We choose the local coordinates around the closed and open string punctures as
\begin{equation}
w=\lambda_o y=\lambda_o\frac{i-z}{i+z}\,,
\end{equation}
and
\begin{equation}
w=\mu z\,.
\end{equation}

If both the closed and the open string punctures are inserted with NS states, we need to insert a PCO. Ideally, we would choose to insert a PCO at a symmetric location. Treating the anti-holomorphic part of the closed string vertex at $i$ as a holomorphic vertex located at $-i,$ we can attempt to place a PCO at the origin where the open string puncture is located. This is a valid choice only if there is no singularity, as PCO comes close to the open string puncture. We can average over PCO locations for generic cases, e.g., $\pm \sqrt{3}/3.$ 

\subsection{Disk with C-C}\label{app:disk CC}
A disk with two closed string punctures has a one-dimensional moduli space. We shall place one closed string puncture at $z=i,$ and the other at $z=ix$ where $0\leq x\leq 1.$ As the calculations we will perform at this order is almost on-shell, we don't need the precise form of the local coordinates but just the range of the parameter $x$ within the string vertex. We shall determine this bound by using the plumbing fixture. 

The disk with two closed string punctures has two degeneration channels: first, it divides into two disks with C-O, and second, into one disk with C and one sphere with C-C-C. 

Let us first study the degeneration channel into two disks with C-O. We shall denote the global coordinates of disks with C-O by $x_i,$ for $i=1,~2.$ We glue those two disks by a plumbing fixture
\begin{equation}
    \mu^2 x_1x_2=-q\,,
\end{equation}
where $0\leq q\leq1.$ As one can check, $x_2=i$ is mapped to $x_1=i q\mu^{-2}.$ We shall declare therefore that the global coordinate $z$ is given by $x_1.$

We shall now study the degeneration channel into one disk with C and one sphere with C-C-C. We shall denote the global coordinate on a disk with C by $x$ and a global coordinate on the sphere by $y.$ We shall glue the closed string puncture at $x=i$ to a closed string puncture at $y=\infty$ via the plumbing fixture
\begin{equation}
    \lambda_c\lambda_o \frac{i-x}{i+x} \frac{2}{1-2y}=q\,,
\end{equation}
where $|q|\leq1.$ We find
\begin{equation}
    x=-i\frac{2+q(\lambda_c\lambda_o)^{-1}-2y+qy(\lambda_c\lambda_o)^{-1}}{-2+q(\lambda_c\lambda_o)^{-1}+2y+qy(\lambda_c\lambda_o)^{-1}}\,,
\end{equation}
and therefore punctures at $y=0,~1$ are mapped to
\begin{equation}
    i\frac{2-q(\lambda_c\lambda_o)^{-1}}{2+q(\lambda_c\lambda_o)^{-1}}\,,
\end{equation}
and
\begin{equation}
    i\frac{2+q(\lambda\lambda_o)^{-1}}{2-q(\lambda\lambda_o)^{-1}}\,,
\end{equation}
respectively. We shall define the global coordinate $z$ as
\begin{equation}
    z:= x \frac{2-q(\lambda_c\lambda_o)^{-1}}{q(\lambda_C\lambda_o)^{-1}+2}\,,
\end{equation}
such that the punctures at $y=0$ and $y=1$ are mapped to
\begin{equation}
    z=i\frac{(q(\lambda_c\lambda_o)^{-1}-2)^2}{(q(\lambda_c\lambda_o)^{-1}+2)^2}\,,
\end{equation}
and
\begin{equation}
    z=i\,,
\end{equation}
respectively. Using the $SL(2;R),$ we can fix $q=e^{-s}.$ 

We, therefore, find that in the vertex region, the region of the movable closed string vertex is given by
\begin{equation}
    z=ix\,,
\end{equation}
with
\begin{equation}
    \mu^{-2}\leq x\leq \frac{(2-(\lambda_c\lambda_o)^{-1})^2}{(2+(\lambda_c\lambda_o)^{-1})^2}\,.
\end{equation}

We also find that PCOs at $y=(1+\pm\sqrt{3})/2$ are mapped to
\begin{equation}
    W= i-i(3\pm2\sqrt{3})(\lambda_c\lambda_o)^{-1}+\dots\,.
\end{equation}

\newpage
\bibliographystyle{JHEP}
\bibliography{refs}
\end{document}